\documentclass[11pt]{article}
\usepackage[T1]{fontenc}
\usepackage{color}
\usepackage{url}
\usepackage{bm}
\usepackage{amsmath}
\usepackage{amssymb,hyperref}
\usepackage{graphicx}
\usepackage[authoryear]{natbib}

\makeatletter
\usepackage[colorinlistoftodos,textwidth=2.1cm]{todonotes}

\providecommand{\tabularnewline}{\\}

\@ifundefined{date}{}{\date{}}

\usepackage{latexsym}
\usepackage{multirow}\usepackage{bm}
\usepackage{url}

\usepackage{enumerate}

\usepackage{colortbl}
\usepackage{subcaption}

\usepackage{dcolumn}
\newcolumntype{.}{D{.}{.}{-1}}
\newcolumntype{d}[1]{D{.}{.}{#1}}
\usepackage{theorem}
\theoremstyle{definition}
\newtheorem{assumption}{Assumption}\newtheorem{example}{Example}\newtheorem{corollary}{Corollary}\newtheorem{theorem}{Theorem}\newtheorem{lemma}{Lemma}

\newcommand{\ind}{\mbox{$\perp\!\!\!\perp$}}

\usepackage{rotating}

\usepackage{arydshln}

\usepackage[compact]{titlesec}

\allowdisplaybreaks

\newcommand{\spacingset}[1]{\renewcommand{\baselinestretch}%
{#1}\small\normalsize}

\newcommand*{\QEDB}{\hfill\ensuremath{\square}}

\newcommand{\logit}{\text{logit}}

\newcommand{\pr}{\mathbb{P}}

\newcommand{\E}{\mathbb{E}}
\newcommand{\LOTE}{\textup{LOTE}}

\newcommand{\ipw}{\textnormal{tp-ps}}

\newcommand{\regta}{\textnormal{tp-om}}
\newcommand{\regps}{\textnormal{ps-om}}

\newcommand{\T}{\mathrm{\scriptscriptstyle T}}

\newcommand{\Mom}{\mathcal{M}_{\textnormal{om}}}
\newcommand{\Mps}{\mathcal{M}_{\textnormal{ps}}}
\newcommand{\Mta}{\mathcal{M}_{\textnormal{tp}}}
\newcommand{\Momps}{\mathcal{M}_{\textnormal{ps}+\textnormal{om}}}
\newcommand{\Momta}{\mathcal{M}_{\textnormal{tp}+\textnormal{om}}}
\newcommand{\Mpsta}{\mathcal{M}_{\textnormal{tp}+\textnormal{ps}}}
\newcommand{\Mompsta}{\mathcal{M}_{\textnormal{tp}+\textnormal{ps}+\textnormal{om} }}

\makeatother

\begin{document}
\title{\textbf{Multiply robust estimation of causal effects under principal ignorability}} 
 
\author{Zhichao Jiang\thanks{Department of Biostatistics and Epidemiology, University of Massachusetts,
Amherst, MA 01003. Email: \href{mailto:zhichaojiang@umass.edu}{zhichaojiang@umass.edu}} \hspace{1.5cm} Shu Yang\thanks{Department of Statistics, North Carolina State University, Raleigh,
NC 27695. Email: \href{mailto:syang24@ncsu.edu}{syang24@ncsu.edu} } \hspace{1.5cm} Peng Ding\thanks{University of California, Berkeley, California, CA 94720. Email: \href{mailto:pengdingpku@berkeley.edu}{pengdingpku@berkeley.edu}}
}

\maketitle

\spacingset{1.5} 
\begin{abstract}
Causal inference concerns not only the average effect of the treatment on the outcome but also the underlying mechanism through an intermediate variable of interest. Principal stratification characterizes such a mechanism by targeting subgroup causal effects within principal strata, which are defined by the joint potential values of an intermediate variable. Due to the fundamental problem of causal inference, principal strata are inherently latent, rendering it challenging to identify and estimate subgroup effects within them. A line of research leverages the principal ignorability assumption that the latent principal strata are mean independent of the potential outcomes conditioning on the observed covariates. Under principal ignorability, we derive various nonparametric identification formulas for causal effects within principal strata in observational studies, which motivate estimators relying on the correct specifications of different parts of the observed-data distribution. Appropriately combining these estimators yields triply robust estimators for the causal effects within principal strata. These triply robust estimators are consistent if two of the treatment, intermediate variable, and outcome models are correctly specified, and moreover, they are locally efficient if all three models are correctly specified. We show that these estimators arise naturally from either the efficient influence functions in the semiparametric theory or the model-assisted estimators in the survey sampling theory. We evaluate different estimators based on their finite-sample performance through simulation and apply them to two observational studies.

\end{abstract}
\newpage{}

\section{Introduction}
Researchers are often interested in understanding the
underlying causal mechanism from the treatment to the
outcome when an intermediate variable is present between them. This requires proper adjustment for the intermediate variable ---
naively conditioning on its observed value does not have a valid causal interpretation unless it is essentially randomized conditional on the treatment and covariates \citep{rosenbaum1984consequences}.
\citet{frangakis2002principal} propose to estimate causal effects
within principal strata, which are defined by the joint potential
values of the intermediate variable under both treatment and control.
Principal strata act as pre-treatment covariates, so the causal effects
within them, often referred to as principal causal effects (PCEs), are conceptually the same as the standard subgroup causal effects. 
PCEs are widely used in
applied statistics to deal with 
noncompliance \citep{angrist1996identification,frumento2012evaluating,mealli2013using},
truncation by death \citep{rubin2006causal,ding2011identifiability,wang2017causal},
missing data \citep{frangakis1999addressing,mattei2014identification},
mediation \citep{rubin2004direct,gallop2009mediation,elliott2010bayesian,mattei2011augmented},
and surrogate evaluation \citep{frangakis2002principal,gilbert2008evaluating,huang2011comparing,li2010bayesian,jiang2016principal}.

Due to the fundamental problem of causal inference, the two potential
values of the intermediate variable are not simultaneously observable,
rendering it challenging to identify
and estimate PCEs without additional assumptions. \citet{angrist1996identification}
establish the nonparametric identification of one PCE, often called the complier average causal effect or the local
average treatment effect, under the monotonicity and the exclusion restriction (ER).
The monotonicity assumes that the treatment changes the intermediate
variable only in one direction for any unit, and the ER assumes that the treatment
affects the outcome only through the intermediate variable. The
identification result of \citet{angrist1996identification} has motivated various estimation methods and
efficiency theories \citep{abadie2003semiparametric,tan2006regression,frolich2007nonparametric,ogburn2015doubly}. Although the ER is a standard assumption, it is not plausible
when the treatment affects the outcome through pathways other than
the intermediate variable. \citet{hirano2000assessing} give an example
of the violation of the ER in a randomized experiment
with noncompliance. Moreover, in mediation, truncation by death, and
principal surrogate evaluation problems, testing the ER
is a scientific question of interest. Thus, we cannot invoke ER
{\it a priori}. Without the ER, \citet{zhang2008evaluating} and \citet{imai2008sharp} derive the large sample bounds on the PCEs, and \citet{li2010bayesian}, \citet{zigler2012bayesian}, and \citet{schwartz2011bayesian} perform model-based Bayesian analyses. Unfortunately, the bounds might be too wide to be informative, whereas the Bayesian analyses could be sensitive to models and priors. Assuming normal linear outcome models within principal strata,
\citet{zhang2009likelihood} and \citet{frumento2012evaluating} estimate 
the PCEs using the likelihood approach, but these analyses can be sensitive to the modeling assumptions and can be unstable even if the models are correctly specified due to the mixture distributions of the observed data \citep{feller2016principal}. Auxiliary covariates or secondary outcomes satisfying additional conditional independence assumptions can help to improve identification and estimation of the PCEs \citep[e.g.,][]{ding2011identifiability,mattei2011augmented,mattei2013exploiting,mealli2013using,jiang2016principal,yang2016using, jiang2020identification}, but those additional assumptions may be hard to justify without prior knowledge.

We focus on an alternative nonparametric identification strategy under
principal ignorability, an assumption in parallel with ignorability
for estimating the average causal effect in observational studies \citep{rosenbaum1983central}.
Principal ignorability assumes that the observed covariates are adequate for controlling for confounding between the principal strata and outcome.
This identification strategy has been popular in applied statistics
\citep{follmann2000effect,hill2002differential,hayden2005estimator,egleston2009estimation,jo2009use,jo2011use,stuart2015assessing,feller2017principal}.

We develop a statistical methodology for estimating the PCEs in both randomized experiments and observational studies under principal ignorability. We first establish three
 identification formulas for each PCE. These formulas motivate
three estimators for each PCE, which rely on correct specifications
of two of the following three models: 
\begin{enumerate}
[(a)] 
\item the model of the treatment conditional on the
covariates, called the \textit{treatment probability}; 
\item the model of the intermediate variable conditional on the treatment
and covariates, called the \textit{principal score} with a little
abuse of terminology; 
\item the model of the mean of the outcome conditional on the treatment,
intermediate variable, and covariates, called the \textit{outcome
mean}. 
\end{enumerate}
The existence of multiple estimators for the same parameter hints at the possibility of a combined estimator for
each PCE. To guide the construction of principled estimators, we derive
the efficient influence functions \citep[EIFs;][]{bickel1993efficient}
for the PCEs under the nonparametric model. These EIFs motivate
novel estimators for PCEs based on the treatment
probability, principal score, and outcome mean. Interestingly, the
novel estimators are triply robust in that they are consistent and
asymptotically Normal if any two of the three models in (a)--(c) are
correctly specified, and locally efficient if all three models are
correctly specified. These results extend the classic doubly robust
estimators for the average causal effect in observational studies
\citep{bang2005doubly} and are similar in spirit to the triply robust
estimators in other contexts of causal inference \citep{tchetgen2012semiparametric,shi2018multiply,wang2018bounded}.
The new triply robust estimators offer additional protection against model misspecification
compared to other non-robust estimators. Finally, we establish an equivalence relationship between the triply robust estimation and the model-assisted estimation, extending the existing results on the average causal effect in observational studies \citep{robins1998jrssbdiscussion, little2004robust, kang2007demystifying, lumley2011connections}.

Previously, \citet{ding2017principal} establish some preliminary results for estimating the PCEs in randomized experiments including an identification formula based on weighting and the corresponding estimators for each PCE with and without adjusting for covariates. Their model-assisted estimator adjusts for covariates but is neither doubly robust nor semiparametrically efficient. So even in randomized experiments, the theory for estimating the PCEs is incomplete. We discuss a broader class of treatment assignments in both randomized experiments and unconfounded observational studies, providing two additional identification formulas for each PCE and proposing more principled estimators based on the EIFs. Our new estimators outperform those in \citet{ding2017principal} and we recommend using them in data analyses.

The rest of this paper proceeds as follows. Section \ref{sec:Notation-and-assumptions}
introduces notation and assumptions for identification. Section \ref{sec:Nonparametric-identification}
presents three different identification formulas and the corresponding
estimators of the PCEs. Section \ref{sec:Semiparametric-efficiency-theory}
derives the EIFs, proposes novel estimators, and shows
the triple robustness of the estimators. 
 Section \ref{sec:Simulation} uses simulation to evaluate
the finite-sample properties of the estimators, and Section \ref{sec:Applications}
applies the novel estimators to two observational studies. 
Section
\ref{sec:Discussion} concludes. The supplementary material contains the technical details including some extensions and the proofs.

\section{Notation and assumptions for principal stratification\label{sec:Notation-and-assumptions}}

Let $Z_{i} \in \{0, 1\}$ be the binary treatment, $S_{i} \in \{0, 1\}$ the binary intermediate
variable, $Y_{i} $ the outcome, and $X_{i} $ a vector of pre-treatment
covariates for unit $i=1,\ldots,n$. We adopt the potential outcomes
framework under the Stable Unit Treatment Value Assumption, and let $S_{iz}$ and $Y_{iz}$ be the potential values
of the intermediate variable and outcome if unit $i$ were to
receive treatment condition $z$ ($z=0,1$). The observed intermediate variable
and outcome are thus $S_{i}=Z_{i}S_{i1}+(1-Z_{i})S_{i0}$ and $Y_{i}=Z_{i}Y_{i1}+(1-Z_{i})Y_{i0}$.
Assume $\{Z_{i},S_{i1},S_{i0},Y_{i1},Y_{i0},X_{i}: i=1,\ldots,n\}$
are independent and identically distributed. Thus, the observed $\{Z_{i},S_{i},Y_{i},X_{i}: i=1,\ldots,n\}$
are also independent and identically distributed. For simplicity,
we drop the subscript $i$ when no confusion arises.

\citet{frangakis2002principal} use the joint potential values of
the intermediate variable to define the principal stratification variable,
$U=(S_{1},S_{0})$. For a binary intermediate variable, $U$ can be $(0,0),(1,0), (0,1),$
and $(1,1)$. For the ease of exposition, we will simplify $(S_{1},S_{0})$ as $S_{1}S_{0}$ throughout the paper. Define the PCE as the average causal effect within a principal
stratum: 
\[
\tau_{s_{1}s_{0}}=\E(Y_{1}-Y_{0}\mid U=s_{1}s_{0}),\quad(s_{1}s_{0}=00,10,11,01).
\]
The scientific meanings of the PCEs vary with the contexts. We review four canonical examples below.

\begin{example}[Noncompliance]\label{eg::noncompliance}
In noncompliance
problems, $Z$ is the treatment assigned, $S$ is the treatment
received, and $Y$ is the outcome. The principal strata $U=(0,0),(1,0), (0,1), (1,1)$ are referred to as
never-takers, compliers, always-takers, and defiers, respectively. 
\citet{angrist1994identification} and \citet{angrist1996identification} propose to estimate $\tau_{10}$, the complier average causal effect, which is also called the local average treatment effect. 
\end{example}

\begin{example}[Truncation by death]\label{eg::truncationbydeath}
In truncation-by-death problems, $Z$ is the treatment, $S$
is the survival status, and $Y$ is often a measure of the quality of life. \citet{rubin2006causal} points out that the
only well-defined causal effect is $\tau_{11}$, which characterizes
the treatment effect for patients who would survive regardless of the
treatment. Other PCEs are not well defined because the quality of life is defined only for survived patients. 
\end{example}

\begin{example}[Mediation]\label{eg::mediation}
In mediation analysis, $S$ is the mediator that lies on the causal pathway from the treatment $Z$ to the outcome $Y$. 
The subgroup effects $\tau_{11}$ and $\tau_{00}$
can assess the direct effect of the treatment on the outcome because the treatment does not change the mediator in these two strata \citep{rubin2004direct,gallop2009mediation,mattei2011augmented}. In contrast, the subgroup effects $\tau_{10}$ and $\tau_{01}$ are less interpretable because they consist of both direct and indirect effects \citep{vanderweele2011principal}. 
\end{example}

\begin{example}[Surrogate evaluation] \label{eg::surrogate}
In surrogate evaluation problems, $S$ is the surrogate candidate for the effect of the treatment $Z$ on the outcome $Y$. \citet{frangakis2002principal} propose the principal surrogate criterion based on ``causal necessity.'' It requires that $Z$ affects $Y$ only if   $Z$ affects $S$, i.e., $\tau_{11}=\tau_{00}=0$. \citet{gilbert2008evaluating} argue that a valid surrogate should also satisfy ``causal sufficiency.'' It requires that if the treatment effect on the surrogate is non-zero, then the treatment effect on the outcome is also non-zero, i.e., $\tau_{10} \neq 0$ and $\tau_{01}\neq 0$. See \citet{jiang2016principal} for a related discussion. 
\end{example}

We will focus on the setting with treatment ignorability for both the intermediate variable and outcome, 
extending the classic treatment ignorability in observational studies.

\begin{assumption}[Treatment ignorability]\label{assump:TAignorability}
$Z\ind(S_{0},S_{1},Y_{0},Y_{1})\mid X$. \end{assumption}

Assumption~\ref{assump:TAignorability} rules out latent confounding
between the treatment and intermediate variable and that between the
treatment and outcome. It holds by the design of a randomized
experiment, where the treatment is independent of all the potential
values and covariates, i.e., $Z\ind(S_{0},S_{1},Y_{0},Y_{1},X)$;
\citet{ding2017principal} focus on this special case. It also holds by the design of a stratified
experiment based on a discrete $X$, where the treatment is independent
of all the potential values within each stratum of $X$. In observational
studies, its plausibility relies on whether or not the observed covariates include all the confounders that affect the treatment as well
as the outcome and intermediate variable.

Since we do not observe $S_{1}$ and $S_{0}$ simultaneously, $U$
is not directly observable. As a result, the PCEs are not identifiable
without additional assumptions. We impose the standard monotonicity
assumption throughout the paper, which helps to identify
the distribution of $U$, even though the individual $U_{i}$'s are
not observed for all units.

\begin{assumption}[Monotonicity]\label{assump:M} $S_{1}\geq S_{0}$. \end{assumption}

Assumption~\ref{assump:M} requires that the treatment has a non-negative
impact on the intermediate variable for all units, which rules out stratum $U=01$. It holds automatically when $S_{0}=0$, e.g., in
one-sided noncompliance problems \citep{sommer1991estimating} and
vaccine trials without immune response under control \citep{follmann2006augmented}.

Under Assumptions~\ref{assump:TAignorability}~and~\ref{assump:M}, two nonparametric identification strategies exist for the PCEs, relying on different additional assumptions. We review them below.

\subsection{Strategy one based on exclusion restriction}

The first strategy assumes the ER:
\begin{eqnarray}
\label{eq::ERassumption}
\tau_{11}=\tau_{00}=0.
\end{eqnarray}
A stronger version of the ER is $Y_1 = Y_0$ for $U=11$ or $00$. 
Under Assumptions~\ref{assump:TAignorability},~\ref{assump:M},~and
\eqref{eq::ERassumption}, 
\citet{angrist1994identification} and 
\citet{angrist1996identification} establish the nonparametric identification
of the complier average causal effect 
\begin{eqnarray*}
\label{eq::CACEidentification}
\tau_{10} = \frac{ \E(Y\mid Z=1) - \E(Y\mid Z=0) }{ \E(S\mid Z=1) - \E(S\mid Z=0) }.
\end{eqnarray*}

By definition, the PCE $\tau_{10}$ represents the effect of the treatment assigned for compliers. Moreover, for compliers with $U=10$, the treatment assigned is identical to the treatment received, so $\tau_{10}$ also measures the effect of the treatment received. This formulation of the noncompliance problem is due to \citet{frangakis2002principal} where the potential outcome $Y_z$ corresponds to the treatment assigned. It corresponds to the intervention $Z$ in the actual experiment without assuming that $S$ is another hypothetical intervention. However, it might cause notational incoherence with \citet{angrist1994identification} and \citet{angrist1996identification}. \citet{angrist1994identification} index the potential outcome $Y_s$ by the treatment received, and thus enforce the ER assumption automatically; \citet{angrist1996identification} index the potential outcome $Y_{zs}$ by both the treatment assigned and received, and reduce it to $Y_s$ under the ER assumption. These different formulations do not cause fundamental differences. An advantage of \citet{frangakis2002principal}'s formulation is its generality to deal with other problems with intermediate variables. In Example \ref{eg::noncompliance} with noncompliance, it allows us to assess the plausibility of the ER by estimating $\tau_{11}$ and $\tau_{00}$; see Section \ref{sec:RTS} for more details.

The ER requires that the treatment has no direct effect
on the outcome, which is sometimes implausible in open-label randomized
experiments. More importantly, it cannot be invoked in problems where
$\tau_{00}$ and $\tau_{11}$ are the quantities of interest, such
as truncation by death in Example \ref{eg::truncationbydeath}, mediation in Example \ref{eg::mediation}, and surrogate evaluation
problems in Example~\ref{eg::surrogate}.

Without the ER, the PCEs are not identifiable. Under weak assumptions, the large-sample bounds on the PCEs are often not informative \citep{zhang2008evaluating, imai2008sharp}. In contrast, Bayesian methods often require specifying strong mixture model assumptions and prior distributions \citep{li2010bayesian, zigler2012bayesian, schwartz2011bayesian}. They are not easy to implement and can be numerically unstable in practice \citep{feller2016principal}. 
Due to these limitations, we focus on another approach assuming
principal ignorability.

\subsection{Strategy two based on principal ignorability}

The principal ignorability can be viewed as the analog of the treatment ignorability assumption in unconfounded observational studies.

\begin{assumption}[Principal ignorability]\label{assump:weak-pi}
$\E(Y_{1}\mid U=11,X)=\E(Y_{1}\mid U=10,X)$ and $\E(Y_{0}\mid U=00,X)=E(Y_{0}\mid U=10,X)$.
\end{assumption}

Assumption~\ref{assump:weak-pi} requires that the expectations of
the potential outcomes do not vary across principal strata conditional
on the covariates. It is widely used in applied statistics \citep{follmann2000effect,hill2002differential,jo2009use,jo2011use,stuart2015assessing}.
Under Assumptions~\ref{assump:TAignorability}~and~\ref{assump:M},
Assumption~\ref{assump:weak-pi} is equivalent to 
\begin{eqnarray}
\E(Y_{1}\mid U=11,Z=1,S=1,X) & = & \E(Y_{1}\mid U=10,Z=1,S=1,X),\label{eqn:pi-1}\\
\E(Y_{0}\mid U=00,Z=0,S=0,X) & = & \E(Y_{0}\mid U=10,Z=0,S=0,X).\label{eqn:pi-0}
\end{eqnarray}
The observed stratum $(Z=1,S=1)$ is a mixture of two principal strata
$U=11,10$. Therefore,~\eqref{eqn:pi-1} means that within the observed
stratum $(Z=1,S=1)$, the expectation of the potential outcome $Y_{1}$
does not vary across the two principal strata conditional on the covariates. So the conditional
expectations in \eqref{eqn:pi-1} simplify to the observable conditional expectation $\E(Y\mid Z=1,S=1,X) $. Similarly,~\eqref{eqn:pi-0} means
that within the observed stratum $(Z=0,S=0)$, the two principal strata
$U=00,10$ are ignorable for the expectation of the potential outcome
$Y_{0}$ conditional on the covariates. So the conditional expectations in~\eqref{eqn:pi-0} simplify to the observable conditional expectation $\E(Y\mid Z=0, S=0, X)$. Intuitively, principal ignorability simplifies a latent mixture problem to an observed mixture problem. With this assumption, we can treat the subpopulation $(Z=z,S=s)$ as a mixture of strata defined by the observed covariates, which is easier to deal with than a mixture of latent principal strata.

We start with Assumptions~\ref{assump:M}~and~\ref{assump:weak-pi} because they allow for deriving simple identification formulas and easy-to-implement estimators. These estimators are numerically stable and statistically robust. They can be benchmark estimators in data analyses. Nevertheless, their plausibility cannot be validated by the observed data, so they should be made with caution. To supplement the theory under Assumptions \ref{assump:M} and \ref{assump:weak-pi}, we also propose corresponding sensitivity analysis techniques for the potential violations of these assumptions. Due to the space limit, we include the theoretical results and numerical examples in the supplementary material. 

\subsection{Principal ignorability and sequential ignorability in mediation analysis}

Before giving the nonparametric identification formulas of the PCEs based on principal ignorability, we comment on its relationship with a commonly-used assumption in mediation analysis. We also make a brief comparison of principal stratification and mediation analysis. 

A stronger version of Assumption~\ref{assump:weak-pi} is $Y_{z}\ind S_{1-z}\mid (S_z, X)$
for $z=0,1$. It assumes that conditional on the covariates, the potential outcome depends only on the potential intermediate variable under the same treatment condition, but not the one under a different treatment condition. Importantly, principal ignorability is different from the \textit{sequential ignorability} between the
intermediate variable and the outcome, which is a common assumption
in mediation analysis \citep{Pearl::2001,imai2010identification, tchetgen2012semiparametric}. In particular, the sequential ignorability assumes away the dependence between the potential outcome and the potential intermediate variable given the covariates, while principal ignorability allows for such dependence, but rules out the dependence between the potential outcome and intermediate variable under different treatment conditions. Hence, the sequential ignorability and principal ignorability focus on different relationships between the potential outcome and the potential intermediate variable and thus do not imply each other. \citet{forastiere2018principal} propose a generalized strong principal ignorability and show that under monotonicity, it is equivalent to the sequential ignorability. Unfortunately, their definition does not imply Assumption~\ref{assump:weak-pi}, and thus it is essentially different from the principal ignorability used in the literature.

In general, principal stratification and mediation analysis can be conceptually different. Principal stratification does not require that $S$ is a well-defined intervention as in Examples \ref{eg::truncationbydeath} and \ref{eg::surrogate}. In contrast, traditional mediation analysis requires that $S$ is a well-defined intervention on the causal pathway from the treatment to the outcome. \citet{vanderweele2011principal} points out this issue, whereas \citet{robins2020interventionist} attempt to relax this assumption with an alternative approach to mediation analysis.

\section{Nonparametric identification and estimation\label{sec:Nonparametric-identification}}
\subsection{Identification formulas} 

To simplify the exposition, define  
\begin{eqnarray*}
\pi(X)=\pr(Z=1\mid X),\quad e_{u}(X)=\pr(U=u\mid X),\quad\mu_{zs}(X)=\E(Y\mid Z=z,S=s,X)
\end{eqnarray*}
for $u=10, 00, 11$ and $z,s=0,1$. The $\pi(X)$ is the treatment
probability given the covariates, also known as the propensity score. The $e_{u}(X)$ is the
principal score which equals the proportion of principal stratum $u$
given the covariates. The $\mu_{zs}(X)$ is the mean of the outcome within the observed group $(Z=z, S=s)$ given the covariates. Let
$\pi=\E\{\pi(X)\}=\pr(Z=1)$ and $e_{u}=\E\{e_{u}(X)\}$ denote the
marginalized treatment probability and principal score over the distribution
of the covariates, respectively. Thus, $\pi$ represents the proportion of treated
units and $e_{u}$ represents the proportion of units with $U=u$.

\begin{table}[t]
\centering \caption{Principal strata in the observed strata defined by $(Z,S)$ under
monotonicity}
\begin{tabular}{cll}
\hline 
 & $S=0$ & $S=1$ \tabularnewline
\cdashline{2-3} $Z=0$ & $U\in\{00,10\}$ & $U=11$\tabularnewline
$Z=1$ & $U=00$ & $U\in\{11,10\}$\tabularnewline
\hline 
\end{tabular}\label{tab:ZS} 
\end{table}

Under Assumption~\ref{assump:M}, Table~\ref{tab:ZS} shows the relationship between
the observed strata defined by $(Z,S)$ and the principal strata.
So under Assumptions~\ref{assump:TAignorability}~and~\ref{assump:M},
the principal scores are identified by 
\[
e_{10}(X)=p_{1}(X)-p_{0}(X), \quad e_{00}(X)=1-p_{1}(X), \quad e_{11}(X)=p_{0}(X),
\]
where $p_{z}(X)=\pr(S=1\mid Z=z,X)$ is the probability of
the intermediate variable conditional on the treatment and covariates. Analogously, the proportions of principal strata are identified
by 
\[
e_{10}=p_{1}-p_{0},\quad e_{00}=1-p_{1},\quad e_{11}=p_{0},
\]
where $p_{z}=\E\{p_{z}(X)\}$ is the marginalized probability of the
intermediate variable over the distribution of the covariates. 
Due to the one-to-one mapping between $\{p_{1}(X),p_{0}(X)\}$ and
$\{e_{11}(X),e_{00}(X),e_{10}(X)\}$, we call both sets the principal
score, and the exact meaning should be clear from the context.
 The following theorem provides three identification formulas for each
PCE.

\begin{theorem}[Nonparametric identification] \label{thm:identification-obs}
 Suppose that Assumptions~\ref{assump:TAignorability}--\ref{assump:weak-pi} hold, $e_u>0$ for $u=10,00,11$, and $0<\pi(x)<1$ for all $x$ in the support of $X$. The following identification formulas hold for the PCEs. 
\begin{enumerate}
	[(a)] 
\item Based on the treatment probability and principal score,
\begin{eqnarray*}
\tau_{10} & = & \E\left\{ \frac{e_{10}(X)}{p_{1}-p_{0}}\frac{S}{p_{1}(X)}\frac{Z}{\pi(X)}Y\right\} -\E\left\{ \frac{e_{10}(X)}{p_{1}-p_{0}}\frac{1-S}{1-p_{0}(X)}\frac{1-Z}{1-\pi(X)}Y\right\},\\
\tau_{00} & = & \E\left\{ \frac{1-S}{1-p_{1}}\frac{Z}{\pi(X)}Y\right\} -\E\left\{ \frac{e_{00}(X)}{1-p_{1}}\frac{1-S}{1-p_{0}(X)}\frac{1-Z}{1-\pi(X)}Y\right\} ,\\
\tau_{11} & = & \E\left\{ \frac{e_{11}(X)}{p_{0}}\frac{S}{p_{1}(X)}\frac{Z}{\pi(X)}Y\right\} -\E\left\{ \frac{S}{p_{0}}\frac{1-Z}{1-\pi(X)}Y\right\}.
\end{eqnarray*}
\item Based on the treatment probability and outcome mean, 
\begin{eqnarray*}
\tau_{10} & = & \E\left[\frac{SZ/\pi(X)-S(1-Z)/\{1-\pi(X)\}}{p_{1}-p_{0}}\left\{ \mu_{11}(X)-\mu_{00}(X)\right\} \right],\label{eq:res2}\\
\tau_{00} & = & \E\left[\frac{1-SZ/\pi(X)}{1-p_{1}}\{\mu_{10}(X)-\mu_{00}(X)\}\right],\nonumber \\
\tau_{11} & = & \E\left[\frac{S(1-Z)/\{1-\pi(X)\}}{p_{0}}\{\mu_{11}(X)-\mu_{01}(X)\}\right].\nonumber 
\end{eqnarray*}
\item Based on the principal score and outcome mean,
\begin{eqnarray*}
\tau_{10} & = & \E\left[\frac{p_{1}(X)-p_{0}(X)}{p_{1}-p_{0}}\{\mu_{11}(X)-\mu_{00}(X)\}\right],\label{eq:res3}\\
\tau_{00} & = & \E\left[\frac{1-p_{1}(X)}{1-p_{1}}\{\mu_{10}(X)-\mu_{00}(X)\}\right],\nonumber \\
\tau_{11} & = & \E\left[\frac{p_{0}(X)}{p_{0}}\{\mu_{11}(X)-\mu_{01}(X)\}\right].\nonumber 
\end{eqnarray*}
\end{enumerate}
\end{theorem}

Theorem~\ref{thm:identification-obs} gives identification formulas
for the PCEs based on three different combinations of the likelihood
components. Theorem~\ref{thm:identification-obs}(a) is an extension of \citet{ding2017principal} with an additional weighting term based on the inverse of the treatment probability, which is also mentioned by \citet{jiang2020identification}. Theorem~\ref{thm:identification-obs}(b)~and~(c) are two additional sets of identification formulas.


Below we give some intuition based on only $\tau_{10}$ since the discussion for the other two PCEs is similar. Theorem~\ref{thm:identification-obs}(a) expresses $\tau_{10}$ as the difference between weighted averages of the outcome under the treatment
and control. The weights in the formula consist of two parts:
 $Z/\pi(X)$
and $(1-Z)/\{1-\pi(X)\}$ correspond to the treatment probability;
$e_{10}(X) S/p_{1}(X)$ and $e_{10}(X)(1-S)/\{1-p_{0}(X)\}$ correspond to the principal score. 
Under Assumptions~\ref{assump:TAignorability}~and~\ref{assump:M}, the conditional expectations of the weights equal
\begin{eqnarray}
\label{eqn:thm1a-weight}\E\left\{ \frac{e_{10}(X)}{p_{1}-p_{0}}\frac{S}{p_{1}(X)}\frac{Z}{\pi(X)} \mid X\right\}\ =\ \E\left\{ \frac{e_{10}(X)}{p_{1}-p_{0}}\frac{1-S}{1-p_{0}(X)}\frac{1-Z}{1-\pi(X)} \mid X\right\} &=& \frac{e_{10}( X) } {e_{10}},
\end{eqnarray} 
i.e., the conditional probability of principal stratum $U=10$ divided by its unconditional probability.

Theorem~\ref{thm:identification-obs}(b) expresses $\tau_{10}$ in
terms of the treatment probability and outcome mean. Under principal
ignorability, the difference between the outcome means equals 
$$
\mu_{11}(X)-\mu_{00}(X)=\E(Y_{1}\mid U=10,X)-\E(Y_{0}\mid U=10,X)
= \E(Y_{1} - Y_0 \mid U=10,X), 
$$
which is the PCE for stratum $U=10$ conditional on $X$. Under Assumptions~\ref{assump:TAignorability}~and~\ref{assump:M},
the conditional expectation of the unnormalized weight equals $\E[SZ/\pi(X)-S(1-Z)/\{1-\pi(X)\}\mid X]=e_{10}(X)$.
Dividing this by the normalizing constant, $p_{1}-p_{0}=e_{10}$,
yields $e_{10}(X)/e_{10}$, which, by Bayes' Theorem, is the density
ratio of $X$ conditional and unconditional on $U=10$. Therefore,
the identification formula for $\tau_{10}$ in Theorem~\ref{thm:identification-obs}(b)
averages the conditional PCE over the distribution of $X$ given
$U=10$, which gives the PCE within stratum $U=10$. 

Theorem~\ref{thm:identification-obs}(c) expresses $\tau_{10}$ in
terms of the principal score and outcome mean. Compared with Theorem~\ref{thm:identification-obs}(b),
the treatment probability weighting is replaced with the principal
score weighting $\{p_{1}(X)-p_{0}(X)\}/(p_{1}-p_{0})$. Under Assumptions~\ref{assump:TAignorability}~and~\ref{assump:M},
the weight again equals $e_{10}(X)/e_{10}$. Therefore, similar to
the discussion of Theorem~\ref{thm:identification-obs}(b), Theorem~\ref{thm:identification-obs}(c)
identifies $\tau_{10}$ by averaging the conditional PCE over
the distribution of $X$ given $U=10$.

\subsection{Estimators based on the nonparametric identification formulas}\label{section::3estimators}

For each PCE, the three identification formulas in Theorem~\ref{thm:identification-obs}
motivate three estimators, which require correct specifications of
different parts of the observed-data distribution. For descriptive convenience,
we introduce additional notation. Let $\mathbb{P}_{n}$ denote the
empirical average, e.g., $\mathbb{P}_{n}h(V)=n^{-1}\sum_{i=1}^{n}h(V_{i})$ for any $h(V)$.
Let $\widehat{\pi}=\mathbb{P}_{n}Z$ be the moment estimator of $\pi$.
Let $\pi(X;\alpha)$ be a working parametric model for the treatment probability
$\pi(X)$, $p_{z}(X;\gamma)$ a working parametric model for the principal score
$p_{z}(X)$ for $z=0,1$, and $\mu_{zs}(X;\beta)$ a working parametric model
for the outcome mean $\mu_{zs}(X)$ for $z,s=0,1$. 
Because $e_{u}(X)$
has a one-to-one mapping to $p_{z}(X)$, we use $e(X;\gamma)$ to
denote a working parametric model for $e_{u}(X)$. We focus on parametric models here and will consider more flexible estimation strategies later. 
 Based on the maximum likelihood
estimation or the method of moments, we obtain estimators $\widehat{\alpha}$,
$\widehat{\gamma}$, and $\widehat{\beta}$. Assume they have probability limits
$\alpha^{*}$, $\gamma^{*}$, and $\beta^{*}$, respectively. We use
$\mathcal{M}$ with subscripts ``tp,'' ``ps,'' and ``om'' to
denote models with the correct specification of the treatment probability,
principal score, and outcome mean, respectively. Therefore, under
$\Mta$, we have $\pi(X;\alpha^{*})=\pi(X)$; under $\Mps$, we have $p_{z}(X;\gamma^{*})=p_{z}(X)$
and $e_{u}(X;\gamma^{*})=e_{u}(X)$ for $z=0,1$ and $u=10,00,11$;
under $\Mom$, we have $\mu_{zs}(X;\beta^{*})=\mu_{zs}(X)$ for $z,s=0,1$.
In addition, we use ``+'' in the subscript to indicate that more
than one model is correctly specified. For example, $\Momps$ denotes
the model with correctly specified $p_{z}(X;\gamma)$ and $\mu_{zs}(X;\beta)$.
We also use the union notation from the standard set theory to denote
the correct specification of at least one model, for example, $\Mta\cup\Momps$
denotes the model with correctly specified $\pi(X;\alpha)$ or $\{p_{z}(X;\gamma),\mu_{zs}(X;\beta)\}$.

To obtain the estimators based on Theorem~\ref{thm:identification-obs},
we need to replace the components in the identification formulas with
their estimated counterparts, and the expectations with the empirical
averages. We use $\{\pi(X;\widehat{\alpha}),p_{z}(X;\widehat{\gamma}),\mu_{zs}(X;\widehat{\beta})\}$
to denote the estimated version of $\{\pi(X;\alpha),p_{z}(X;\gamma),\mu_{zs}(X;\beta)\}$.
For $p_{1}$ and $p_{0}$, we can simply use $\mathbb{P}_{n}\{p_{1}(X;\widehat{\gamma})\}$
and $\mathbb{P}_{n}\{p_{0}(X;\widehat{\gamma})\}$ as the estimators, which are consistent under $\Mps$. 
Moreover, the doubly robust estimators \citep{bang2005doubly}, 
\[
\widehat{p}_{1}=\mathbb{P}_{n}\left[\frac{Z\{S-p_{1}(X;\widehat{\gamma})\}}{\pi(X;\widehat{\alpha})}+p_{1}(X;\widehat{\gamma})\right],\quad\widehat{p}_{0}=\mathbb{P}_{n}\left[\frac{(1-Z)\{S-p_{0}(X;\widehat{\gamma})\}}{1-\pi(X;\widehat{\alpha})}+p_{0}(X;\widehat{\gamma})\right],
\]
improve them, which are consistent for $p_{1}$ and $p_{0}$ under
$\Mta\cup\Mps$.

The identification formulas in Theorem~\ref{thm:identification-obs}(a)
motivate the following weighting estimators based on the treatment
probability and principal score. 

\begin{example}\label{eg1-weighting} The treatment probability--principal
score (tp-ps) estimators are 
\begin{eqnarray*}
\widehat{\tau}_{10,\ipw} & = & \mathbb{P}_{n}\left\{ \frac{e_{10}(X;\widehat{\gamma})}{\widehat{p}_{1}-\widehat{p}_{0}}\frac{S}{p_{1}(X;\widehat{\gamma})}\frac{Z}{\pi(X;\widehat{\alpha})}Y-\frac{e_{10}(X;\widehat{\gamma})}{\widehat{p}_{1}-\widehat{p}_{0}}\frac{1-S}{1-p_{0}(X;\widehat{\gamma})}\frac{1-Z}{1-\pi(X;\widehat{\alpha})}Y\right\} ,\label{eq:w1}\\
\widehat{\tau}_{00,\ipw} & = & \mathbb{P}_{n}\left\{ \frac{1-S}{1-\widehat{p}_{1}}\frac{Z}{\pi(X;\widehat{\alpha})}Y-\frac{e_{00}(X;\widehat{\gamma})}{1-\widehat{p}_{1}}\frac{1-S}{1-p_{0}(X;\widehat{\gamma})}\frac{1-Z}{1-\pi(X;\widehat{\alpha})}Y\right\} ,\nonumber \\
\widehat{\tau}_{11,\ipw} & = & \mathbb{P}_{n}\left\{ \frac{e_{11}(X;\widehat{\gamma})}{\widehat{p}_{0}}\frac{S}{p_1(X;\widehat{\gamma})}\frac{Z}{\pi(X;\widehat{\alpha})}Y-\frac{S}{\widehat{p}_{0}}\frac{1-Z}{1-\pi(X;\widehat{\alpha})}Y\right\} .\nonumber 
\end{eqnarray*}
\end{example}

The weighting estimators in Example~\ref{eg1-weighting} involve
the inverse of the treatment probability. Thus, they may be unstable
if some estimated treatment probabilities are close to zero or one.
A strategy to mitigate this issue is to stabilize the estimators by normalizing the weights \citep{hernan2001marginal}.
For example, the stabilized weighting estimator of $\tau_{11}$ is
\begin{eqnarray*}
\widehat{\tau}'_{11,\ipw} & = & \mathbb{P}_{n}\left\{ e_{11}(X;\widehat{\gamma})\frac{S}{p_1(X;\widehat{\gamma})}\frac{Z}{\pi(X;\widehat{\alpha})}Y\right\} \bigg/\mathbb{P}_{n}\left\{ e_{11}(X;\widehat{\gamma})\frac{S}{p_1(X;\widehat{\gamma})}\frac{Z}{\pi(X;\widehat{\alpha})}\right\} \label{eq:w2}\\
 & & -\mathbb{P}_{n}\left\{ \frac{\textup{s}(1-Z)}{1-\pi(X;\widehat{\alpha})}Y\right\} \bigg/\mathbb{P}_{n}\left\{ \frac{\textup{s}(1-Z)}{1-\pi(X;\widehat{\alpha})}\right\} .\nonumber 
\end{eqnarray*}
The stabilized weighting estimators for $\tau_{10}$ and $\tau_{00}$ have similar forms. 
The estimators $\widehat{\tau}_{u,\ipw}$ are consistent under $\Mpsta$, i.e., correct
specifications of the treatment probability and principal score. However,
if either model is incorrectly specified, they are inconsistent.

The identification formulas in Theorem~\ref{thm:identification-obs}(b)
motivate the following estimators based on the treatment probability
and the outcome mean.

\begin{example}\label{eg2-reg} The treatment probability--outcome
mean (tp-om) estimators are 
\begin{eqnarray*}
\widehat{\tau}_{10,\regta} & = & \mathbb{P}_{n}\left[\frac{ZS/\pi(X;\widehat{\alpha})-(1-Z)S/\{1-\pi(X;\widehat{\alpha})\}}{\widehat{p}_{1}-\widehat{p}_{0}}\left\{ \mu_{11}(X;\widehat{\beta})-\mu_{00}(X;\widehat{\beta})\right\} \right],\label{eq:reg1}\\
\widehat{\tau}_{00,\regta} & = & \mathbb{P}_{n}\left[\frac{Z(1-S)/\pi(X;\widehat{\alpha})}{1-\widehat{p}_{1}}\left\{ \mu_{10}(X;\widehat{\beta})-\mu_{00}(X;\widehat{\beta})\right\} \right],\nonumber \\
\widehat{\tau}_{11,\regta} & = & \mathbb{P}_{n}\left[\frac{(1-Z)S/\{1-\pi(X;\widehat{\alpha})\}}{{\widehat{p}_{0}}}\left\{ \mu_{11}(X;\widehat{\beta})-\mu_{01}(X;\widehat{\beta})\right\} \right].\nonumber 
\end{eqnarray*}
\end{example} 
Similar to the estimators in Example~\ref{eg1-weighting},
we can also obtain the stabilized weighting version of the estimators
in Example~\ref{eg2-reg}. For example, the stabilized version of
$\widehat{\tau}_{11,\regta}$ is 
\begin{eqnarray*}
\widehat{\tau}'_{11,\regta}=\mathbb{P}_{n}\left[\frac{(1-Z)S}{1-\pi(X;\widehat{\alpha})}\left\{ \mu_{11}(X;\widehat{\beta})-\mu_{01}(X;\widehat{\beta})\right\} \right]\bigg/\mathbb{P}_{n}\left\{ \frac{(1-Z)S}{1-\pi(X;\widehat{\alpha})}\right\} .
\end{eqnarray*}
The estimators $\widehat{\tau}_{u,\regta} $ are consistent under $\Momta$.

The identification formulas in Theorem~\ref{thm:identification-obs}(c)
motivate the following estimators based on the principal score and
outcome mean.

\begin{example}\label{eg3-reg} The principal score--outcome mean
(ps-om) estimators are 
\begin{eqnarray*}
\widehat{\tau}_{10,\regps} & = & \mathbb{P}_{n}\left[\frac{p_{1}(X;\widehat{\gamma})-p_{0}(X;\widehat{\gamma})}{\widehat{p}_{1}-\widehat{p}_{0}}\left\{ \mu_{11}(X;\widehat{\beta})-\mu_{00}(X;\widehat{\beta})\right\} \right],\label{eq:reg2}\\
\widehat{\tau}_{00,\regps} & = & \mathbb{P}_{n}\left[\frac{1-p_{1}(X;\widehat{\gamma})}{1-\widehat{p}_{1}}\left\{ \mu_{10}(X;\widehat{\beta})-\mu_{00}(X;\widehat{\beta})\right\} \right],\nonumber \\
\widehat{\tau}_{11,\regps} & = & \mathbb{P}_{n}\left[\frac{p_{0}(X;\widehat{\gamma})}{\widehat{p}_{0}}\left\{ \mu_{11}(X;\widehat{\beta})-\mu_{01}(X;\widehat{\beta})\right\} \right].\nonumber 
\end{eqnarray*}

\end{example} 
Th estimators $\widehat{\tau}_{u,\regps}$ are consistent under $\Momps$.

\section{From the EIFs to triply robust estimators\label{sec:Semiparametric-efficiency-theory}}

Theorem~\ref{thm:identification-obs} presents three identification formulas, which motivate infinitely many estimators for each PCE. This calls for the construction of more principled estimators. In this section, we derive the EIF for each PCE to motivate a new estimator. The EIFs below are derived under the nonparametric model of the observed data distribution, which is a standard strategy in the literature. In particular, the derivation ignores the restrictions implied by the monotonicity assumption \citep[cf.][]{frolich2007nonparametric,hong2010semiparametric}. For simplicity, we use the terminology ``EIF'' throughout.

\subsection{EIFs and the resulting estimators}
\label{sec:score} 

Because the PCEs have a ratio form $\tau_u = \E\{(Y_1-Y_0)\bm{1}(U=u)\}/\pr(U=u)$, we will first define a general quantity to represent the EIFs of the numerators and denominators, and then combine them to have the EIFs for the PCEs.

Define the following
quantity for any function $f(Y,S,X)$:
\begin{eqnarray}\label{eqn:generic}
\psi_{f(Y_{z},S_{z},X)} = \frac{\bm{1}(Z=z) \left[f(Y,S,X)-\E\{f(Y,S,X)\mid X,Z=z\}\right] }{\pr(Z=z\mid X)} 
+\E\{f(Y,S,X)\mid X,Z=z\}.  
\end{eqnarray} 
Under Assumption~\ref{assump:TAignorability}, we can show that $\E\{\psi_{f(Y_{z},S_{z},X)}\}=\E\{f(Y_{z},S_{z},X)\}$.
In fact, $\psi_{f(Y_{z},S_{z},X)}-\E\{f(Y_{z},S_{z},X)\}$ is the
EIF for $\E\{f(Y_{z},S_{z},X)\}$; see Lemma \ref{lem:marginalp0} in the supplementary material.
With $f(Y,S,X)=S$,~\eqref{eqn:generic} reduces to 
\begin{eqnarray*}
\psi_{S_{z}} & = & \frac{\bm{1}(Z=z) \{S-p_{z}(X)\} }{\pr(Z=z\mid X)} +p_{z}(X),
\end{eqnarray*}
and $\psi_{S_{z}}-\E(S_{z})$ is the EIF for $\E(S_{z})$. This reduces
to a standard result in observational studies \citep{hahn1998role},
which is the foundation for constructing the doubly robust estimator
for $\E(S_{z})$ \citep{bang2005doubly}. With $f(Y,S,X)=YS$ and
$z=0$,~\eqref{eqn:generic} reduces to 
\begin{eqnarray*}
\psi_{Y_{0}S_{0}} & = & \frac{\bm{1}(Z=0) \{YS-\mu_{01}(X)p_{0}(X)\}}{1-\pi(X)} +\mu_{01}(X)p_{0}(X),
\end{eqnarray*}
and $\psi_{Y_{0}S_{0}}-\E(Y_{0}S_{0})$ is the EIF for $\E(Y_{0}S_{0})$,
which equals $\E(Y_{0}\mid U=11)\pr(U=11)$ because $S_{0}=1$ is
equivalent to $U=11$ under monotonicity. Based on the $\psi$ notation in \eqref{eqn:generic},
the following theorem gives the EIFs for the PCEs. 

\begin{theorem}[EIFs]
 \label{thm:SET} 
Suppose $\tau_u$'s are identified in Theorem \ref{thm:identification-obs}.
The EIF for $\tau_{10}$ is $\phi_{10}=\{\phi_{1,10}-\phi_{0,10}-\tau_{10}(\psi_{S_{1}}-\psi_{S_{0}})\}/(p_{1}-p_{0})$,
where 
\begin{eqnarray*}
\phi_{1,10} & = & \frac{e_{10}(X)}{p_{1}(X)}\psi_{Y_{1}S_{1}}-\mu_{11}(X)\left\{ \psi_{S_{0}}-\frac{p_{0}(X)}{p_{1}(X)}\psi_{S_{1}}\right\} ,\\
\phi_{0,10} & = & \frac{e_{10}(X)}{1-p_{0}(X)}\psi_{Y_{0}(1-S_{0})}-\mu_{00}(X)\left\{ \psi_{1-S_{1}}-\frac{1-p_{1}(X)}{1-p_{0}(X)}\psi_{1-S_{0}}\right\} .
\end{eqnarray*}
The EIF for $\tau_{00}$ is $\phi_{00}=\left(\phi_{1,00}-\phi_{0,00}-\tau_{00}\psi_{1-S_{1}}\right)/(1-p_{1})$,
where 
\begin{eqnarray*}
\phi_{1,00} = \psi_{Y_{1}(1-S_{1})},\quad\phi_{0,00} = \frac{e_{00}(X)}{1-p_{0}(X)}\psi_{Y_{0}(1-S_{0})}+\mu_{00}(X)\left\{ \psi_{1-S_{1}}-\frac{1-p_{1}(X)}{1-p_{0}(X)}\psi_{1-S_{0}}\right\} .
\end{eqnarray*}
The EIF for $\tau_{11}$ is $\phi_{11}=\left(\phi_{1,11}-\phi_{0,11}-\tau_{11}\psi_{S_{0}}\right)/p_{0}$,
where 
\begin{eqnarray*}
\phi_{1,11} = \frac{e_{11}(X)}{p_{1}(X)}\psi_{Y_{1}S_{1}}+\mu_{11}(X)\left\{ \psi_{S_{0}}-\frac{p_{0}(X)}{p_{1}(X)}\psi_{S_{1}}\right\} ,\quad\phi_{0,11} = \psi_{Y_{0}S_{0}}.
\end{eqnarray*}
\end{theorem}

From Theorem~\ref{thm:SET}, the semiparametric efficiency bounds
for the PCEs are $\E(\phi_{u}^{2})$ for $u=10,00,11$ \citep{bickel1993efficient}. 
The EIFs have mean zero, so we can obtain another set of identification formulas by solving $\E(\phi_{u})=0$.

\begin{corollary} \label{cor:est} Under Assumptions~\ref{assump:TAignorability}--\ref{assump:weak-pi},
\begin{eqnarray}
\tau_{10}=\frac{\E(\phi_{1,10}-\phi_{0,10})}{\E(\psi_{S_{1}}-\psi_{S_{0}})},\quad\tau_{00}=\frac{\E(\phi_{1,00}-\phi_{0,00})}{\E(1-\psi_{S_{1}})},\quad\tau_{11}=\frac{\E(\phi_{1,11}-\phi_{0,11})}{\E(\psi_{S_{0}})}.\label{eqn:identification-EIF}
\end{eqnarray}
\end{corollary}

As a sanity check of~\eqref{eqn:identification-EIF}, we can verify
that the denominator of $\tau_{u}$ in~\eqref{eqn:identification-EIF}
equals $\pr(U=u)$, and the numerator equals $\E\{(Y_{1}-Y_{0})\bm{1}(U=u)\}$,
for $u=10,00,11$. 
Based on Corollary~\ref{cor:est}, we can improve the estimators in
Examples~\ref{eg1-weighting}--\ref{eg3-reg}. Denote the estimator
for $\psi_{f(Y_{z},S_{z},X)}$ by 
\begin{eqnarray*}
\widehat{\psi}_{f(Y_{z},S_{z},X)}=\frac{\bm{1}(Z=z) [f(Y,S,X)-\widehat{\E}\{f(Y,S,X)\mid X,Z=z\}] }{\pi^z(X;\widehat{\alpha})\{1-\pi(X;\widehat{\alpha})\}^{1-z} }+\widehat{\E}\{f(Y,S,X)\mid X,Z=z\},
\end{eqnarray*}
where $\widehat{\E}\{f(Y,S,X)\mid X,Z=z\}$ is the fitted conditional
expectation of $f(Y,S,X)$ given $X$ and $Z=z$. When $f(Y,S,X)=S$, we have $\widehat{\E}\{f(Y,S,X)\mid X,Z=z\}=p_{z}(X;\widehat{\gamma})$, which reduces to the estimated principal score and results in the estimator
$\mathbb{P}_n(\widehat{\psi}_{S_{z}})=\widehat{p}_{z}$. 
When $f(Y,S,X)=YS$, we have $\widehat{\E}\{f(Y,S,X)\mid X,Z=z\}=\mu_{z1}(X;\widehat{\beta})p_{z}(X;\widehat{\gamma})$,
which relies on both the principal score and outcome mean.

Corollary~\ref{cor:est} motivates the following estimators:
\begin{equation}
\label{eqn::tr}
\widehat{\tau}_{10}=\frac{ \mathbb{P}_n (\widehat{\phi}_{1,10}-\widehat{\phi}_{0,10})}{ \mathbb{P}_n (\widehat{\psi}_{S_{1}}-\widehat{\psi}_{S_{0}})},\quad\widehat{\tau}_{00}=\frac{ \mathbb{P}_n (\widehat{\phi}_{1,00}-\widehat{\phi}_{0,00})}{ \mathbb{P}_n (1-\widehat{\psi}_{S_{1}})},\quad\widehat{\tau}_{11}=\frac{ \mathbb{P}_n (\widehat{\phi}_{1,11}-\widehat{\phi}_{0,11})}{ \mathbb{P}_n (\widehat{\psi}_{S_{0}})},
\end{equation}
where 
\begin{eqnarray*}
\widehat{\phi}_{1,10} & = &\frac{e_{10}(X;\widehat{\gamma})}{p_{1}(X;\widehat{\gamma})}\widehat{\psi}_{Y_{1}S_{1}}-\mu_{11}(X;\widehat{\beta})\left\{ \widehat{\psi}_{S_{0}}-\frac{p_{0}(X;\widehat{\gamma})}{p_{1}(X;\widehat{\gamma})}\widehat{\psi}_{S_{1}}\right\},\\
\widehat{\phi}_{0,10} & = & \frac{e_{10}(X;\widehat{\gamma})}{1-p_{0}(X;\widehat{\gamma})}\widehat{\psi}_{Y_{0}(1-S_{0})}-\mu_{00}(X;\widehat{\beta})\left\{ \widehat{\psi}_{1-S_{1}}-\frac{1-p_{1}(X;\widehat{\gamma})}{1-p_{0}(X;\widehat{\gamma})}\widehat{\psi}_{1-S_{0}}\right\} ,\nonumber \\
\widehat{\phi}_{0,00} & = &\frac{e_{00}(X;\widehat{\gamma})}{1-p_{0}(X;\widehat{\gamma})}\widehat{\psi}_{Y_{0}(1-S_{0})}+\mu_{00}(X;\widehat{\beta})\left\{ \widehat{\psi}_{1-S_{1}}-\frac{1-p_{1}(X;\widehat{\gamma})}{1-p_{0}(X;\widehat{\gamma})}\widehat{\psi}_{1-S_{0}}\right\} ,\nonumber \\
\widehat{\phi}_{1,11} & = &\frac{e_{11}(X;\widehat{\gamma})}{p_{1}(X;\widehat{\gamma})}\widehat{\psi}_{Y_{1}S_{1}}+\mu_{11}(X;\widehat{\beta})\left\{ \widehat{\psi}_{S_{0}}-\frac{p_{0}(X;\widehat{\gamma})}{p_{1}(X;\widehat{\gamma})}\widehat{\psi}_{S_{1}}\right\},\nonumber \\
\widehat{\phi}_{1,00}& =&\widehat{\psi}_{Y_{1}(1-S_{1})},\\
 \widehat{\phi}_{0,11} &=&\widehat{\psi}_{Y_{0}S_{0}}. 
\end{eqnarray*}
These estimators for the PCEs are all in ratio forms, similar to the
classic Wald estimator for the complier average causal effect under the
monotonicity and the ER \citep{angrist1996identification}.

Motivating estimators based on EIFs is a standard approach in semiparametric statistics. This approach, however, involves advanced statistical theory. To add more intuition for the estimators above, we offer an alternative perspective in the supplementary material based on model-assisted estimation from the classic survey sampling theory. This extends the results on doubly-robust and model-assisted estimation for the average causal effect in unconfounded observational studies \citep{robins1998jrssbdiscussion, little2004robust, kang2007demystifying, lumley2011connections}.

Interestingly, although the $\widehat{\tau}_u$'s involve models for the treatment probability, principal score, and the outcome, their consistency does not require the correct specification
of all three models. We will characterize this {\it triple robustness} property in the next subsection.

\subsection{Triple robustness}
\label{sec:robustness}

The following theorem shows the triple robustness and local efficiency
of the estimators constructed based on the EIFs.

\begin{theorem}[Triple robustness and local efficiency]\label{thm:triple}
	Suppose that Assumptions \ref{assump:TAignorability}--\ref{assump:weak-pi}
	hold, $\delta<\{\pi(x;\alpha^{*}),\pi(x;\hat{\alpha})\}<1-\delta$,
		and $\{p_{1}(x;\gamma^{*}),p_{1}(x;\widehat{\gamma}),1-p_{0}(x;\gamma^{*}),1-p_{0}(x;\widehat{\gamma})\} > \delta$
		for some $\delta\in(0,1)$ and all $x$ in the support of $X$. Each estimator $\widehat{\tau}_{u}$ in~\eqref{eqn::tr} is triply robust in the sense
	that it is consistent for $\tau_{u}$ under $\Mpsta\cup\Momta\cup\Momps$.
	Moreover, $\widehat{\tau}_{u}$ has the influence function $\phi_{u}$
	and therefore achieves the semiparametric efficiency bound under $\Mompsta$.
\end{theorem}

The regularity condition in Theorem \ref{thm:triple} is similar to the classic overlap condition \citep{rosenbaum1983central, d2020overlap}, which rules out small quantities in the denominators of the estimators. 
Theorem \ref{thm:triple} states that $\widehat{\tau}_{u}$ is consistent
if any two of the three models are correctly specified, and locally
efficient if all three models are correctly specified. For the variance calculation of these estimators, we use the nonparametric bootstrap.

To gain more intuition, we then give the sketch of the proof for the triple robustness of $\widehat{\tau}_{10}=( \mathbb{P}_n \widehat{\phi}_{1,10}- \mathbb{P}_n\widehat{\phi}_{0,10})/( \mathbb{P}_n\widehat{\psi}_{S_{1}}- \mathbb{P}_n\widehat{\psi}_{S_{0}})$
and relegate additional technical details to the supplementary material. For simplicity in this paragraph, let $ \mathcal{M}_\text{triple} = \Mpsta\cup\Momta\cup\Momps$ denote the set containing at least two correct models. 
The denominator is consistent for $\E(S_{1}-S_{0})=\pr(U=10)$ under
$\Mta\cup\Mps \supseteq \mathcal{M}_\text{triple} $.
For the terms in the numerator, we calculate their asymptotic biases in Section~\ref{app:triple}. In particular, 
$ \mathbb{P}_n \widehat{\phi}_{1,10} - \E\{Y_1\bm{1}(U=10)\}$ has the probability limit $B_1 + B_2 - B_3$, where
\begin{eqnarray*}
B_1 &=& \E\left[\frac{\{\mu_{11}(X)p_1(X) - \mu_{11}(X;\beta^*)p_1(X;\beta^*)\}\{\pi(X) - \pi(X;\alpha^*)\}}{\pi(X;\alpha^*)}\right], \\
B_2 &=& \E\left[\frac{\{\pi(X) p_0(X;\gamma^*)p_1(X)-\pi(X;\alpha^*) p_0(X)p_1(X;\gamma^*)\}\{\mu_{11}(X)-\mu_{11}(X;\beta^*)\}}{\pi(X;\alpha^*) p_1(X;\gamma^*) } \right] ,\\
B_3 &=& \E\left[\frac{\{\pi(X)-\pi(X;\alpha^*)\}\{p_0(X)-p_0(X;\gamma^*)\}\mu_{11}(X;\beta^*)}{1-\pi(X;\alpha^*)}\right].
\end{eqnarray*}
The bias $B_1$ equals $0$ under $\Momps\cup\Mta$; the bias $B_2$ equals $0$ under $\Mpsta\cup\Mom$; the bias $B_3$ equals $0$ under $\Mta\cup\Mps$. Each of these three sets contains $ \mathcal{M}_\text{triple}$. As a result, $\mathbb{P}_n(\widehat{\phi}_{1,10})$ is consistent
for $\E\{Y_{1}\bm{1}(U=10)\}$ under $ \mathcal{M}_\text{triple}$.
Similarly, we can show $\mathbb{P}_n(\widehat{\phi}_{0,10})$ is consistent for
$\E\{Y_{0}\bm{1}(U=10)\}$ under $ \mathcal{M}_\text{triple}$. So the triple robustness of $\widehat{\tau}_{10}$ holds.

The bias formulas above suggest that
the proposed triply robust estimator would remain consistent and asymptotically normal under some regularity conditions when using nonparametric or machine learning estimation 
for the nuisance functions $\pi(X)$, $p_z(X)$, and $\mu_{zs}(X)$, denoted by \textcolor{black}{$\widehat{\pi}(X)$, $\widehat{p}_{z}(X)$, and $\widehat{\mu}_{zs}(X)$}. This property would be similar to that of the doubly robust estimator for estimating the average causal effect in unconfounded observational studies \citep{chernozhukov2018double}. In other contexts involving intermediate variables, \citet{zheng2017longitudinal} and \citet{miles2020semiparametric} have established similar results for multiply robust estimators. Theorem \ref{thm:triple-1} formalizes
the results for the proposed estimators using nonparametric or machine
learning estimation. 

\begin{theorem}[Triple machine learning estimation]\label{thm:triple-1}
	\textcolor{black}{Suppose that Assumptions \ref{assump:TAignorability}--\ref{assump:weak-pi}
		hold,}
\begin{enumerate}
	[(a)]
	\item \textcolor{black}{$\{\widehat{\pi}(x),\widehat{p}_{z}(x),\text{\ensuremath{\widehat{\mu}}}_{zs}(x)\}\rightarrow\{\pi(x),p_{z}(x),\text{\ensuremath{\mu}}_{zs}(x)\}$
		in probability for all $x$ in the support of $X$,} 
	\item \textcolor{black}{$\{\widehat{\pi}(x),\widehat{p}_{z}(x),\text{\ensuremath{\widehat{\mu}}}_{zs}(x)\}\ \text{and }\{\pi(x),p_{z}(x),\text{\ensuremath{\mu}}_{zs}(x)\}$
		are in a Donsker class,}
	\item \textcolor{black}{$\delta<\{\pi(x),\widehat{\pi}(x)\}<1-\delta$,
		 $\{p_{1}(x),\widehat{p}_{1}(x),1-p_{0}(x),1-\widehat{p}_{0}(x)\} > \delta$
		and $\{|\widehat{\mu}_{zs}(x)|,|\mu_{zs}(x)|\}<C$ for some $\delta\in(0,1) , C>0$, and all $x$ in the support of $X$, and }
	\item \textcolor{black}{$||\widehat{g}(X)-g(X)||_{2}\times||\widehat{h}(X)-h(X)||_{2}=o_{\pr}(n^{-1/2})$,
		for any $g\neq h\in(\pi,p_{z},\mu_{zs})$, where $||\cdot||_{2}$
		denotes the $L_{2}$-norm, i.e. $||f(X)||_{2}^{2}=\int f(x)^{2}{\textup{d} F_X(x)}$.} 
\end{enumerate}
	\textcolor{black}{Then $\widehat{\tau}_{u}$ in (\ref{eqn::tr}) 
		is asymptotically normal, has the influence function $\phi_{u}$, and
		achieves the semiparametric efficiency bound.}
	
\end{theorem}

Conditions (a)--(d) are analogous to those
	for double machine learning estimation of average causal
	effects \citep[e.g.,][]{kennedy2016semiparametric,bradic2019sparsity}. The consistency
	in (a) and the rates of convergence in (d) are well studied for commonly
	used flexible models. Condition (b) restricts the complexity of
	the spaces of the nuisance functions and their estimators. The cross-fitting
	technique can be used to relax this condition \citep{chernozhukov2018double}.
	The conditions in (c) may not be necessary but enable bounding the error
	$|\widehat{\tau}_{u}-\pr_{n}\phi_{u}|$ by the summation of the
	terms in the form of $||\widehat{g}(X)-g(X)||_{2}\times||\widehat{h}(X)-h(X)||_{2}$
	with $g\neq h\in(\pi,p_{z},\mu_{zs})$. Thus, by Condition (d), the results
	in Theorem \ref{thm:triple-1} follow. 

Section~\ref{app:extension} in the supplementary material extends the identification and estimation framework to two important scenarios under randomization, i.e., $Z\ind(S_{1},S_{0},Y_{1},Y_{0},X)$, and strong monotonicity, i.e., $S_1 \geq S_0$, respectively. We also establish robustness properties of the corresponding estimators there.

\section{Simulation\label{sec:Simulation}}
We evaluate the finite-sample properties of various estimators at sample size $n=500$. Generate covariate $X\in\mathbb{R}^{5}$ by $X_{j}\sim\text{N}(0.25,1)$ for $j=1,\dots,4$,
and $X_{5}\sim\mathrm{Bernoulli}\ (0.5)$. We use linear predictors,
$C_{j}=X_{j}-0.25$, or quadratic predictors, $\widetilde{C}_{j}=(X_{j}^{2}-1)/\sqrt{2}$,
for $j=1,\ldots,4$. Generate the treatment by $Z\mid X\sim\text{Bernoulli}\{\pi(X)\}$,
the intermediate variable by $S\mid(Z=z,X)\sim\text{Bernoulli}\{p_{z}(X)\}$,
and the outcome by $Y\mid(Z=z,S=s,X)\sim\text{ N}\{\mu_{zs}(X),1\}$. 
 To assess the robustness of the estimators to
model misspecification, 
we consider two different choices for each of $\pi(X)$,
$p_{z}(X)$, and $\mu_{zs}(X)$, summarized in Table~\ref{tab:simulation-setup}. 
We indicate the models by the name
of the dependent variable and whether or not the predictors are linear.
For example, ``tp:no'' is the model with $\pi(X)=2\sum_{j=1}^{4}\widetilde{C}_{j}/5$,
and ``ps:yes'' is the model with $p_{z}(X)=2\{(2z-1)-\sum_{j=1}^{4}C_{j}\}/5$.

\begin{table}[t]
\caption{Models for simulation with two specifications
for each of $\logit\{\pi(X)\}$, $\logit\{p_{z}(X)\}$, and $\mu_{zs}(X)$,
indicated by ``yes'' and ``no.'' }

\begin{centering}
\begin{tabular}{cccl}
\hline 
 & $\logit\{\pi(X)\}$ & $\logit\{p_{z}(X)\}$ & \multicolumn{1}{c}{$\mu_{zs}(X)$}\tabularnewline
\cdashline{2-4} yes & 0 & $2\{(2z-1)-\sum_{j=1}^{4}C_{j}\}/5$ & $\sum_{j=1}^{5}C_{j}\left(1+z+s\right)/4$\tabularnewline
no & $2\sum_{j=1}^{4}\widetilde{C}_{j}/5$ & $2\{(2z-1)-\sum_{j=1}^{4}\widetilde{C}_{j}\}/5$ & $\sum_{j=1}^{5}\widetilde{C}_{j}\left(1+z+s\right)/4$ \tabularnewline
\hline 
\end{tabular}
\par\end{centering}
\label{tab:simulation-setup} 
\end{table}

 We calculate the true value of $\tau_{u}$ based on the identification formulas in Theorem \ref{thm:identification-obs} and the true models. 
We then compare the following estimators for $\tau_{u}$, 
\begin{enumerate}
\item weighting estimators: $\widehat{\tau}_{u,\ipw}$ and $\widehat{\tau}'_{u,\ipw}$
given in Example~\ref{eg1-weighting}; 
\item regression estimators: $\widehat{\tau}_{u,\regta}$ given in Example~\ref{eg2-reg}
and $\widehat{\tau}_{u,\regps}$ given in Example~\ref{eg3-reg}; 
\item triply robust estimators: $\widehat{\tau}_{u}$ and $\widehat{\tau}_{u,{\rm ml}}$ with parametric models and with flexible generalized additive models for nuisance functions, respectively. 
\end{enumerate}
We also consider the weighting estimator and the regression estimator
in \citet{ding2017principal}, which are proposed under randomized
experiments. 
Under ``ps:yes'' and ``ps:no,'' 
we estimate the principal score by logistic
regressions with linear predictors $X$ and $(X_{1},X_{2})$, respectively; 
we estimate the outcome mean by linear regressions with the linear predictor $X$. 
Therefore, under generative models with the label ``yes,'' the fitting
models are correctly specified, while under generative models with
the label ``no,'' the fitting models are misspecified.

\begin{figure} 
\includegraphics[width=1\textwidth,height=0.3\textheight]{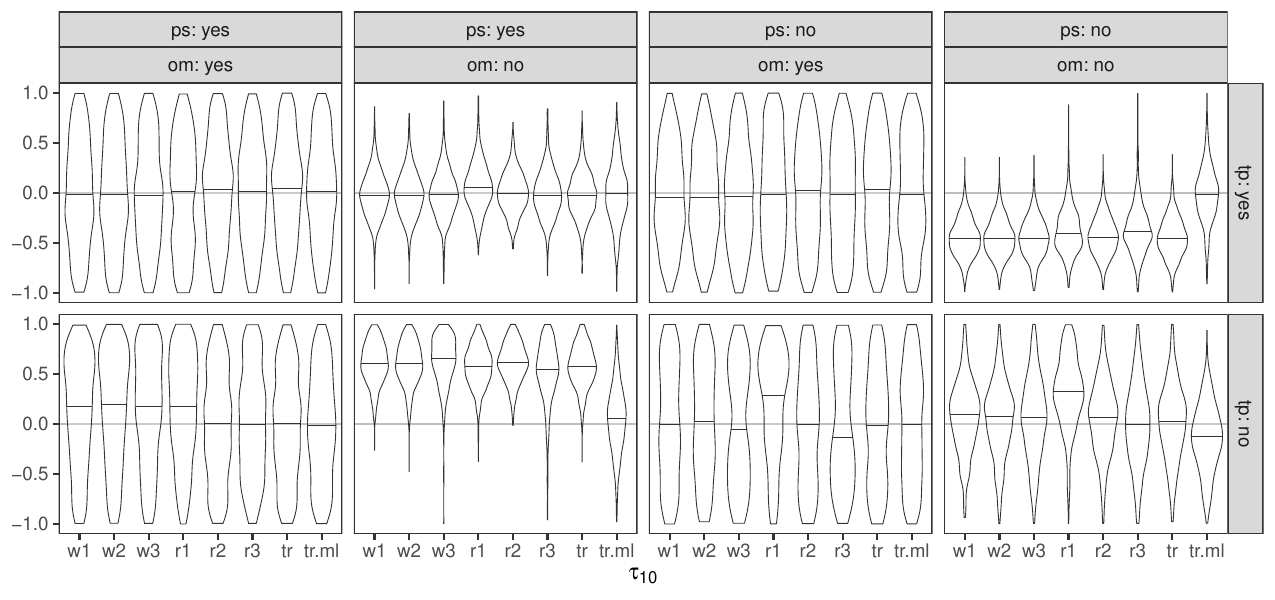}
\includegraphics[width=1\textwidth,height=0.3\textheight]{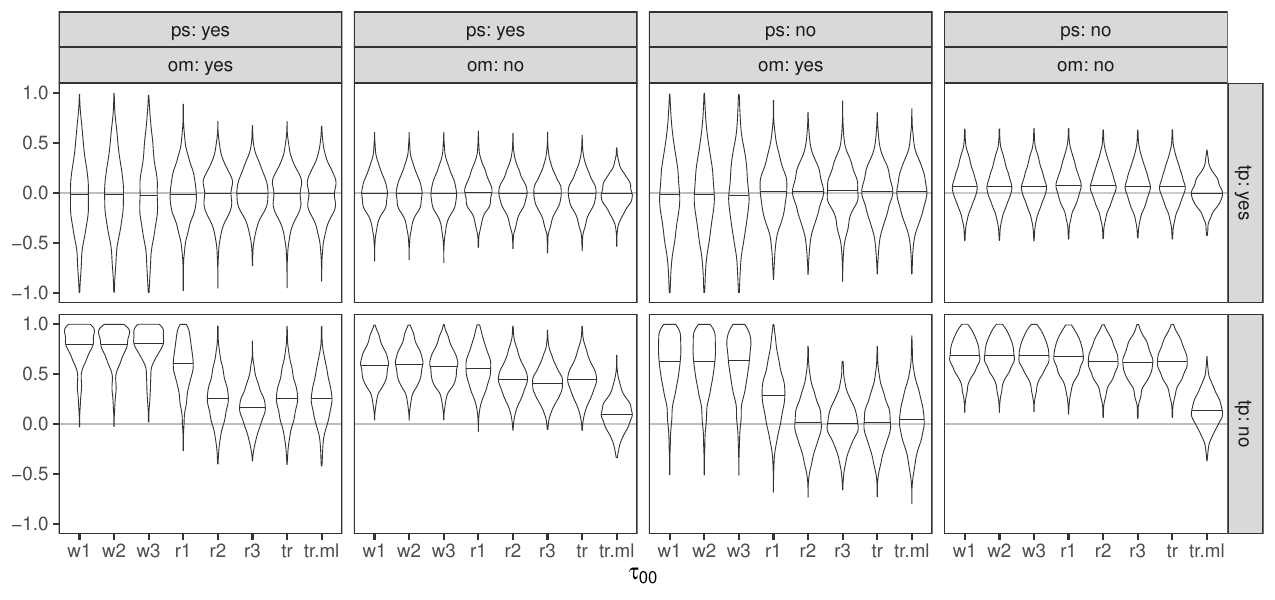}
\includegraphics[width=1\textwidth,height=0.3\textheight]{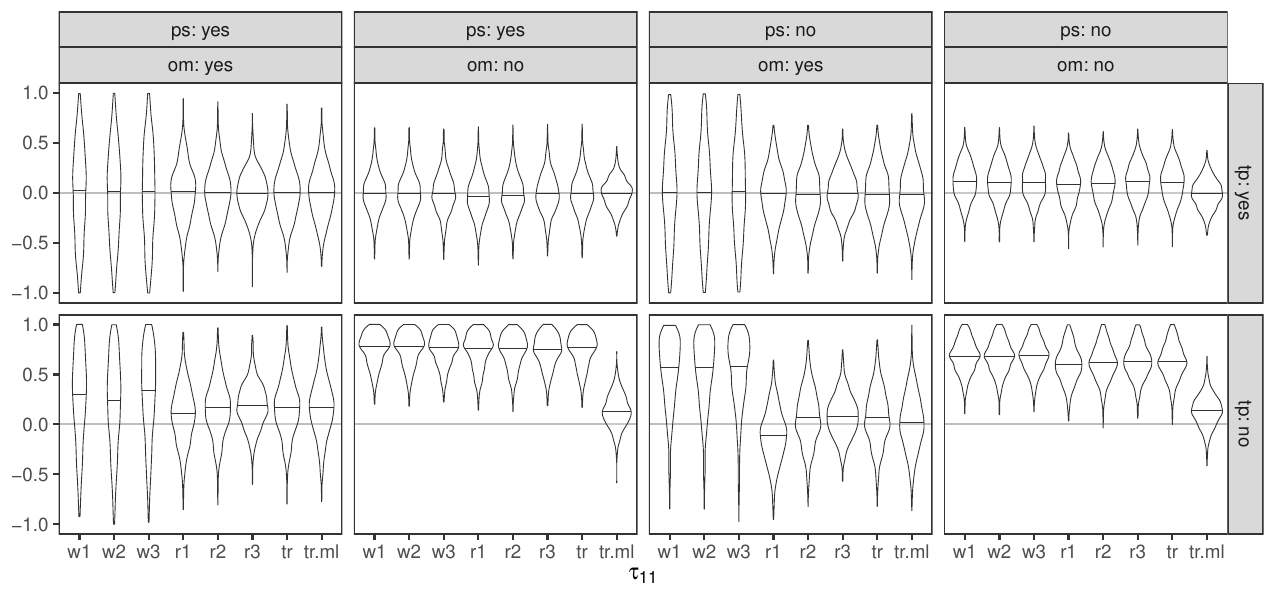}
\caption{Violin plots of estimators in eight scenarios. Labels: ``w1'' for
$\widehat{\tau}_{u,\ipw}$, ``w2'' for $\widehat{\tau}'_{u,\ipw}$, and
``w3'' for the weighting estimator in \citet{ding2017principal};
``r1'' for $\widehat{\tau}_{u,\regta}$; ``r2'' for $\widehat{\tau}_{u,\regps}$;
``r3'' for the regression estimator in \citet{ding2017principal};
``tr'' for the triply robust estimator $\widehat{\tau}_{u}$ and
	``tr.ml'' for the triply robust estimator $\widehat{\tau}_{u,{\rm ml}}$.} 
\label{fig:continuous} 
\end{figure}

We compare the estimators in $2^{3}=8$ scenarios depending on whether the treatment probability, principal score, or outcome model is correctly specified. Figure~\ref{fig:continuous} presents
the violin plots based on 1000 repeated sampling of the estimators. For all the three PCEs, the weighting estimators $\widehat{\tau}_{u,\ipw}$ and
$\widehat{\tau}'_{u,\ipw}$ (indicated by ``w1'' and ``w2'' in
the figures) are biased when the treatment probability or principal
score model is misspecified. The bias with a misspecified treatment
probability is larger than that with a misspecified principal score because the weights corresponding to the treatment
probability are unbounded while the weights corresponding
to the principal score are bounded within $[0,1]$. The weighting estimator
in \citet{ding2017principal} (indicated by ``w3'' in the figures)
performs similarly to $\widehat{\tau}_{u,\ipw}$ and $\widehat{\tau}'_{u,\ipw}$,
because the treatment is randomized under ``tp:yes.'' As our theory predicts, the regression estimator $\widehat{\tau}_{u,\regta}$
(indicated by ``r1'' in the figures) is unbiased under $\Momta$;
the regression estimator $\widehat{\tau}_{u,\regps}$ (indicated
by ``r2'' in the figures) is unbiased under $\Momps$. The regression
estimator in \citet{ding2017principal} (indicated by ``r3'' in
the figures) performs similarly to $\widehat{\tau}_{u,\regta}$ in
terms of bias. The triply robust estimator $\widehat{\tau}_{u}$ (indicated by ``tr''
in the figures) is unbiased under $\Mpsta\cup\Momta\cup\Momps$, verifying 
its triple robustness.
With flexible models for the nuisance functions, $\widehat{\tau}_{u,{\rm ml}}$ is unbiased under $\Mpsta\cup\Momta\cup\Momps$ and is less biased than other estimators in most scenarios.

\section{Applications to two observational studies\label{sec:Applications}}


\subsection{Return to schooling\label{sec:RTS}}

The dataset from the U.S. National Longitudinal Survey of Young
Men contains $3010$ men with age between $14$ and $24$ in
the year 1966. \citet{card1993using} uses it to estimate the causal
effect of education on earnings, utilizing the geographic variation
in college proximity as an instrumental variable for education. 
Thus, the treatment $Z$ is the indicator of growing up near a four-year college;
the intermediate variable $S$ is the indicator of whether the individual
receives education beyond high school; and the outcome $Y$ is the
log wage in the year 1976, ranging from $4.6$ to $7.8$. Monotonicity
is plausible because living close to a college
would make an individual more likely to receive higher education.
To make principal ignorability plausible, we include the following covariates:
race, age and squared age, a categorical variable indicating living with both parents, single mom, or both parents, and variables
summarizing the living areas in the past. 
Unlike \citet{card1993using}, we do not invoke the ER
that living near a college affected the earnings only through education. 
Rather, under principal ignorability, our analysis can assess the plausibility of the ER. \citet{guo2014using} and \citet{yang2014dissonant} used similar strategies to test the ER.

We use a linear model for the outcome mean and logistic models
for the treatment probability and principal score, and estimate the asymptotic variances
by the nonparametric bootstrap. Table \ref{t:education-wage} presents the results for the estimated proportions of principal strata ($\widehat e_u$) and PCEs using the weighting estimators ($\widehat{\tau}_{u,\ipw}$ and $\widehat{\tau}'_{u,\ipw}$), regression estimators ($\widehat{\tau}_{u,\regta}$ and $\widehat{\tau}_{u,\regps}$), and the triply robust estimator ($\widehat \tau_u$). We omit $\widehat{\tau}_{u,\rm{tr.ml}}$ because it produces similar results as $\widehat{\tau}_{u}$. All estimators are close, except for the unstabilized weighting estimator. This is due to the extreme fitted treatment probabilities. The estimators for $\tau_{00}$ and $\tau_{11}$ are not significant,
suggesting no significant evidence of violating the ER.
The estimated $\tau_{10}$ is positive and statistically significant,
implying education has a positive effect on earnings. This finding
corroborates with previous analyses.

\begin{table}[htbp]
\caption{Analysis of the National Longitudinal Survey of Young Men.}
 
\begin{centering}
\begin{tabular}{lrrr}
\hline 
 & \multicolumn{1}{c}{$u=10$ } & \multicolumn{1}{c}{$u=00$ } & \multicolumn{1}{c}{$u=11$ }\tabularnewline
\cdashline{2-4} 
$\widehat{e}_{u}$ & $7\%$ $(3\%,10\%)$ & $48\%$ $(46\%,50\%)$ & $45\%$ $(42\%,49\%)$ \tabularnewline
$\widehat{\tau}_{u,\ipw}$ & \bm{$-0.87$} \bm{$(-1.69,-0.05)$} & $0.10$ $(-0.25,0.44)$ & 0.50 $(-0.06,1.05)$\tabularnewline
$\widehat{\tau}'_{u,\ipw}$ & \bm{$0.15$} \bm{$(0.00,0.30)$} & $0.01$ $(-0.04,0.06)$ & $0.02$ $(-0.04,0.08)$ \tabularnewline
$\widehat{\tau}_{u,\regta}$ & $0.09$ $(-0.03,0.21)$ & $0.02$ $(-0.02,0.07)$ & $0.01$ $(-0.05,0.08)$\tabularnewline
$\widehat{\tau}_{u,\regps}$ & \bm{$0.12$} \bm{$(0.03,0.21)$} & $0.02$ $(-0.03,0.07)$ & $0.01$ $(-0.05,0.07)$\tabularnewline
$\widehat{\tau}_{u}$ & $0.10$ $(-0.01,0.23)$ & $0.02$ $(-0.03,0.07)$ & $0.01$ $(-0.05,0.07)$\tabularnewline
\hline 
\end{tabular}
\par\end{centering}
\label{t:education-wage} 
\end{table}

%
%

\subsection{Causal effect of flooding on health \label{subsec:flooding}}

We re-analyze a dataset from \citet{guo2018mediation} with $774$
households in Bangladesh to investigate the effect of flooding on children's diarrhea.
The treatment $Z$ is the indicator of whether a household was severely
affected by the flood; the intermediate variable $S$ is the indicator
of whether the per capita calorie consumption of the household was
less than $2000$ calories; and the outcome $Y$ is the number of
days a child had diarrhea. Monotonicity is plausible because the calorie
consumption would be negatively affected if the household was severely
affected by the flood. To ensure principal ignorability, we include
the following covariates: gender, age, the size of the household,
mother's education, father's education, mother's age, and father's
age. As pointed out by \citet{del20011998}, the ER might
be violated due to an alternative pathway through mother's health.
We use our method to evaluate this assumption by estimating $\tau_{00}$
and $\tau_{11}$.

Again we use a linear model for the outcome mean and logistic models
for the treatment probability and principal score. Table \ref{tab:The-flood-health-data}
presents the results for the estimated proportions of principal strata and PCEs. 
The estimated $\tau_{00}$ and $\tau_{11}$ are both positive and
the estimated $\tau_{00}$ is statistically significant, indicating
that being affected by the flood tends to directly increase the number
of days of diarrhea. This also confirms the suspicion of the violation
of the ER in \citet{del20011998}. Although the estimated
$\tau_{10}$ is positive, it is imprecisely estimated and not statistically
significant, due to the small proportion of stratum $U=10$.

\begin{table}[htbp]
\caption{\label{tab:The-flood-health-data} Analysis of the flood data. }

\begin{centering}
\begin{tabular}{lrrr}
\hline 
 & \multicolumn{1}{c}{ $u=10$ } & \multicolumn{1}{c}{ $u=00$ } & \multicolumn{1}{c}{ $u=11$ }\tabularnewline
\cdashline{2-4} 
$\widehat{e}_{u}$ & $9\%$ $(2\%,15\%)$ & $45\%$ $(41\%,50\%)$ & $46\%$ $(41\%,51\%)$ \tabularnewline
$\widehat{\tau}_{u,\ipw}$ & $0.74$ $(-3.60,5.09)$ & \bm{$0.98$} \bm{$(0.19,1.77)$} & $1.97$ $(-0.52,2.47)$\tabularnewline
$\widehat{\tau}'_{u,\ipw}$ & $0.86$ $(-3.35,5.07)$ & \bm{$0.92$} \bm{$(0.13,1.71)$} & $1.11$ $(-0.36,2.58)$ \tabularnewline
$\widehat{\tau}_{u,\regta}$ & $1.74$ $(-3.55,7.03)$ & \bm{$0.93$} \bm{$(0.10,1.77)$} & $1.01$ $(-0.45,2.47)$\tabularnewline
$\widehat{\tau}_{u,\regps}$ & $1.50$ $(-3.01,6.01)$ & \bm{$0.92$} \bm{$(0.09,1.75)$} & $1.10$ $(-0.33,2.53)$\tabularnewline
$\widehat{\tau}_{u}$ & $1.51$ $(-3.68,6.71)$ & \bm{$0.88$} \bm{$(0.05,1.71)$} & $1.10$ $(-0.33,2.53)$\tabularnewline
\hline 
\end{tabular}
\par\end{centering}
\label{tab:The-flood-health-data} 
\end{table}

\section{Discussion\label{sec:Discussion}}
PCEs characterize subgroup causal effects of important scientific meanings, providing insights into the underlying causal mechanism between the treatment and outcome. We develop an identification and estimation framework for PCEs
under principal ignorability. The proposed estimators are analogous
to those for the average causal effect in unconfounded observational studies. They are easy to implement which involve
the model fitting of the treatment, intermediate variable, and outcome
conditional on baseline covariates. They are
triply robust and locally efficient, naturally extending
the classic doubly robust estimator for the average causal effect.


 In mediation analysis, \citet{tchetgen2012semiparametric} develop a general semiparametric framework for the direct and indirect effects. They focus on two scalar estimands, the natural direct and indirect effects, in the overall population. In contrast, we focus on the treatment effects within  principal strata, resulting in more estimands in different subpopulations. Although 
the PCEs and the direct and indirect effects are related in certain scenarios \citep{vanderweele2008simple,vanderweele2011principal, forastiere2018principal}, there is no universal relationship between them, and the PCEs are applicable to a number of applications other than mediation analysis. Moreover, as discussed in Section~\ref{sec:Notation-and-assumptions}, the identification assumptions in the two methods concern different aspects of the relationship between the potential outcome and the potential intermediate variable. 

We can generalize the theory to other causal estimands within principal
strata by invoking a stronger version of principal ignorability: $Y_{1}\ind U\mid(Z=1,S=1,X)$
and $Y_{0}\ind U\mid(Z=0,S=0,X)$. Under this assumption, we can identify
the effects on a transformation of the outcome: $\E\{h(Y_{1},X)\mid U=u\}-\E\{h(Y_{0},X)\mid U=u\}$
for any function $g(\cdot)$. This further ensures the identification
of distributional or quantile causal effects within principal strata.
Similar to the main paper, we can also derive EIFs and propose robust estimators for these causal estimands.

More generally, we can extend the results to deal with  a continuous $S$, where the number of principal strata is infinity. The principal score becomes the conditional density of $U=(S_1,S_0)$ given covariates, which is not identifiable even under monotonicity. Therefore, the extension requires more sophisticated identification and estimation strategies. We leave the technical investigation to future research.

\section*{Acknowledgments}
We thank the Associate Editor, three reviewers, Anqi Zhao, and Sizhu Lu for helpful comments.  Peng Ding was partially funded by the U.S. National Science Foundation (grant \# 1945136).

%

 \bibliographystyle{Chicago}
\bibliography{DR_PSACE_ref}

\newpage{}
\begin{center}
\textbf{\Large{}Supplementary Material} 
\par\end{center}

\pagenumbering{arabic} 
\renewcommand*{\thepage}{S\arabic{page}}

\setcounter{lemma}{0} 
\global\long\def\thelemma{\textup{S}\arabic{lemma}}%
 \setcounter{equation}{0} 
\global\long\def\theequation{S\arabic{equation}}%
 \setcounter{section}{0} 
\global\long\def\thesection{S\arabic{section}}%
 \setcounter{assumption}{0} 
\global\long\def\theassumption{S\arabic{assumption}}%
 \setcounter{theorem}{0} 
\global\long\def\thetheorem{S\arabic{theorem}}%
 \setcounter{figure}{0} 
\global\long\def\thefigure{S\arabic{figure}}%
 
\def\thesubsection{S\arabic{section}.\arabic{subsection}}%

Section~\ref{app:covbalancing} establishes the balancing properties of the treatment probability and principal score, and discusses an alternative estimation strategy for them based on these balancing properties.

Section \ref{sec::model-assisted} provides an alternative motivation for the triply robust estimators based on model-assisted estimation in survey sampling.

Section~\ref{app:extension} extends the identification and estimation framework for the PCEs to two scenarios under randomization and strong monotonicity, respectively.

Section~\ref{app:sensitivity} proposes sensitivity analyses for principal ignorability and monotonicity, respectively.

Section~\ref{app:identification} proves the identification
results, including Theorem~\ref{thm:identification-obs} in the main text, Theorem~\ref{thm:balancing} Section~\ref{app:covbalancing}, and Theorems~\ref{thm:identification-sens}~and~\ref{thm:identification-sens-mon} in Section~\ref{app:sensitivity}.

Section~\ref{app:EIF} proves the EIFs, including
Theorem~\ref{thm:SET} in the main text, Theorems~\ref{thm:SET-rand},~\ref{thm:double-rand},~and~\ref{thm:SET-strmon} in Section~\ref{app:extension}, and Theorems~\ref{thm:SET-sens}~and~\ref{thm:SET-sens-mon} in Section~\ref{app:sensitivity}.

Section~\ref{app:triple} proves the multiple robustness
and local efficiency, including Theorems~\ref{thm:triple}~and~\ref{thm:triple-1} in the main text and Theorems~\ref{thm:sens-dr}~and~\ref{thm:sens-mon-tr} in Section~\ref{app:extension}.

\section{The role of covariate balancing}

\label{app:covbalancing} 

\subsection{Balancing properties}

In observational studies, \citet{rosenbaum1983central} prove the
balancing property of the treatment probability: the covariate distributions are the same
 in 
 the treatment
and control groups conditional
on it or weighted by its inverse. 
Now we generalize this result to the balancing
properties of both the treatment probability and principal score.

\begin{theorem}[Balancing properties] \label{thm:balancing} For
any function $h(X)$ that has finite moments $\E\{h(X)\mid U=u\}<\infty$
for $u=10,00,11$, the following three sets of balancing conditions
hold.  
\begin{enumerate}[(a)] 
\item \label{eq::balance-a} Corresponding to stratum $U=10$,
we have 
\begin{eqnarray*}
\E\left\{ \frac{p_{1}(X)-p_{0}(X)}{p_{1}-p_{0}}\frac{S}{p_{1}(X)}\frac{Z}{\pi(X)}h(X)\right\} & = & \E\left\{ \frac{p_{1}(X)-p_{0}(X)}{p_{1}-p_{0}}\frac{1-S}{1-p_{0}(X)}\frac{1-Z}{1-\pi(X)}h(X)\right\} \nonumber \\
 & = & \E\left[\frac{SZ/\pi(X)-S(1-Z)/\{1-\pi(X)\}}{p_{1}-p_{0}}h(X)\right]\nonumber \\
 & = & \E\left\{ \frac{p_{1}(X)-p_{0}(X)}{p_{1}-p_{0}}h(X)\right\}.
\end{eqnarray*}
\item \label{eq::balance-b} Corresponding to stratum $U=00$,
we have 
\begin{eqnarray*}
\E\left\{ \frac{(1-S)Z}{(1-p_{1})\pi(X)}h(X)\right\} & = & \E\left\{ \frac{1-p_{1}(X)}{1-p_{1}}\frac{1-S}{1-p_{0}(X)}\frac{1-Z}{1-\pi(X)}h(X)\right\} \\
 & = & \E\left\{ \frac{1-SZ/\pi(X)}{1-p_{1}}h(X)\right\} \\
 & = & \E\left\{ \frac{1-p_{1}(X)}{1-p_{1}}h(X)\right\}.
\end{eqnarray*}
\item \label{eq::balance-c} Corresponding to stratum $U=11$,
we have 
\begin{eqnarray*}
\E\left\{ \frac{p_{0}(X)}{p_{0}}\frac{S}{p_{1}(X)}\frac{Z}{\pi(X)}h(X)\right\} & = & \E\left[\frac{S(1-Z)}{p_{0}\{1-\pi(X)\}}h(X)\right]\ =\ \E\left\{ \frac{p_{0}(X)}{p_{0}}h(X)\right\}.
\end{eqnarray*}
\end{enumerate}
Under Assumptions~\ref{assump:TAignorability}~and~\ref{assump:M},
the three formulas in~\eqref{eq::balance-a}--\eqref{eq::balance-c}
equal $\E\{h(X)\mid U=u\}$ for $u=10,00,11$, respectively. \end{theorem}

Theorem~\ref{thm:balancing} generalizes the two-way balancing properties
in \citet{ding2017principal} 
to the four-way balancing properties.  
The terms involving $p_{1}$ and $p_{0}$ cancel in the above formulas. Nevertheless, we keep them to aid interpretation.

For each principal stratum, the balancing conditions consist of the expectations of $h(X)$ in pseudo-populations defined by four different weights. Because all except one of the weights involve the principal score, we would still have all these balancing conditions even in randomized experiments. Therefore, Theorem~\ref{thm:balancing} provides more balancing conditions than those in \citet{ding2017principal}. Although these weights take different forms, the conditional expectations of them given $X$ are the same. For example, based on the discussion under
Theorem~\ref{thm:identification-obs}, the conditional expectations of the weights in Theorem~\ref{thm:balancing}(a) all equal $e_{10}(X)/e_{10}$. A similar discussion applies to the weights in Theorem~\ref{thm:balancing}(b)~and~(c).

The identities in Theorem~\ref{thm:balancing} follow from the multiple
expressions of $\E\{h(X)\mid U=u\}$. Because any functions of covariates
are unaffected by the treatment within principal strata, the PCEs
on $h(X)$ are all zero. Then, the first identity in each of the three
formulas can be viewed as special cases of Theorem~\ref{thm:identification-obs}(a)
with $Y$ replaced by $h(X)$. For example, by replacing $Y$ with
$h(X)$ in Theorem \ref{thm:identification-obs}(a), the two terms
on the right-hand side both equal $\E\{h(X)\mid U=u\}$. This yields
the first identity in Theorem~\ref{thm:balancing}(a). On the other
hand, $\E\{h(X)\mid U=u\}$ is the expectation of $h(X)$ over the
distribution of $X$ within stratum $U=u$. Therefore,
we can obtain the second and third identities in each of the three
formulas by replacing the outcome mean difference with $h(X)$ in
Theorem~\ref{thm:identification-obs}(b)~and~(c). Theorem~\ref{thm:balancing}(c)
has one fewer identity than (a) and (b), because the identity obtained
from Theorem~\ref{thm:identification-obs}(a) coincides with that
obtained from Theorem~\ref{thm:identification-obs}(b). Although
the heuristics come from Theorem~\ref{thm:identification-obs},
the formulas in Theorem~\ref{thm:balancing} hold as long as the treatment probability and principal
score are correctly specified; they do not require Assumptions~\ref{assump:TAignorability}--\ref{assump:weak-pi}.
If Assumptions~\ref{assump:TAignorability}~and~\ref{assump:M}
hold, then we can connect them with the expectations
of $h(X)$ within principal strata.

\subsection{Applications of balancing properties}

First, the balancing properties in Theorem~\ref{thm:balancing} are the theoretical foundation for model checking of the treatment probability and principal
score without touching the outcome data. Theorem~\ref{thm:balancing} holds with the true treatment probability and principal
score. Therefore, any empirical violations of the balancing properties provide the basis for refuting the models of the treatment probability and principal
score. Checking for these balance conditions does not involve modeling the outcome mean. 
With such a balance checking procedure, it
is reasonable to favor the estimators that are consistent under $\Mpsta$
(e.g., $\widehat{\tau}_{u,\ipw}$ and $\widehat{\tau}_{u}$) over
those that are consistent only under $\Mom$ (e.g., $\widehat{\tau}_{u,\regta}$
and $\widehat{\tau}_{u,\regps}$).  

Second, by choosing different $h(X)$ in Theorem~\ref{thm:balancing}, the balancing properties imply infinitely many moment conditions. This allows for an alternative approach to constructing more efficient estimators. In the context of missing data, 
 \citet{graham2011efficiency} proposes a strategy to utilize all the moment conditions from balancing properties, which turns out to be equivalent to the approach based on the EIF. Similarly, for each PCE, we can leverage the moment conditions implied by Theorem~\ref{thm:balancing} to improve the estimators in Examples~\ref{eg1-weighting}--\ref{eg3-reg}. The resulting estimator will be the same as the proposed triply robust estimator.

Lastly, the balancing properties provide an alternative estimation strategy
for $\pi(X;\alpha)$ and $p_{z}(X;\gamma)$ by constructing estimating
equations based on the identities in Theorem \ref{thm:balancing}. For example, the identities in
Theorem~\ref{thm:balancing}(c) imply estimating equations
\begin{eqnarray*}
\E\left\{ p_{0}(X;\gamma)\frac{S}{p_{1}(X;\gamma)}\frac{Z}{\pi(X;\alpha)}h(X)\right\} = \E\left\{ \frac{\textup{s}(1-Z)}{1-\pi(X;\alpha)}h(X)\right\} =\E\left\{ p_{0}(X;\gamma)h(X)\right\} 
\end{eqnarray*}
for any $h(X)$.  
The identities in Theorem~\ref{thm:balancing}(a) and (b) imply other estimating equations. 
We
can choose a set of $h(X)$, and then solve the corresponding estimating
equations to obtain estimators of $\alpha$ and $\gamma$. This
strategy is similar to the covariate balancing propensity score
\citep{imai2014covariate} for estimating the average causal effect
in observational studies.
We can also construct balancing weights for estimating the PCEs based on Theorem~\ref{thm:balancing}, similar to 
the strategies for estimating the average causal effect \citep{zubizarreta2015stable,tan2020regularized}.
We do not pursue this direction in the current paper and leave it
to future work.

\section{Connection with model-assisted estimation}\label{sec::model-assisted}

To gain more intuition about the EIFs and the identification formulas,
we provide an interpretation of~\eqref{eqn:identification-EIF}
based on the model-assisted estimation for the PCEs.
In observational studies for the average causal effects, 
augmented inverse probability weighting and some model-assisted estimation strategies are equivalent, both leveraging outcome modeling to improve efficiency and robustness \citep{robins1998jrssbdiscussion, little2004robust, kang2007demystifying, lumley2011connections}. 
However, as far as we know, analogous connections are missing between multiply robust estimation and model-assisted estimation.
As an extension, we make a modest contribution to show that this equivalence also holds between \eqref{eqn:identification-EIF} and certain forms of the model-assisted estimation for the PCEs.

We focus on $\tau_{10}$. It has the following ratio form: 
\begin{eqnarray}
\tau_{10} = 
\frac{\E\{Y_{1}\bm{1}(U=10)\}-\E\{Y_{0}\bm{1}(U=10)\}}{\E(S_{1}-S_{0})}.\label{eqn:modelassist-10}
\end{eqnarray}
We first discuss the model-assisted estimation for the denominator
$\E(S_{1}-S_{0})$.
The potential outcome $S_z$ decomposes into two terms based on its model given $X$, i.e., the principal score $p_{z}(X)=\E(S_{z}\mid X)$:
\begin{eqnarray}
\E(S_{z})=\E\{S_{z}-p_{z}(X)\}+\E\{p_{z}(X)\},\quad(z=0,1).\label{eqn:decompositionnSz}
\end{eqnarray}
Applying the inverse probability weighting formula
to $\E\{S_{z}-p_{z}(X)\}$ yields
\[
\E(S_{1})=\E\left[ \frac{Z\{S-p_{1}(X)\}}{\pi(X)}+p_{1}(X)\right], \quad\E(S_{0})=\E\left[ \frac{(1-Z)\{S-p_{0}(X)\}}{1-\pi(X)}+p_{0}(X)\right]. 
\]
This leads to 
$
\E(S_{1}-S_{0})=\E(\psi_{S_{1}}-\psi_{S_{0}}),
$
the model-assisted estimation formula for $\E(S_{1}-S_{0})$, which involves the treatment probability and principal score. 
Subtracting $p_{z}(X)$
from $S_{z}$ reduces the variation in $S_{z}$, and thus the model-assisted estimator improves the efficiency over the simple weighting estimator. This recovers the classic doubly robust estimator \citep{bang2005doubly}. Similarly, $\E\{\psi_{f(Y_{z},S_{z},X)}\}$ is the
model-assisted estimation formula for $\E\{f(Y_{z},S_{z},X)\}$ with
the model of $f(Y_{z},S_{z},X)$.

We then discuss the model-assisted estimation for the numerator of~\eqref{eqn:modelassist-10}. Decompose it based on the outcome mean: 
\begin{eqnarray}
\E[\{Y_{1}-\mu_{11}(X)\}\bm{1}(U=10)]+\E\{\mu_{11}(X)\bm{1}(U=10)\}.\label{eqn:modelassist-1-10}
\end{eqnarray}
Applying the inverse probability weighting formula in Theorem~\ref{thm:identification-obs}(a)
to the first term of~\eqref{eqn:modelassist-1-10} yields
\begin{eqnarray*}
\E[\{Y_{1}-\mu_{11}(X)\}\bm{1}(U=10)] & = & \E\left[e_{10}(X)\frac{S}{p_{1}(X)}\frac{Z}{\pi(X)}\{Y-\mu_{11}(X)\}\right];\label{eqn:modelassist-1}
\end{eqnarray*}
applying the model-assisted
estimation formula to the second term of \eqref{eqn:modelassist-1-10} yields 
\begin{eqnarray*}
\E\{\mu_{11}(X)\bm{1}(U=10)\} 
= \E\{\mu_{11}(X) (S_{1}-S_{0}) \} 
= \E\left\{ \mu_{11}(X)(\psi_{S_{1}}-\psi_{S_{0}})\right\} .\label{eqn:modelassist-2}
\end{eqnarray*}
Importantly, the quantities inside the above two expectations sum to 
$\phi_{1,10}$: 
\begin{eqnarray}
e_{10}(X)\frac{S}{p_{1}(X)}\frac{Z}{\pi(X)}\{Y-\mu_{11}(X)\}+\mu_{11}(X)(\psi_{S_{1}}-\psi_{S_{0}})=\phi_{1,10},\label{eqn:phi110}
\end{eqnarray}
which proves that $\E(\phi_{1,10})$ is the model-assisted estimation
formula for $\E\{Y_{1}\bm{1}(U=10)\}$. Similarly,  
$\E(\phi_{0,10})$ is the model-assisted estimation formula for $\E\{Y_{0}\bm{1}(U=10)\}$.
As a result, the model-assisted estimation formula for $\tau_{10}$
coincides with the one derived in~\eqref{eqn:identification-EIF}. 

The above perspective explains the efficiency gain of the estimator $\hat{\tau}_u$ $(u=10,00,11)$ over the simple weighting estimators. More surprisingly, they also improve robustness over all estimators in Examples \ref{eg1-weighting}--\ref{eg3-reg}, as shown in Section \ref{sec:robustness} in the main paper.

\section{Extensions}
\label{app:extension}

We extend the general identification and
estimation framework for the PCEs in Section~\ref{sec:Semiparametric-efficiency-theory}
to two important scenarios with different assumptions.

\subsection{Randomized experiments}

In randomized experiments, a stronger version
of Assumption~\ref{assump:TAignorability} holds. 

\begin{assumption}[Randomization]
\label{assump:rand} $Z\ind(S_{1},S_{0},Y_{1},Y_{0},X)$. 
\end{assumption}

Under Assumption~\ref{assump:rand}, the identification formulas
in Theorem~\ref{thm:identification-obs} hold by replacing $\pi(X)$
with $\pi$. Theorem~\ref{thm:identification-obs}(a) reduces to
the identification formulas in \citet{ding2017principal}. Moreover,
Theorem~\ref{thm:identification-obs}(b)~and~(c) provide two additional
identification formulas for each of the PCEs. Under randomization,
the EIFs have simpler forms, as shown in the following theorem.

\begin{theorem}[EIFs under randomization]
\label{thm:SET-rand} 
Suppose $\tau_u$'s are identified by the formulas in Theorem~\ref{thm:identification-obs} with  $\pi(X)$ replacing
by $\pi$. Under Assumption~\ref{assump:rand}, 
the EIFs for the PCEs are the same as those in Theorem~\ref{thm:SET}
with $\pi(X)$ replaced by $\pi$.  
\end{theorem}

Theorem~\ref{thm:SET-rand} demonstrates that knowing the treatment probability does not change the EIFs for the PCEs, which is similar to \citet{hahn1998role}'s result  that knowing the treatment probability does not change the EIF for the average causal effect in uncounfounded observational studies. 
We recommend adopting our estimators $\widehat{\tau}_{u}$
even in randomized experiments. 
The following theorem shows the double robustness property of these estimators, which is a direct application of Theorem~\ref{thm:triple}.

\begin{theorem}[Double robustness and local efficiency under randomization]\label{thm:double-rand}
Suppose Assumptions~\ref{assump:M},~\ref{assump:weak-pi},~and~\ref{assump:rand} hold, 
$\delta< \{\pi,\pi(x;\hat{\alpha})\}<1-\delta$,
	and $\{p_{1}(x;\gamma^{*}),p_{1}(x;\widehat{\gamma}),1-p_{0}(x;\gamma^{*}),1-p_{0}(x;\widehat{\gamma})\} > \delta$
	for some $\delta\in(0,1)$ and all $x$ in the support of $X$. 
The estimator
$\widehat{\tau}_{u}$ $(u=10,00,11)$ is doubly robust in the sense
that it is consistent for $\tau_{u}$ under $\Mps \cup \Mom$. Moreover, $\widehat{\tau}_{u}$ achieves the semiparametric
efficiency bound under $\Momps$. 
\end{theorem}

In randomized experiments, the treatment
probability is always correctly specified by including the null model and thus the triple
robustness simplifies to the double robustness with respect to the principal
score and outcome mean. 
Because the treatment probability does not depend on $X$ under randomization, we can simplify $\widehat{\tau}_{u}$ by replacing the estimated propensity score with the estimated treated proportion.  
However, if $\Momps$ does not hold, then 
using the estimated propensity score might improve the efficiency of the estimator. This has been pointed out by \citet{shen2014inverse} for the inverse probability weighting estimator of the average causal effect under randomization.

 \citet{ding2017principal} propose model-assisted estimators for the PCEs. However, their estimators are not based on the EIFs and use only the principal score model, without using the outcome model. As a result, their estimators are
 neither doubly robust nor semiparametrically efficient. Even in randomized experiments, their estimators are suboptimal, and we propose improved estimators.

\subsection{Strong monotonicity}

When strong monotonicity holds, we have stronger results below.

\begin{assumption}[Strong monotonicity] \label{assump:SM}$S_{0}=0$. \end{assumption}

Assumption~\ref{assump:SM} implies $S_{1}\geq S_{0}$ and is thus
stronger than Assumption~\ref{assump:M}. Assumption~\ref{assump:SM} eliminates principal strata $U=11$ and $U=01$,
and restricts $p_{0}(X)=p_{0}=0$. Therefore, there are only two principal
strata $U=10$ and $U=00$. The identification formulas for $\tau_{10}$
and $\tau_{00}$ can be obtained by applying these restrictions in
Theorem~\ref{thm:identification-obs}. The following theorem gives
their EIFs  and shows the triple robustness property of the corresponding estimators.

\begin{theorem}[EIFs and triple robustness under strong monotonicity]
\label{thm:SET-strmon}
Suppose $\tau_{10}$ and $\tau_{00}$  are identified by the formulas in Theorem~\ref{thm:identification-obs} with $p_{0}(X)=p_{0}=0$.
Under Assumption~\ref{assump:SM}, the EIFs for $\tau_{10}$ and $\tau_{00}$ are the same as those given
in Theorem~\ref{thm:SET} with $p_{0}(X)$, $p_{0}$ and $\psi_{S_{0}}$
set to $0$. 
 Suppose further that $\delta< \{\pi(x;\alpha^{*}),\pi(x;\hat{\alpha})\}<1-\delta$,
	and $ \{p_{1}(x;\gamma^{*}),p_{1}(x;\widehat{\gamma})\} > \delta$
	for some $\delta\in(0,1)$ and all $x$ in the support of $X$. 
The estimator $\widehat{\tau}_{u}$ $(u=10,00)$ is triply
robust in the sense that it is consistent for $\tau_{u}$ under
$\Mpsta \cup \Momta \cup \Momps$. Moreover, $\widehat{\tau}_{u}$
achieves the semiparametric efficiency bound under $\Mompsta$.
\end{theorem}

We use the same estimators $\widehat{\tau}_{10}$ and $\widehat{\tau}_{00}$
for $\tau_{10}$ and $\tau_{00}$, because empirically the data would
inform that $p_{0}(X;\widehat{\gamma})=\widehat{p}_{0}=0$ and $\widehat{\psi}_{S_{0}}=0$.
The triple robustness of $\widehat{\tau}_{10}$ and $\widehat{\tau}_{00}$
is the same as that given in Theorem~\ref{thm:triple}, except that
the models of $p_{0}(X)$ and $\mu_{01}(X)$ are no longer needed.

\section{Sensitivity analysis}
\label{app:sensitivity} 
Assumptions~\ref{assump:M}~and~\ref{assump:weak-pi} are crucial for the nonparametric
identification of the PCEs. However, these two assumptions cannot be easily justified by prior knowledge and may be violated in some applications.
Therefore, we
extend the sensitivity analysis in \citet{ding2017principal} from
randomized experiments to observational studies and derive the semiparametric efficiency theory for the sensitivity analysis.

\subsection{Sensitivity analysis for principal ignorability: method}
\label{sec:sensitivity-ps} 

Assumptions~\ref{assump:weak-pi} is violated if there are latent confounders
between the principal strata and outcome. For the sensitivity analysis, we assume the following
tilting model:
\begin{equation}
\frac{E(Y_{1}\mid U=10,X)}{E(Y_{1}\mid U=11,X)}=\epsilon_{1}(X),\quad\frac{E(Y_{0}\mid U=10,X)}{E(Y_{0}\mid U=00,X)}=\epsilon_{0}(X).\label{eq:sensitivityM-1}
\end{equation}
Under~\eqref{eq:sensitivityM-1}, we can treat $\epsilon_{1}(X)$
and $\epsilon_{0}(X)$ as sensitivity parameters. If $\epsilon_{1}(X)=\epsilon_{0}(X)=1$,
then Assumption~\ref{assump:weak-pi} holds. 
When the outcome is non-negative and $\epsilon_{1}(X)$ and $\epsilon_{0}(X)$ are constant not depending on $X$, \eqref{eq:sensitivityM-1} effectively assumes log-linear models on the outcome on the latent $U$ and $X$, similar to \citet{scharfstein1999adjusting}. 
Model~\eqref{eq:sensitivityM-1} also
extends \citet{ding2017principal} by allowing the sensitivity parameters
to depend on $X$, e.g., $\epsilon_{z}(X)=\exp(-X^{\T}\eta_{z})$
for $z=0,1$. The following theorem establishes the nonparametric
identification of the PCEs when the sensitivity parameters are known.

\begin{theorem}
\label{thm:identification-sens}
 Under Assumptions~\ref{assump:TAignorability},
\ref{assump:M}, and \eqref{eq:sensitivityM-1} with known $\epsilon_{1}(X)$
and $\epsilon_{0}(X)$, $e_u>0$ for $u=10,00,11$, and $0<\pi(x)<1$ for all $x$ in the support of $X$. Define
\begin{align*}
\omega_{1,10}(X) & \ =\ \frac{\epsilon_{1}(X)e_{10}(X)+\epsilon_{1}(X)e_{11}(X)}{\epsilon_{1}(X)e_{10}(X)+e_{11}(X)}, & \omega_{0,10}(X) & \ =\ \frac{\epsilon_{0}(X)e_{10}(X)+\epsilon_{0}(X)e_{00}(X)}{\epsilon_{0}(X)e_{10}(X)+e_{00}(X)},\\
\omega_{0,00}(X) & \ =\ \frac{e_{10}(X)+e_{00}(X)}{\epsilon_{0}(X)e_{10}(X)+e_{00}(X)}, & \omega_{1,11}(X) & \ =\ \frac{e_{10}(X)+e_{11}(X)}{\epsilon_{1}(X)e_{10}(X)+e_{11}(X)}.
\end{align*}
The following identification formulas hold for the PCEs.
\begin{enumerate}
	[(a)] 
\item Based on the treatment probability and principal score,
\begin{eqnarray*}
\tau_{10} & = &  \E\left\{ \frac{\omega_{1,10}(X)e_{10}(X)}{p_{1}-p_{0}}\frac{S}{p_{1}(X)}\frac{Z}{\pi(X)}Y\right\} -\E\left\{ \frac{\omega_{0,10}(X)e_{10}(X)}{p_{1}-p_{0}}\frac{1-S}{1-p_{0}(X)}\frac{1-Z}{1-\pi(X)}Y\right\},\\
\tau_{00} & = & \E\left\{ \frac{1-S}{1-p_{1}}\frac{Z}{\pi(X)}Y\right\} -\E\left\{ \frac{\omega_{0,00}(X)e_{00}(X)}{1-p_{1}}\frac{1-S}{1-p_{0}(X)}\frac{1-Z}{1-\pi(X)}Y\right\},\\
\tau_{11} & = &\E\left\{ \frac{\omega_{1,11}(X)e_{11}(X)}{p_{0}}\frac{S}{p_{1}(X)}\frac{Z}{\pi(X)}Y\right\} -\E\left\{ \frac{S}{p_{0}}\frac{1-Z}{1-\pi(X)}Y\right\}.
\end{eqnarray*}
\item Based on the treatment probability and outcome mean, 
\begin{eqnarray*}
\tau_{10} & = & \E\left[\left\{ \frac{SZ}{\pi(X)}-\frac{\textup{s}(1-Z)}{1-\pi(X)}\right\} \frac{\omega_{1,10}(X)\mu_{11}(X)-\omega_{0,10}(X)\mu_{00}(X)}{p_{1}-p_{0}}\right],\\
\tau_{00} & = & \E\left[\left\{ 1-\frac{SZ}{\pi(X)}\right\} \frac{\mu_{10}(X)-\omega_{0,00}(X)\mu_{00}(X)}{p_{1}-p_{0}}\right],\nonumber \\
\tau_{11} & = & \E\left[\frac{S(1-Z)}{1-\pi(X)}\frac{\omega_{1,11}(X)\mu_{11}(X)-\mu_{01}(X)}{p_{1}-p_{0}}\right].\nonumber 
\end{eqnarray*}
\item Based on the principal score and outcome mean,
\begin{eqnarray*}
\tau_{10} & = &  \E\left[\frac{p_{1}(X)-p_{0}(X)}{p_{1}-p_{0}}\{\omega_{1,10}(X)\mu_{11}(X)-\omega_{0,10}(X)\mu_{00}(X)\}\right],\\
\tau_{00} & = & \E\left[\frac{1-p_{1}(X)}{1-p_{1}}\{\mu_{10}(X)-\omega_{0,00}(X)\mu_{00}(X)\}\right],\nonumber \\
\tau_{11} & = & \E\left[\frac{p_{0}(X)}{p_{0}}\{\omega_{1,11}(X)\mu_{11}(X)-\mu_{01}(X)\}\right].\nonumber 
\end{eqnarray*}
\end{enumerate}
\end{theorem}

Theorem \ref{thm:identification-sens} is analogous to Theorem~\ref{thm:identification-obs}.
These formulas motivate the estimators for the PCEs by replacing the components
with their empirical versions.
Similar to Section~\ref{sec:score}, we can derive the EIFs for the
PCEs under~\eqref{eq:sensitivityM-1} as a guidance to propose more
principled estimators.

\begin{theorem} \label{thm:SET-sens}
Suppose $\tau_u$'s are identified in Theorem \ref{thm:identification-sens}. 
The EIF for $\tau_{10}$  is $\phi'_{10}=\{ \phi'_{1,10}-\phi'_{0,10}-\tau_{10}(\psi_{S_{1}}-\psi_{S_{0}})\} /(p_{1}-p_{0}),$
where 
\begin{eqnarray*}
\phi'_{1,10} & = & \frac{\omega_{1,10}(X)e_{10}(X)}{p_{1}(X)}\psi_{Y_{1}S_{1}}-\frac{\omega_{1,10}^{2}(X)\mu_{11}(X)}{\epsilon_{1}(X)}\left\{ \psi_{S_{0}}-\frac{p_{0}(X)}{p_{1}(X)}\psi_{S_{1}}\right\} ,\\
\phi'_{0,10} & = & \frac{\omega_{0,10}(X)e_{10}(X)}{1-p_{0}(X)}\psi_{Y_{0}(1-S_{0})}-\frac{\omega_{0,10}^{2}(X)\mu_{00}(X)}{\epsilon_{0}(X)}\left\{ \psi_{1-S_{1}}-\frac{1-p_{1}(X)}{1-p_{0}(X)}\psi_{1-S_{0}}\right\} .
\end{eqnarray*}
The EIF for $\tau_{00}$ is $\phi'_{00}=\left(\phi'_{1,00}-\phi'_{0,00}-\tau_{00}\psi_{1-S_{1}}\right)/(1-p_{1})$,
where $\phi'_{1,00}=\phi_{1,00}$ and 
\begin{eqnarray*}
\phi'_{0,00} & = & \frac{\omega_{0,00}(X)e_{00}(X)}{1-p_{0}(X)}\psi_{Y_{0}(1-S_{0})}+\frac{\omega_{0,00}^{2}(X)\mu_{00}(X)}{\epsilon_{0}(X)}\left\{ \psi_{1-S_{1}}-\frac{1-p_{1}(X)}{1-p_{0}(X)}\psi_{1-S_{0}}\right\} .
\end{eqnarray*}
The EIF for $\tau_{11}$ is $\phi'_{11}=\left(\phi'_{1,11}-\phi'_{0,11}-\tau_{11}\psi_{S_{0}}\right)/p_{0}$,
where $\phi'_{0,11}=\phi_{0,11}$ and 
\begin{eqnarray*}
\phi'_{1,11} & = & \frac{\omega_{1,11}(X)e_{11}(X)}{p_{1}(X)}\psi_{Y_{1}S_{1}}+\frac{\omega_{1,11}^{2}(X)\mu_{11}(X)}{\epsilon_{1}(X)}\left\{ \psi_{S_{0}}-\frac{p_{0}(X)}{p_{1}(X)}\psi_{S_{1}}\right\} .
\end{eqnarray*}
\end{theorem}

We can construct estimators for the PCEs by solving the estimating
equation $\mathbb{P}_{n}\phi'_{u}=0$ with the components replaced
by their empirical versions:
\begin{equation*}
\widehat{\tau}'_{10}=\frac{ \mathbb{P}_n (\widehat{\phi}'_{1,10}-\widehat{\phi}'_{0,10})}{ \mathbb{P}_n (\widehat{\psi}_{S_{1}}-\widehat{\psi}_{S_{0}})},\quad\widehat{\tau}'_{00}=\frac{ \mathbb{P}_n (\widehat{\phi}'_{1,00}-\widehat{\phi}'_{0,00})}{ \mathbb{P}_n (1-\widehat{\psi}_{S_{1}})},\quad\widehat{\tau}'_{11}=\frac{ \mathbb{P}_n (\widehat{\phi}'_{1,11}-\widehat{\phi}'_{0,11})}{ \mathbb{P}_n (\widehat{\psi}_{S_{0}})} . 
\end{equation*}
When $\epsilon_{1}(X)=\epsilon_{0}(X)=1$,
these estimators reduce to $\widehat{\tau}_{u}$ given in Section~\ref{sec:score}.
Unlike $\widehat{\tau}_{u}$, they are not triply robust but only doubly robust, similar to \citet{tchetgen2012semiparametric}'s sensitivity analysis for the natural direct and indirect effects. 


\begin{theorem}\label{thm:sens-dr}
Suppose Assumptions~\ref{assump:TAignorability}, \ref{assump:M},
and \eqref{eq:sensitivityM-1} hold with known $\epsilon_{1}(X) \neq 0$ and $\epsilon_{0}(X) \neq 0$.  Suppose further that $\delta<\{\pi(x;\alpha^{*}),\pi(x;\hat{\alpha})\}<1-\delta$, 
	and  $\{ p_{1}(x;\gamma^{*}),p_{1}(x;\widehat{\gamma}),1-p_{0}(x;\gamma^{*}),1-p_{0}(x;\widehat{\gamma})\} > \delta$
	for some $\delta\in(0,1)$ and all $x$ in the support of $X$. 
The estimator $\widehat{\tau}'_{u}$ $(u=10,00,11)$
is doubly robust in the sense that it is consistent for $\tau_{u}$
under $\Mpsta \cup \Momps$. 
Moreover, $\widehat{\tau}'_{u}$ has the influence function $\phi_{u}$
and therefore achieves the semiparametric efficiency bound under $\Mompsta$.
\end{theorem}

Our sensitivity analysis estimators require a correct principal score model, weakening the triple robustness to double robustness. 

\subsection{Sensitivity analysis for principal ignorability: examples}

\begin{figure}[t]
	\centering
	\includegraphics[width=0.8\textwidth,height=0.25\textheight]{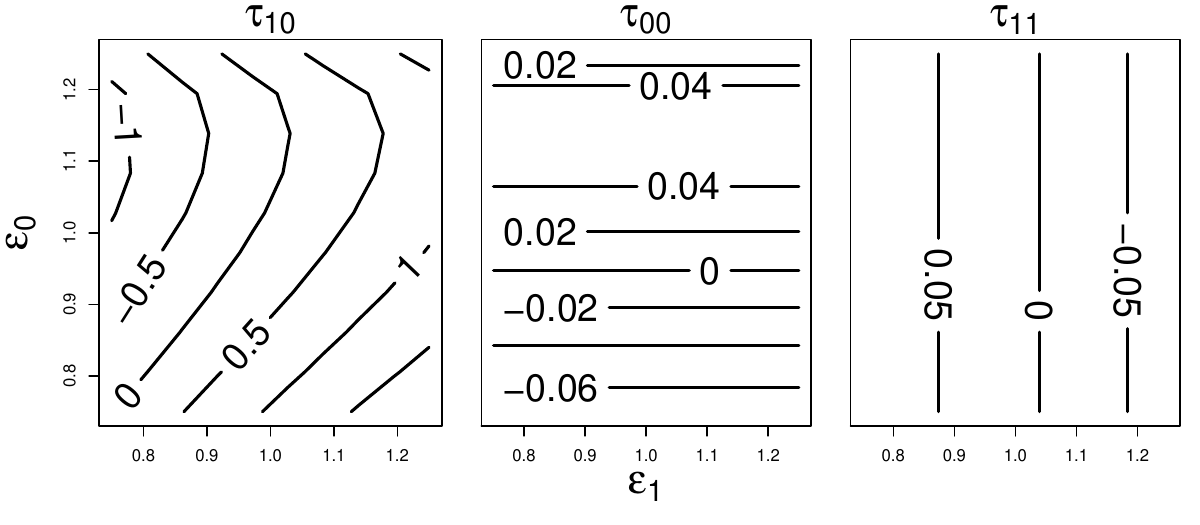}
	\includegraphics[width=0.8\textwidth,height=0.25\textheight]{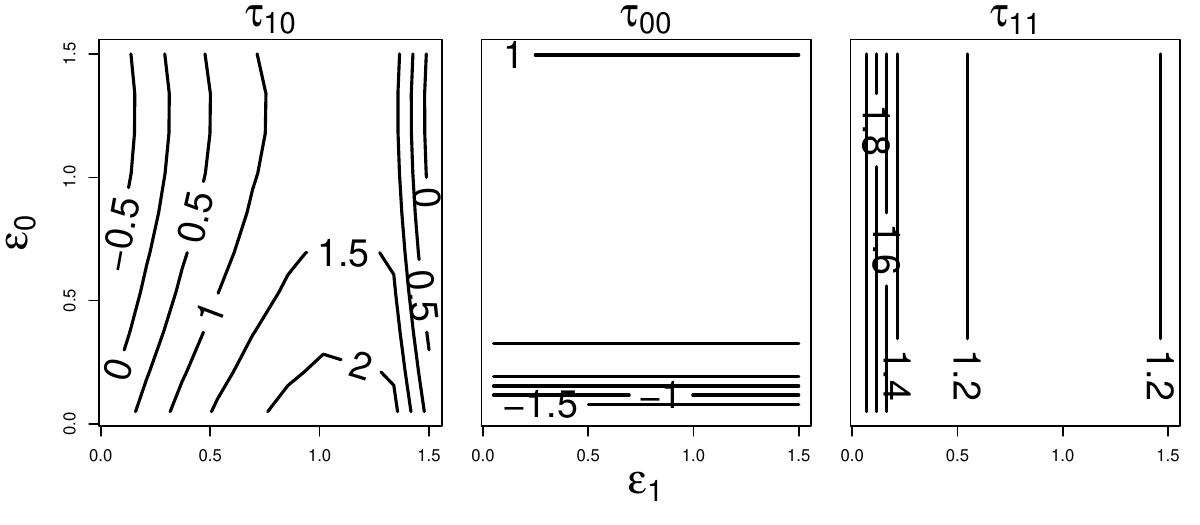}
	\caption{Sensitivity analysis for Sections~\ref{sec:RTS} (top)~and~\ref{subsec:flooding} (bottom).} 
	\label{fig:sen} 
\end{figure}

We first re-visit the example in Section \ref{sec:RTS} to assess the robustness of the conclusions to the violation of principal ignorability. For the ease of presentation, we assume the sensitivity parameters not dependent on $X$, i.e, $\epsilon_0(X)=\epsilon_0$ and $\epsilon_1(X)=\epsilon_1$, and vary them from $0.75$ to $1.25$. The upper panel of
Figure~\ref{fig:sen} displays the contour plots of the estimated $\tau_{10}$, $\tau_{00}$, and $\tau_{11}$ with different values of $(\epsilon_0,\epsilon_1)$. When 
principal ignorability holds ($\epsilon_0=1,\epsilon_1=1$), $\widehat\tau_{10}$ is positive. However, the estimate varies from negative to positive with small changes of ($\epsilon_0,\epsilon_1$) from $0.75$ to $1.25$. This implies that the result is sensitive to the violation of principal ignorability.
Similar discussions apply to $\widehat\tau_{00}$ and $\widehat\tau_{11}$.

We then re-visit the example in Section \ref{subsec:flooding}.
The lower panel of
Figure~\ref{fig:sen} displays the contour plots of the estimated parameters. 
Unlike the example in Section~\ref{sec:RTS}, the result is not very sensitive to the violation of principal ignorability. In particular, $\widehat\tau_{10}$ changes the sign only when $\epsilon_1$ becomes close to $1.25$, $\widehat\tau_{00}$ remains positive in a wide range of $\epsilon_0$, and $\widehat\tau_{11}$ remains positive for all values of $\epsilon_1$.  

\subsection{Sensitivity analysis for monotonicity}
\label{sec:sensitivity-mon} 
Without monotonicity, we have an additional stratum $U=01$. In this case, the proportions of principal strata are not identifiable without further assumptions. Define the sensitivity parameter as the ratio between the proportion of stratum $01$ and stratum $10$ conditional on the covariates,
\begin{eqnarray}
\label{eq:sensitivity-mon}
\xi(X) \ = \ \frac{\Pr(U=01\mid X)}{\Pr(U=10\mid X)}.
\end{eqnarray}
It characterizes the deviation from monotonicity. If $\xi(X) = 0 $, then monotonicity holds. \citet{ding2017principal} discuss a special case with $\xi(X) $ being a constant.

 For a fixed sensitivity parameter $\xi(X) \neq 1$, we can identify the principal scores by
 \begin{eqnarray}
 \label{eq:ps-sens-mon}
\begin{array}{ll}
e_{\xi,10}(X)  = \frac{p_1(X)- p_0(X)}{1-\xi(X)}, & e_{\xi,00}(X)  =  1-p_0(X)- \frac{p_1(X)- p_0(X)}{1-\xi(X)},\\
e_{\xi,11}(X)  = p_1(X)- \frac{p_1(X)- p_0(X)}{1-\xi(X)}, &  e_{\xi,01}(X)  =  \frac{\xi(X)\{p_1(X)- p_0(X)\}}{1-\xi(X)}
\end{array}
\end{eqnarray}
%
and the proportions of principal strata by $e_{\xi,u}=\E\{e_{\xi,u}(X)\}$ for all $u$.  
When $\xi(X)=1$, strata $10$ and $01$ have equal proportions. This corresponds to the boundary case when the treatment has zero average causal effect on $S$. We rule out this boundary case because the principal scores are not identifiable.

Without monotonicity, we need to extend principal ignorability to include stratum $01$.

\begin{assumption}[Principal ignorability without monotonicity]\label{assump:weak-pi-nonmon}
$\E(Y_{1}\mid U=s_11,X)=\E(Y_{1}\mid U=s_10,X)$ for $s_1=0,1$ and $\E(Y_{0}\mid U=1s_0,X)=E(Y_{0}\mid U=0s_0,X)$ for $s_0=0,1$.
\end{assumption}

The following theorem establishes the nonparametric identification of the PCEs with a known $\xi(X)$.

\begin{theorem}
\label{thm:identification-sens-mon} Under Assumptions~\ref{assump:TAignorability},~\ref{assump:weak-pi-nonmon}, and~\eqref{eq:sensitivity-mon} with a known $\xi(X) \neq 1$, 
$e_{\xi,u}>0$ for $u=10,00,11$, and $0<\pi(x)<1$ for all $x$ in the support of $X$. The following identification formulas hold for the PCEs.
\begin{enumerate}
	[(a)] 
\item Based on the treatment probability and principal score,
\begin{eqnarray*}
\tau_{10} & = &   \E\left\{ \frac{e_{\xi,10}(X)}{e_{\xi,10}}\frac{S}{p_{1}(X)}\frac{Z}{\pi(X)}Y\right\} - \E\left\{ \frac{e_{\xi,10}(X)}{e_{\xi,10}}\frac{1-S}{1-p_{0}(X)}\frac{1-Z}{1-\pi(X)}Y\right\},\\
\tau_{00} & = &\E\left\{ \frac{e_{\xi,10}(X)}{e_{\xi,10}}\frac{S}{p_{1}(X)}\frac{Z}{\pi(X)}Y\right\} -\E\left\{ \frac{e_{\xi,10}(X)}{e_{\xi,10}}\frac{1-S}{1-p_{0}(X)}\frac{1-Z}{1-\pi(X)}Y\right\},\\
\tau_{11} & = &\E\left\{ \frac{e_{\xi,11}(X)}{e_{\xi,11}}\frac{S}{p_{1}(X)}\frac{Z}{\pi(X)}Y\right\} -\E\left\{ \frac{e_{\xi,11}(X)}{e_{\xi,11}}\frac{S}{p_{0}(X)}\frac{1-Z}{1-\pi(X)}Y\right\},\\
\tau_{01} & =&  \E\left\{ \frac{e_{\xi,01}(X)}{e_{\xi,01}}\frac{1-S}{1-p_{1}(X)}\frac{Z}{\pi(X)}Y\right\} -\E\left\{ \frac{e_{\xi,01}(X)}{e_{\xi,01}}\frac{1-S}{1-p_{0}(X)}\frac{1-Z}{1-\pi(X)}Y\right\}.
\end{eqnarray*}
\item Based on the treatment probability and outcome mean, 
\begin{eqnarray*}
\tau_{10} & = & \E\left[\left\{ \frac{SZ}{\{1-\xi(X)\}\pi(X)}-\frac{S(1-Z)}{\{1-\xi(X)\}\{1-\pi(X)\}}\right\} \frac{\mu_{11}(X)-\mu_{00}(X)}{e_{\xi,10}}\right],\\
\tau_{00} & = & \E\left[\left\{ 1-\frac{SZ}{\{1-\xi(X)\}\pi(X)}-\frac{\xi(X)S(1-Z)}{\{1-\xi(X)\}\{1-\pi(X)\}}\right\} \frac{\mu_{10}(X)-\mu_{00}(X)}{e_{\xi,00}}\right],\nonumber \\
\tau_{11} & = & \E\left[\left\{ -\frac{\xi(X)SZ}{\{1-\xi(X)\}\pi(X)}+\frac{S(1-Z)}{\{1-\xi(X)\}\{1-\pi(X)\}}\right\} \frac{\mu_{11}(X)-\mu_{01}(X)}{e_{\xi,11}}\right], \\
\tau_{01} &=&\E\left[\left\{ \frac{\xi(X)SZ}{\{1-\xi(X)\}\pi(X)}-\frac{\xi(X)S(1-Z)}{\{1-\xi(X)\}\{1-\pi(X)\}}\right\} \frac{\mu_{10}(X)-\mu_{01}(X)}{e_{\xi,01}}\right].
\end{eqnarray*}
\item Based on the principal score and outcome mean,
\begin{eqnarray*}
\tau_{10} & = &  \E\left[\frac{e_{\xi,10}(X)}{e_{\xi,10}}\{\mu_{11}(X)-\mu_{00}(X)\}\right],\\
\tau_{00} & = &  \E\left[\frac{e_{\xi,00}(X)}{e_{\xi,00}}\{\mu_{10}(X)-\mu_{00}(X)\}\right],\nonumber \\
\tau_{11} & = & \E\left[\frac{e_{\xi,11}(X)}{e_{\xi,11}}\{\mu_{11}(X)-\mu_{01}(X)\}\right],\nonumber  \\
\tau_{01} & =& \E\left[\frac{e_{\xi,01}(X)}{e_{\xi,01}}\{\mu_{10}(X)-\mu_{01}(X)\}\right].
\end{eqnarray*}
\end{enumerate}
\end{theorem}

Theorem \ref{thm:identification-sens-mon} is analogous to Theorem~\ref{thm:identification-obs}.
These formulas motivate the estimators for the PCEs by replacing the components
with their empirical versions.
 Similar to Section~\ref{sec:score}, we can derive the EIFs for the
PCEs under~\eqref{eq:sensitivity-mon} as guidance to propose more
principled estimators. For simplicity, we only give the result for $\tau_{10}$.

\begin{theorem} \label{thm:SET-sens-mon}
Suppose $\tau_u$'s are identified in Theorem \ref{thm:identification-sens-mon}. 
The EIF for $\tau_{10}$  is 
$$\phi^\ast_{10}=\frac{1}{e_{\xi,10} } \left\{ (\phi^\ast_{1,10}-\phi^\ast_{0,10}) - \frac{\tau_{10}(\psi_{S_{1}}-\psi_{S_{0}})}{1-\xi(X)} \right\},$$
where 
\begin{eqnarray*}
\phi^\ast_{1,10} & = & \frac{p_{1}(X)-p_0(X)}{\{1-\xi(X)\}p_{1}(X)}\psi_{Y_{1}S_{1}}-\frac{\mu_{11}(X)}{1-\xi(X)}\left\{ \psi_{S_{0}}-\frac{p_{0}(X)}{p_{1}(X)}\psi_{S_{1}}\right\} ,\\
\phi^\ast_{0,10} & = &\frac{p_{1}(X)-p_0(X)}{ \{1-\xi(X)\}\{1-p_{0}(X)\}}\psi_{Y_{0}(1-S_{0})}-\frac{\mu_{00}(X)}{1-\xi(X)}\left\{ \psi_{1-S_{1}}-\frac{1-p_{1}(X)}{1-p_{0}(X)}\psi_{1-S_{0}}\right\} .
\end{eqnarray*}
\end{theorem}

We can construct estimators for the PCEs by solving the estimating
equation $\mathbb{P}_{n}\phi^\ast_{u}=0$ with the components replaced
by their empirical versions, e.g.,
\begin{equation*}
\widehat{\tau}'_{10}=\frac{ \mathbb{P}_n (\widehat{\phi}^\ast_{1,10}-\widehat{\phi}^\ast_{0,10})}{ \mathbb{P}_n (\widehat{\psi}_{S_{1}}-\widehat{\psi}_{S_{0}})} . 
\end{equation*}
When $\xi(X)=0$,
these estimators reduce to $\widehat{\tau}_{u}$ given in Section~\ref{sec:score}.
The following theorem shows the triple robustness and local efficiency of the estimators constructed based on the EIFs.

\begin{theorem}\label{thm:sens-mon-tr}
Suppose Assumptions~\ref{assump:TAignorability}, \ref{assump:weak-pi-nonmon},
and~\eqref{eq:sensitivity-mon} hold with a known $\xi(X) \neq 1$, $\delta< \{\pi(x;\alpha^{*}),\pi(x;\hat{\alpha})\}<1-\delta$,
and  $\{p_{1}(x;\gamma^{*}),p_{1}(x;\widehat{\gamma}),1-p_{0}(x;\gamma^{*}),1-p_{0}(x;\widehat{\gamma})\} > \delta$
for some $\delta\in(0,1)$ and all $x$ in the support of $X$. 
The estimator $\widehat{\tau}^\ast_{u}$ $(u=10,00,11,01)$
is triply robust in the sense that it is consistent for $\tau_{u}$
under $\Mpsta \cup \Momta \cup \Momps$. 
Moreover, $\widehat{\tau}^\ast_{u}$ has the influence function $\phi^\ast_{u}$
and therefore achieves the semiparametric efficiency bound under $\Mompsta$.
\end{theorem}

\section{Proof of the identification results}

\label{app:identification}

Throughout the proofs, we will use $f(\cdot)$ to denote the probability
density functions for continuous random variables and the probability mass
functions for discrete random variables. We will use the law of total
expectation repeatedly and will mark the steps with $\LOTE$. We will
use the following lemma repeatedly, the proof of which is straightforward. 

\begin{lemma}\label{lem:1} Let $X$ and $Y$ be two random variables
with densities $f_{1}(x)$ and $f_{2}(y)$. For any function $h(\cdot)$
with $\E\{h(X)\}<\infty$, we have 
\begin{eqnarray*}
\E\{h(X)\}=\E\left\{ \frac{f_{1}(Y)}{f_{2}(Y)}h(Y)\right\} .
\end{eqnarray*}
\end{lemma}

\subsection{Proof of Theorem \ref{thm:balancing}}

It follows from the LOTP and Lemma \ref{lem:1} by conditioning on $X$. We omit the details due to the length of the supplementary material. The earliest ArXiv version of our paper contains the details.

\subsection{Proof of Theorem~\ref{thm:identification-obs}}

\label{app:identification-obs} We prove only the identification formulas
for $\tau_{10}$ and omit the similar proofs of the identification
formulas for $\tau_{00}$ and $\tau_{11}$. We have 
\begin{eqnarray}
\tau_{10} & = & \E(Y_{1}\mid U=10)-\E(Y_{0}\mid U=10)\nonumber \\
 & = & \E\left\{ \E(Y_{1}\mid U=10,X)\mid U=10\right\} -\E\left\{ \E(Y_{0}\mid U=10,X)\mid U=10\right\} \nonumber \\
 & = & \E\left\{ \E\left(Y_{1}\mid U=10\ {\rm or}\ 11,X\right)\mid U=10\right\} \nonumber \\
 & & -\E\left\{ \E\left(Y_{0}\mid U=10\ {\rm or}\ 00,X\right)\mid U=10\right\} \quad(\text{Assumption}~\ref{assump:weak-pi})\nonumber \\
 & = & \E\left\{ \E\left(Y\mid Z=1,U=10\ {\rm or}\ 11,X\right)\mid U=10\right\} \nonumber \\
 & & -\E\left\{ \E\left(Y\mid Z=0,U=10\ {\rm or}\ 00,X\right)\mid U=10\right\} \quad(\text{Assumption}~\ref{assump:TAignorability})\nonumber \\
 & = & \E\left\{ \E\left(Y\mid Z=1,S=1,X\right)\mid U=10\right\} -\E\left\{ \E\left(Y\mid Z=0,S=0,X\right)\mid U=10\right\} \nonumber \\
 & = & \E\{\mu_{11}(X)\mid U=10\}-\E\{\mu_{00}(X)\mid U=10\}.\label{eqn:thm1-10-mu}
\end{eqnarray}
Theorem~\ref{thm:balancing}  ensures that for any
$h(X)$, 
\begin{eqnarray}
\E\{h(X)\mid U=10\} & = & \E\left[\left\{ \frac{SZ}{\pi(X)}-\frac{S(1-Z)}{1-\pi(X)}\right\} \frac{h(X)}{p_{1}-p_{0}}\right],\label{eqn:thm1-10-1}\\
 & = & \E\left\{ \frac{p_{1}(X)-p_{0}(X)}{p_{1}-p_{0}}h(X)\right\} .\label{eqn:thm1-10-2}
\end{eqnarray}
Applying~\eqref{eqn:thm1-10-1} to the two terms in~\eqref{eqn:thm1-10-mu},
we obtain 
\begin{eqnarray*}
\tau_{10}=\E\left[\left\{ \frac{SZ}{\pi(X)}-\frac{S(1-Z)}{1-\pi(X)}\right\} \frac{\mu_{11}(X)-\mu_{00}(X)}{p_{1}-p_{0}}\right],
\end{eqnarray*}
which is the identification formula in Theorem~\ref{thm:identification-obs}(b).
Applying~\eqref{eqn:thm1-10-2} to the two terms in~\eqref{eqn:thm1-10-mu},
we obtain 
\begin{eqnarray}
\tau_{10}=\E\left[\frac{p_{1}(X)-p_{0}(X)}{p_{1}-p_{0}}\{\mu_{11}(X)-\mu_{00}(X)\}\right],\label{eqn:thm1c-10}
\end{eqnarray}
which is the identification formula in Theorem~\ref{thm:identification-obs}(c).

We finally prove the identification formula in Theorem~\ref{thm:identification-obs}(a).
We have, 
\begin{eqnarray*}
\E\left\{ \frac{e_{10}(X)}{p_{1}-p_{0}}\frac{S}{p_{1}(X)}\frac{Z}{\pi(X)}Y\right\} & = & \E\left[\E\left\{ \frac{e_{10}(X)}{p_{1}-p_{0}}\frac{S}{p_{1}(X)}\frac{Z}{\pi(X)}Y \mid X\right\} \right]\quad(\LOTE)\nonumber \\
 & = & \E\left\{ \frac{e_{10}(X)}{p_{1}-p_{0}}\frac{\pr(Z=1,S=1\mid X)}{p_{1}(X)\pi(X)}\mu_{11}(X)\right\} \nonumber \\
 & = & \E\left\{ \frac{p_{1}(X)-p_{0}(X)}{p_{1}-p_{0}}\mu_{11}(X)\right\} ,\label{eqn:thm1a-10-1}\\
\E\left\{ \frac{e_{10}(X)}{p_{1}-p_{0}}\frac{1-S}{1-p_{0}(X)}\frac{1-Z}{1-\pi(X)}Y\right\} & = & \E\left[\E\left\{ \frac{e_{10}(X)}{p_{1}-p_{0}}\frac{1-S}{1-p_{0}(X)}\frac{1-Z}{1-\pi(X)}Y\mid X\right\} \right]\quad(\LOTE)\nonumber \\
 & = & \E\left\{ \frac{e_{10}(X)}{p_{1}-p_{0}}\frac{\pr(Z=0,S=0\mid X)}{\{1-p_{0}(X)\}\{1-\pi(X)\}}\mu_{00}(X)\right\} \nonumber \\
 & = & \E\left\{ \frac{p_{1}(X)-p_{0}(X)}{p_{1}-p_{0}}\mu_{00}(X)\right\} ,\label{eqn:thm1a-10-2}
\end{eqnarray*}
which, coupled with \eqref{eqn:thm1c-10}, imply the identification
formula in Theorem~\ref{thm:identification-obs}(a). \QEDB

\subsection{Proof of Theorem~\ref{thm:identification-sens}}

\label{app:identification-sens} 
We prove only the identification formulas
for $\tau_{10}$ and omit the similar proofs of the identification
formulas for $\tau_{00}$ and $\tau_{11}$.
We have, 
\begin{eqnarray*}
\mu_{11}(X) & = & \E(Y_{1}\mid Z=1,S=1,U=10,X)\pr(U=10\mid Z=1,S=1,X)\\
 & & +\E(Y_{1}\mid Z=1,S=1,U=11,X)\pr(U=11\mid Z=1,S=1,X)\quad(\LOTE)\\
 & = & \E(Y_{1}\mid U=10,X)\pr(U=10\mid Z=1,S=1,X)\\
 & & +\E(Y_{1}\mid U=11,X)\pr(U=11\mid Z=1,S=1,X)\quad(\text{Assumption}~\ref{assump:weak-pi})\\
 & = & \E(Y_{1}\mid U=10,X)\frac{e_{10}(X)}{e_{10}(X)+e_{11}(X)}+\E(Y_{1}\mid U=11,X)\frac{e_{11}(X)}{e_{10}(X)+e_{11}(X)}\\
 & = & \E(Y_{1}\mid U=10,X)\frac{\epsilon_{1}(X)e_{10}(X)+e_{11}(X)}{\epsilon_{1}(X)\{e_{10}(X)+e_{11}(X)\}}.\quad(\text{by }\eqref{eq:sensitivityM-1})
\end{eqnarray*}
Therefore, 
\begin{eqnarray}
\E(Y_{1}\mid U=10,X) & = & \frac{\epsilon_{1}(X)\{e_{10}(X)+e_{11}(X)\}}{\epsilon_{1}(X)e_{10}(X)+e_{11}(X)}\mu_{11}(X)=\omega_{1,10}(X)\mu_{11}(X).\label{eq:y1-10}
\end{eqnarray}
Similarly, 
\begin{eqnarray*}
\E(Y_{0}\mid U=10,X) & = & \frac{\epsilon_{0}(X)\{e_{10}(X)+e_{00}(X)\}}{\epsilon_{0}(X)e_{10}(X)+e_{00}(X)}\mu_{00}(X)=\omega_{0,10}(X)\mu_{00}(X).\label{eq:y0-10}
\end{eqnarray*}
The above two formulas imply
\begin{eqnarray}
\tau_{10} & = & \E\{\E(Y_{1}\mid U=10,X)\mid U=10\}-\E\{\E(Y_{0}\mid U=10,X)\mid U=10\}\nonumber \\
 & = & \E\left\{ \omega_{1,10}(X)\mu_{11}(X)\mid U=10\right\} -\E\left\{ \omega_{0,10}(X)\mu_{00}(X)\mid U=10\right\} .\label{eqn:thm9-10}
\end{eqnarray}
Applying~\eqref{eqn:thm1-10-1} to the two terms in~\eqref{eqn:thm9-10},
we obtain 
\begin{eqnarray}
\tau_{10}=\E\left[\left\{ \frac{SZ}{\pi(X)}-\frac{S(1-Z)}{1-\pi(X)}\right\} \frac{\omega_{1,10}(X)\mu_{11}(X)-\omega_{0,10}(X)\mu_{00}(X)}{p_{1}-p_{0}}\right],\label{eqn:thm9b-10}
\end{eqnarray}
which is the identification formula in Theorem~\ref{thm:identification-sens}(b).
Applying~\eqref{eqn:thm1-10-2} to the two terms in~\eqref{eqn:thm9-10},
we obtain 
\begin{eqnarray}
\tau_{10}=\E\left[\frac{p_{1}(X)-p_{0}(X)}{p_{1}-p_{0}}\{\omega_{1,10}(X)\mu_{11}(X)-\omega_{0,10}(X)\mu_{00}(X)\}\right],\label{eqn:thm9c-10}
\end{eqnarray}
which is the identification formula in Theorem~\ref{thm:identification-sens}(c).

We finally prove the identification formula in Theorem~\ref{thm:identification-sens}(a). We have
\begin{eqnarray*}
 & & \E\left\{ \frac{\omega_{1,10}(X)e_{10}(X)}{p_{1}-p_{0}}\frac{S}{p_{1}(X)}\frac{Z}{\pi(X)}Y\right\} \nonumber \\
 & = & \E\left[\E\left\{ \frac{\omega_{1,10}(X)e_{10}(X)}{p_{1}-p_{0}}\frac{S}{p_{1}(X)}\frac{Z}{\pi(X)}Y\right\} \mid X\right]\quad(\LOTE)\nonumber \\
 & = & \E\left\{ \frac{\omega_{1,10}(X)e_{10}(X)}{p_{1}-p_{0}}\frac{\pr(Z=1,S=1\mid X)}{p_{1}(X)\pi(X)}\mu_{11}(X)\right\} \nonumber \\
 & = & \E\left\{ \frac{p_{1}(X)-p_{0}(X)}{p_{1}-p_{0}}\omega_{1,10}(X)\mu_{11}(X)\right\} ,\label{eqn:thm9a-10-1}
\end{eqnarray*}
and 
\begin{eqnarray*}
 & & \E\left\{ \frac{\omega_{0,10}(X)e_{10}(X)}{p_{1}-p_{0}}\frac{1-S}{1-p_{0}(X)}\frac{1-Z}{1-\pi(X)}Y\right\} \nonumber \\
 & = & \E\left[\E\left\{ \frac{\omega_{0,10}(X)e_{10}(X)}{p_{1}-p_{0}}\frac{1-S}{1-p_{0}(X)}\frac{1-Z}{1-\pi(X)}Y\mid X\right\} \right]\quad(\LOTE)\nonumber \\
 & = & \E\left\{ \frac{\omega_{0,10}(X)e_{10}(X)}{p_{1}-p_{0}}\frac{\pr(Z=0,S=0\mid X)}{\{1-p_{0}(X)\}\{1-\pi(X)\}}\mu_{00}(X)\right\} \nonumber \\
 & = & \E\left\{ \frac{p_{1}(X)-p_{0}(X)}{p_{1}-p_{0}}\omega_{0,10}(X)\mu_{00}(X)\right\},\label{eqn:thm9a-10-2}
\end{eqnarray*}
which, coupled with~\eqref{eqn:thm9c-10}, imply the identification formula in Theorem~\ref{thm:identification-sens}(a).
\QEDB

\subsection{Proof of Theorem~\ref{thm:identification-sens-mon}}
\label{app:identification-sens-mon} 
We prove only the identification formulas
for $\tau_{10}$ and omit the similar proofs of the identification
formulas for $\tau_{00}$, $\tau_{11}$, and $\tau_{01}$.
Under Assumptions~\ref{assump:TAignorability}~and~\ref{assump:weak-pi-nonmon},
we have 
\begin{eqnarray*}
	\E(Y_{1}\mid U=10,X) & = & \E(Y_{1}\mid S_{1}=1,X)=\mu_{11}(X).
\end{eqnarray*}
Therefore, 
\begin{eqnarray*}
	\E(Y_{1}\mid U=10) & = & \E\{\E(Y_{1}\mid U=10,X)\mid U=10\}\\
	& = & \E\{\mu_{11}(X)\mid U=10\}\\
	& = & \E\left\{ \frac{\Pr(U=10\mid X)}{\Pr(U=10)}\mu_{11}(X)\right\} \quad(\text{Lemma}~\ref{lem:1})\\
	& = & \E\left\{ \frac{e_{\xi,10}(X)}{e_{\xi,10}}\mu_{11}(X)\right\} .
\end{eqnarray*}
Similarly, 
\begin{eqnarray*}
	\E(Y_{0}\mid U=10) & = & \E\left\{ \frac{e_{\xi,10}(X)}{e_{\xi,10}}\mu_{00}(X)\right\} .
\end{eqnarray*}
The above two formulas imply 
\begin{eqnarray}
\tau_{10} & = & \E\left[\frac{e_{\xi,10}(X)}{e_{\xi,10}}\{\mu_{11}(X)-\mu_{00}(X)\}\right],\label{eqn:sens-mon-10-1}
\end{eqnarray}
which is the identification formula in Theorem~\ref{thm:identification-sens-mon}(c).
Applying Theorem~\ref{thm:balancing}(a) to~\eqref{eqn:sens-mon-10-1}, we have 
\begin{eqnarray*}
\tau &=&  \E\left\{ \frac{p_{1}(X)-p_{0}(X)}{1-\xi(X)}\frac{\mu_{11}(X)-\mu_{00}(X)}{e_{\xi,10}}\right\} \nonumber \\
& =& \E\left[\left\{ \frac{SZ}{\pi(X)}-\frac{S(1-Z)}{1-\pi(X)}\right\} \frac{\mu_{11}(X)-\mu_{00}(X)}{\{1-\xi(X)\}e_{\xi,10}}\right],\label{eqn:sens-mon-10-2}
\end{eqnarray*}
which is the identification formula in Theorem~\ref{thm:identification-sens-mon}(b).

We finally prove the identification formula in Theorem~\ref{thm:identification-sens-mon}(a). We have
\begin{eqnarray}
& & \E\left\{ \frac{e_{\xi,10}(X)}{e_{\xi,10}}\frac{S}{p_{1}(X)}\frac{Z}{\pi(X)}Y\right\} \nonumber \\
& = & \E\left[\E\left\{ \frac{e_{\xi,10}(X)}{e_{\xi,10}}\frac{S}{p_{1}(X)}\frac{Z}{\pi(X)}Y\right\} \mid X\right]\quad(\LOTE)\nonumber \\
& = & \E\left\{ \frac{e_{\xi,10}(X)}{e_{\xi,10}}\frac{\pr(Z=1,S=1\mid X)}{p_{1}(X)\pi(X)}\mu_{11}(X)\right\} \nonumber \\
& = & \E\left\{ \frac{e_{\xi,10}(X)}{e_{\xi,10}}\mu_{11}(X)\right\} \label{eqn:sens-mon-10-3}
\end{eqnarray}
and 
\begin{eqnarray}
& & \E\left\{ \frac{e_{\xi,10}(X)}{e_{\xi,10}}\frac{1-S}{1-p_{0}(X)}\frac{1-Z}{1-\pi(X)}Y\right\} \nonumber \\
& = & \E\left[\E\left\{ \frac{e_{\xi,10}(X)}{e_{\xi,10}}\frac{1-S}{p_{0}(X)}\frac{1-Z}{1-\pi(X)}Y\right\} \mid X\right]\quad(\LOTE)\nonumber \\
& = & \E\left\{ \frac{e_{\xi,10}(X)}{e_{\xi,10}}\frac{\pr(Z=0,S=1\mid X)}{\{1-p_{0}(X)\}\{1-\pi(X)\}}\mu_{00}(X)\right\} \nonumber \\
& = & \E\left\{ \frac{e_{\xi,10}(X)}{e_{\xi,10}}\mu_{00}(X)\right\}, \label{eqn:sens-mon-10-4}
\end{eqnarray}
which,  coupled with~\eqref{eqn:sens-mon-10-1}, imply the identification formula in Theorem~\ref{thm:identification-sens-mon}(a).
\QEDB

\section{Proof of the EIFs}

In this section, we prove Theorems~\ref{thm:SET},~\ref{thm:SET-rand},~\ref{thm:double-rand},~\ref{thm:SET-strmon},~\ref{thm:SET-sens},~and~\ref{thm:SET-sens-mon}. \label{app:EIF} 

\subsection{Proof of Theorem~\ref{thm:SET}}

\label{app:SET} 

\subsubsection{Preliminaries}

\label{app:SET-lemma} 
We will use the semiparametric theory in \citet{bickel1993efficient}
to derive the EIFs. Let $V=(X,Z,S,Y)$ be the vector of all observed
variables with the likelihood factorized as 
\begin{eqnarray}
f(V) & = & f(X)f(Z\mid X)f(S\mid Z,X)f(Y\mid Z,S,X).\label{eq:CR-lik}
\end{eqnarray}
For $z=0,1$ and $u=10,00,11$, let $\mu_{z,u}$ be the identification of  $\E(Y_{z}\mid U=u)$ given in Theorem~\ref{thm:identification-obs}.
Then, we have $\tau_{u}=\mu_{1,u}-\mu_{0,u}$. To derive the EIFs,
we consider a one-dimensional parametric submodel, $f_{\theta}(V)$,
which contains the true model $f(V)$ at $\theta=0$, i.e., $f_{\theta}(V)\vert_{\theta=0}=f(V)$.
We use $\theta$ in the subscript to denote the quantity
with respect to the submodel, e.g., $\mu_{z,u,\theta}$ is the value
of $\mu_{z,u}$ in the submodel. We use dot to denote
the partial derivative with respect to $\theta$, e.g., $\dot{\mu}_{z,u,\theta}=\partial\mu_{z,u,\theta}/\partial\theta$,
and use $\textup{s}_\theta(\cdot)$ to denote the score function of the submodel. From~\eqref{eq:CR-lik},
the score function under the submodel of can be decomposed as 
\[
\textup{s}_{\theta}(V)=\textup{s}_{\theta}(X)+\textup{s}_{\theta}(Z\mid X)+\textup{s}_{\theta}(S\mid Z,X)+\textup{s}_{\theta}(Y\mid S,Z,X),
\]
where $\textup{s}_{\theta}(X)=\partial\log f_{\theta}(X)/\partial\theta$ ,
$\textup{s}_{\theta}(Z\mid X)=\partial\log f_{\theta}(Z\mid X)/\partial\theta$,
$\textup{s}_{\theta}(S\mid Z,X)=\partial\log f_{\theta}(S\mid Z,X)/\partial\theta$,
and $\textup{s}_{\theta}(Y\mid S,Z,X)=\partial\log f_{\theta}(Y\mid S,Z,X)/\partial\theta$
are the score functions corresponding to the four components of the
likelihood. Analogous to $f_{\theta}(V)\vert_{\theta=0}=f(V)$, we 
write $\textup{s}_{\theta}(\cdot)\rvert_{\theta=0}$ as $\textup{s}(\cdot)$, which is the score function evaluated at the true parameter under the one-dimensional submodel.

From the semiparametric
theory, the tangent space 
\begin{equation}
\Lambda=H_{1}\oplus H_{2}\oplus H_{3}\oplus H_{4}\label{eq:tangent space}
\end{equation}
is the direct sum of 
\begin{eqnarray*}
H_{1} & = & \{h(X):\E\{h(X)\}=0\},\\
H_{2} & = & \{h(Z,X):\E\{h(Z,X)\mid X\}=0\},\\
H_{3} & = & \{h(S,Z,X):\E\{h(S,Z,X)\mid Z,X\}=0\},\\
H_{4} & = & \{h(Y,Z,S,X):\E\{h(Y,Z,S,X)\mid Z,S,X\}=0\},
\end{eqnarray*}
where $H_{1}$, $H_{2}$, $H_{3}$, and $H_{4}$ are orthogonal to
each other. The EIF for $\mu_{z,u}$, denoted by $\varphi_{z,u}(V)\in\Lambda$,
must satisfy 
\[
\left.\dot{\mu}_{z,u,\theta}\right\rvert _{\theta=0}=\E\{\varphi_{z,u}(V)\textup{s}(V)\}.
\]
We will derive the EIFs by calculating $\left.\dot{\mu}_{z,u,\theta}\right\rvert _{\theta=0}$.
To simplify the proof, we introduce some lemmas.


\begin{lemma}\label{lem:ratio} Consider a ratio-type parameter $R=N/D$.
If $\dot{N}_{\theta}|_{\theta=0}=\E\{\varphi_{N}(V)\textup{s}(V)\}$ and $\dot{D}_{\theta}|_{\theta=0}=\E\{\varphi_{D}(V)\textup{s}(V)\}$,
then $\dot{R}_{\theta}|_{\theta=0}=\E\{\varphi_{R}(V)\textup{s}(V)\}$ where
\begin{equation}
\varphi_{R}(V)=\frac{1}{D}\varphi_{N}(V)-\frac{R}{D}\varphi_{D}(V).\label{eq:ratio-ses}
\end{equation}
In particular, if $\varphi_{N}(V)$ and $\varphi_{D}(V)$ are the
EIFs for $N$ and $D$, then $\varphi_{R}(V)$ is the EIF for $R.$
\end{lemma}

\noindent 
\textit{Proof of Lemma \ref{lem:ratio}.} Let $R_{\theta},$ $N_{\theta}$,
and $D_{\theta}$ denote the quantities $R$, $N$, and $D$ evaluated
with respect to the parametric submodel $f_{\theta}(V)$. By the chain
rule, we have 
\begin{eqnarray*}
\left.\dot{R}_{\theta}\right\rvert _{\theta=0} & = & \left.\frac{\dot{N}_{\theta}}{D}\right\rvert _{\theta=0}-R_{\theta}\left.\frac{\dot{D}_{\theta}}{D}\right\rvert _{\theta=0}\\
 & = & \frac{1}{D}\E\{\varphi_{N}(V)\textup{s}(V)\}-\frac{R}{D}\E\{\varphi_{D}(V)\textup{s}(V)\}\\
 & = & \E\left[\left\{ \frac{1}{D}\varphi_{N}(V)-\frac{R}{D}\varphi_{D}(V)\right\} \textup{s}(V)\right],
\end{eqnarray*}
which yields \eqref{eq:ratio-ses}. \QEDB \medskip{}

\begin{lemma} \label{lemma:score of E} For any $h(V)$ that does
not depend on $\theta$, $\partial\E_{\theta}\{h(V)\}/\partial\theta\mid_{\theta=0}=\E\{h(V)\textup{s}(V)\}$.
\end{lemma} The proof is straightforward and thus omitted. \medskip{}

\noindent \begin{lemma}\label{lem:p0dot} Define $\mu_{0f}(X)=\E\{f(Y,S,X)\mid Z=0,X\}$
and $\mu_{1f}(X)=\E\{f(Y,S,X)\mid Z=1,X\}$ for any $f(Y,S,X)$. We
have 
\begin{eqnarray*}
\dot{\mu}_{0f,\theta}(X)|_{\theta=0} & = & \E\left[\{\psi_{f(Y_{0},S_{0},X)}-\mu_{0f}(X)\}\textup{s}(Y,S\mid Z,X)\mid X\right],\\
\dot{\mu}_{1f,\theta}(X)|_{\theta=0} & = & \E\left[\{\psi_{f(Y_{1},S_{1},X)}-\mu_{1f}(X)\}\textup{s}(Y,S\mid Z,X)\mid X\right].
\end{eqnarray*}
As a special case, for $p_{0}(X)=\E(S\mid Z=0,X)$ and $p_{1}(X)=\E(S\mid Z=1,X)$,
we have 
\begin{eqnarray*}
\dot{p}_{0,\theta}(X)\mid_{\theta=0} & = & \E\left[\{\psi_{S_{0}}-p_{0}(X)\}\textup{s}(S\mid Z,X)\mid X\right],\\
\dot{p}_{1,\theta}(X)\mid_{\theta=0} & = & \E\left[\{\psi_{S_{1}}-p_{1}(X)\}\textup{s}(S\mid Z,X)\mid X\right].
\end{eqnarray*}
 \end{lemma}

\noindent \textit{Proof of Lemma \ref{lem:p0dot}.} We first prove
the general result: 
\begin{eqnarray*}
\dot{\mu}_{0f,\theta}(X)|_{\theta=0} & = & \frac{\partial}{\partial\theta}\E_{\theta}\{f(Y,S,X)\mid Z=0,X\}\mid_{\theta=0}\\
 & = & \E\left\{ f(Y,S,X)\text{\ensuremath{\times}}\textup{s}(Y,S\mid Z=0,X)\mid Z=0,X\right\} \quad(\text{Lemma}~\ref{lemma:score of E})\\
 & = & \E\left[\{f(Y,S,X)-\mu_{0f}(X)\}\textup{s}(Y,S\mid Z=0,X)\mid Z=0,X\right]\\
 & = & \E\left[\frac{(1-Z)\{f(Y,S,X)-\mu_{0f}(X)\}}{1-\pi(X)}\textup{s}(Y,S\mid Z,X)\mid X\right]\\
 & = & \E\left[\{\psi_{f(Y_{0},S_{0},X)}-\mu_{0f}(X)\}\textup{s}(Y,S\mid Z,X)\mid X\right]
\end{eqnarray*}
where the third equality follows from $\mu_{0f}(X)\E\{\textup{s}(Y,S\mid Z=0,X)\mid Z=0,X\}=0$.
Similarly, we can prove the result for $\mu_{1f}(X)$.

Choosing $f(Y,S,X)=S$, the results for $p_{z}(X)$ follow because
$$
\E\left[\{\psi_{S_{z}}-p_{z}(X)\}\textup{s}(Y\mid Z,S,X)\mid X\right]=0.
$$ 
\QEDB \medskip{}

\begin{lemma}\label{lem:marginalp0} 
Define $\mu_{0f}=\E\{\mu_{0f}(X) \}$ and $\mu_{1f}=\E\{\mu_{1f}(X)\}$. We
have 
\begin{eqnarray*}
\dot{\mu}_{0f,\theta}|_{\theta=0} & = & \E\left[\{\psi_{f(Y_{0},S_{0},X)}-\mu_{0f}\}\textup{s}(V)\right],\\
\dot{\mu}_{1f,\theta}|_{\theta=0} & = & \E\left[\{\psi_{f(Y_{1},S_{1},X)}-\mu_{1f}\}\textup{s}(V)\right].
\end{eqnarray*}
Moreover, $\psi_{f(Y_{0},S_{0},X)}-\mu_{0f}$ and $\psi_{f(Y_{1},S_{1},X)}-\mu_{1f}$ are EIFs for $\mu_{0f}$ and $\mu_{1f}$, respectively. As a special case, 
for $p_{0}=\E\{p_{0}(X)\}$
and $p_{1}=\E\{p_{1}(X)\}$, we have 
\begin{eqnarray*}
\dot{p}_{0,\theta}\mid_{\theta=0} & = & \E\left\{ \left(\psi_{S_{0}}-p_{0}\right)\textup{s}(V)\right\}, \\
\dot{p}_{1,\theta}\mid_{\theta=0} & = & \E\left\{ \left(\psi_{S_{1}}-p_{1}\right)\textup{s}(V)\right\} ,
\end{eqnarray*}
and $\psi_{S_{0}}-p_{0}$ and $\psi_{S_1}-p_1$ are the EIFs
for $p_0$ and $p_1$, respectively. \end{lemma}

\noindent \textit{Proof
of Lemma \ref{lem:marginalp0}.} 
We prove the general result:
\begin{eqnarray*}
\dot{\mu}_{0f,\theta}\mid_{\theta=0} & = & \E\{\mu_{0f}(X)\textup{s}(V)\}+\E\{\dot{\mu}_{0f,\theta}(X)\mid_{\theta=0}\}\quad(\text{Lemma}~\ref{lemma:score of E})\\
 & = & \E\{\mu_{0f}(X)\textup{s}(V)\}+\E\left[\{\psi_{f(Y_{0},S_{0},X)}-\mu_{0f}(X)\}\textup{s}(Y,S\mid Z,X)\right]\quad(\text{Lemma}~\ref{lem:p0dot})\\
 & = & \E\{(\psi_{f(Y_{0},S_{0},X)}-\mu_{0f})\textup{s}(Y,S\mid Z,X)\}\\
 & = & \E\{(\psi_{f(Y_{0},S_{0},X)}-\mu_{0f})\textup{s}(V)\},
\end{eqnarray*}
where the last equality follows from $\E\{(\psi_{f(Y_{0},S_{0},X)}-\mu_{0f})\textup{s}(Z\mid X)\}=\E\{(\psi_{f(Y_{0},S_{0},X)}-\mu_{0f})\textup{s}(X)\}=0$.
Because $\psi_{f(Y_{0},S_{0},X)}-\mu_{0f}$ lies in the tangent space, it is the EIF for $\mu_{0f}$. Similarly, we can prove the result for $\mu_{1f}$. The results for $p_1$ and $p_0$ follow by taking $f(Y,S,X)=S$. 
\citet{hahn1998role}
gives an alternative proof of the EIFs of $p_1$ and $p_0$. \QEDB
\medskip{}

\begin{lemma}\label{lem:mudot} For $\mu_{zs}(X)$, we have 
\begin{eqnarray*}
\dot{\mu}_{11,\theta}(X)\mid_{\theta=0} & = & \E\left\{ \frac{\psi_{Y_{1}S_{1}}-\mu_{11}(X)\psi_{S_{1}}}{p_{1}(X)}\textup{s}(Y\mid Z,S,X)\mid X\right\} ,\\
\dot{\mu}_{01,\theta}(X)\mid_{\theta=0} & = & \E\left\{ \frac{\psi_{Y_{0}S_{0}}-\mu_{01}(X)\psi_{S_{0}}}{p_{0}(X)}\textup{s}(Y\mid Z,S,X)\mid X\right\} ,\\
\dot{\mu}_{10,\theta}(X)\mid_{\theta=0} & = & \E\left\{ \frac{\psi_{Y_{1}(1-S_{1})}-\mu_{10}(X)\psi_{1-S_{1}}}{1-p_{1}(X)}\textup{s}(Y\mid Z,S,X)\mid X\right\} ,\\
\dot{\mu}_{00,\theta}(X)\mid_{\theta=0} & = & \E\left\{ \frac{\psi_{Y_{0}(1-S_{0})}-\mu_{00}(X)\psi_{1-S_{0}}}{1-p_{0}(X)}\textup{s}(Y\mid Z,S,X)\mid X\right\} .
\end{eqnarray*}
\end{lemma} \textit{Proof of Lemma \ref{lem:mudot}.} We prove only
the result for $\mu_{11,\theta}(X)$ and the proofs for other parameters
are similar. A key observation is the ratio representation: 
\[
\mu_{11}(X)=\E(Y\mid Z=1,S=1,X)=\frac{\E(YS\mid Z=1,X)}{\E(S\mid Z=1,X)}=\frac{\E(YS\mid Z=1,X)}{p_{1}(X)}.
\]
From Lemma \ref{lem:p0dot}, the numerator satisfies 
\[
\frac{\partial}{\partial\theta}\E(YS\mid Z=1,X)\mid_{\theta=0}=\E\left[\{\psi_{Y_{1}S_{1}}-p_{1}(X)\mu_{11}(X)\}\textup{s}(Y,S\mid Z,X)\mid X\right],
\]
and the denominator satisfies 
\[
\dot{p}_{1}(X)\mid_{\theta=0}=\E\left[\{\psi_{S_{1}}-p_{1}(X)\}\textup{s}(Y,S\mid Z,X)\mid X\right].
\]
We can then use Lemma \ref{lem:ratio} to calculate the path derivative
of $\mu_{11,\theta}(X)$ with all distributions conditional on $X$,
yielding 
\[
\dot{\mu}_{11,\theta}(X)\mid_{\theta=0}=\E\left\{ \frac{\psi_{Y_{1}S_{1}}-\mu_{11}(X)\psi_{S_{1}}}{p_{1}(X)}\textup{s}(Y,S\mid Z,X)\mid X\right\} .
\]
The conclusion follows by using $\E[\{\psi_{Y_{1}S_{1}}-\mu_{11}(X)\psi_{S_{1}}\}\textup{s}(S\mid Z,X)\mid X]=0$.
\QEDB 
\medskip{}

\subsubsection{EIF for $\tau_{10}$}
 \label{app:SET-10}
Below, we derive the EIF for $\tau_{10}$. The EIFs for $\tau_{00}$
and $\tau_{11}$ can be derived similarly.

First, from Theorem~\ref{thm:identification-obs}(c), we can write
$\mu_{1,10}=N/D$, where 
\begin{eqnarray*}
N=\E\left[\{p_{1}(X)-p_{0}(X)\}\mu_{11}(X)\right],\quad D=p_{1}-p_{0}.
\end{eqnarray*}
Lemma~\ref{lem:marginalp0} implies 
\begin{eqnarray}
\varphi_{D}(V)=(\psi_{S_{1}}-\psi_{S_{0}})-(p_{1}-p_{0})=(\psi_{S_{1}}-\psi_{S_{0}})-D,\label{eq:phi2}
\end{eqnarray}
so based on Lemma \ref{lem:ratio}, the key is to derive $\varphi_{N}(V)$.
From the chain rule, we have 
\begin{eqnarray}
\left.\dot{N}_{\theta}\right\rvert _{\theta=0} & = & \left.\frac{\partial}{\partial\theta}\E_{\theta}\left[\{p_{1,\theta}(X)-p_{0,\theta}(X)\}\mu_{11,\theta}(X)\right]\right\rvert _{\theta=0}\nonumber \\
 & = & \E[\{p_{1}(X)-p_{0}(X)\}\mu_{11}(X)\textup{s}(X)] \quad (\text{Lemma}~\ref{lemma:score of E}) \nonumber \\
 && +\left.\E_{\theta}\left[\{\dot{p}_{1,\theta}(X)-\dot{p}_{0,\theta}(X)\}\mu_{11,\theta}(X)\right]\right\rvert _{\theta=0}\nonumber \\
 & & +\left.\E_{\theta}\left[\{p_{1,\theta}(X)-p_{0,\theta}(X)\}\dot{\mu}_{11,\theta}(X)\right]\right\rvert _{\theta=0}. \label{eqn:proofthm3-10-phi1}
\end{eqnarray}
Because $\E\{N\textup{s}(X)\}=0$, the first term in \eqref{eqn:proofthm3-10-phi1}
equals 
\begin{eqnarray*}
\E\{\{p_{1}(X)-p_{0}(X)\}\mu_{11}(X)\textup{s}(X)\}=\E\left(\left[\{p_{1}(X)-p_{0}(X)\}\mu_{11}(X)-N\right]\textup{s}(X)\}\right).\label{eqn:proofthm3-10-phi2}
\end{eqnarray*}
From Lemma~\ref{lem:p0dot}, the second term in \eqref{eqn:proofthm3-10-phi1}
reduces to 
\begin{eqnarray*}
 & & \left.\E_{\theta}\left[\{\dot{p}_{1,\theta}(X)-\dot{p}_{0,\theta}(X)\}\mu_{11,\theta}(X)\right]\right\rvert _{\theta=0}\nonumber \\
 & = & \E\left(\E\left[\{\psi_{S_{1}}-\psi_{S_{0}}-p_{1}(X)+p_{0}(X)\}\textup{s}(S\mid Z,X)\mu_{11}(X)\mid X\right]\right)\nonumber \\
 & = & \E\left[\{\psi_{S_{1}}-\psi_{S_{0}}-p_{1}(X)+p_{0}(X)\}\mu_{11}(X)\textup{s}(S\mid Z,X)\right].\label{eqn:proofthm3-10-phi3}
\end{eqnarray*}
From Lemma~\ref{lem:mudot}, the third term in \eqref{eqn:proofthm3-10-phi1}
reduces to 
\begin{eqnarray*}
 & & \left.\E_{\theta}\left[\{p_{1,\theta}(X)-p_{0,\theta}(X)\}\dot{\mu}_{11,\theta}(X)\right]\right\rvert _{\theta=0}\nonumber \\
 & = & \E\left[\{p_{1}(X)-p_{0}(X)\}\frac{\psi_{Y_{1}S_{1}}-\mu_{11}(X)\psi_{S_{1}}}{p_{1}(X)}\textup{s}(Y\mid Z,S,X)\right].\label{eqn:proofthm3-10-phi4}
\end{eqnarray*}
Plugging the above three formulas
into~\eqref{eqn:proofthm3-10-phi1} gives 
\begin{eqnarray}
\left.\dot{N}_{\theta}\right\rvert _{\theta=0} & = & \E\left(\left[\{p_{1}(X)-p_{0}(X)\}\mu_{11}(X)-N\right]\textup{s}(X)\}\right)\nonumber \\
 & & +\E\left[\{\psi_{S_{1}}-\psi_{S_{0}}-p_{1}(X)+p_{0}(X)\}\mu_{11}(X)\textup{s}(S\mid Z,X)\right]\nonumber \\
 & & +\E\left[\{p_{1}(X)-p_{0}(X)\}\frac{\psi_{Y_{1}S_{1}}-\mu_{11}(X)\psi_{S_{1}}}{p_{1}(X)}\textup{s}(Y\mid Z,S,X)\right].\label{eqn:proofthm3-10-phi}
\end{eqnarray}
We can verify that 
\begin{eqnarray}
\{p_{1}(X)-p_{0}(X)\}\mu_{11}(X)-N & \in & H_{1},\label{eqn:SET-part1}\\
\{\psi_{S_{1}}-\psi_{S_{0}}-p_{1}(X)+p_{0}(X)\}\mu_{11}(X) & \in & H_{3},\label{eqn:SET-part2}\\
\{p_{1}(X)-p_{0}(X)\}\frac{\psi_{Y_{1}S_{1}}-\mu_{11}(X)\psi_{S_{1}}}{p_{1}(X)} & \in & H_{4}.\label{eqn:SET-part3}
\end{eqnarray}
Because $H_{1}$, $H_{2}$, $H_{3}$, and $H_{4}$ are orthogonal to
each other, we can write~\eqref{eqn:proofthm3-10-phi} as 
\begin{eqnarray*}
\left.\dot{N}_{\theta}\right\rvert _{\theta=0} & = & \E\left(\left[\{p_{1}(X)-p_{0}(X)\}\mu_{11}(X)-N\right]\textup{s}(V)\}\right)\\
&&+\E\left[\{\psi_{S_{1}}-\psi_{S_{0}}-p_{1}(X)+p_{0}(X)\}\mu_{11}(X)\textup{s}(V)\right]\\
 & & +\E\left[\{p_{1}(X)-p_{0}(X)\}\frac{\psi_{Y_{1}S_{1}}-\mu_{11}(X)\psi_{S_{1}}}{p_{1}(X)}\textup{s}(V)\right].
\end{eqnarray*}
As a result, we obtain the EIF for $N$: 
\begin{eqnarray*}
\varphi_{N}(V) & = & \{p_{1}(X)-p_{0}(X)\}\mu_{11}(X)-N+\{\psi_{S_{1}}-\psi_{S_{0}}-p_{1}(X)+p_{0}(X)\}\mu_{11}(X)\\
 & & +\{p_{1}(X)-p_{0}(X)\}\frac{\psi_{Y_{1}S_{1}}-\mu_{11}(X)\psi_{S_{1}}}{p_{1}(X)}\\
 & = & (\psi_{S_{1}}-\psi_{S_{0}})\mu_{11}(X)-N+\left\{ 1-\frac{p_{0}(X)}{p_{1}(X)}\right\} \left\{ \psi_{Y_{1}S_{1}}-\mu_{11}(X)\psi_{S_{1}}\right\} \\
 & = & \frac{e_{10}(X)}{p_{1}(X)}\psi_{Y_{1}S_{1}}-N-\mu_{11}(X)\left\{ \psi_{S_{0}}-\frac{p_{0}(X)}{p_{1}(X)}\psi_{S_{1}}\right\} ,
\end{eqnarray*}
which equals $\phi_{1,10}-N$ (see Theorem~\ref{thm:SET} for the
expression of $\phi_{1,10}$). From Lemma \ref{lem:ratio}, the EIF
for $\mu_{1,10}$ is 
\begin{eqnarray*}
\varphi_{1,10}(V)=\frac{1}{p_{1}-p_{0}}\phi_{1,10}-\frac{\mu_{1,10}(\psi_{S_{1}}-\psi_{S_{0}})}{p_{1}-p_{0}}.
\end{eqnarray*}
Similarly, the EIF for $\mu_{0,10}$ is 
\begin{eqnarray*}
\varphi_{0,10}(V)=\frac{1}{p_{1}-p_{0}}\phi_{0,10}-\frac{\mu_{0,10}(\psi_{S_{1}}-\psi_{S_{0}})}{p_{1}-p_{0}}.
\end{eqnarray*}
Therefore, the EIF for $\tau_{10}$ is 
\begin{eqnarray*}
\frac{(\phi_{1,10}-\phi_{0,10})-\tau_{10}(\psi_{S_{1}}-\psi_{S_{0}})}{p_{1}-p_{0}}.
\end{eqnarray*}
\QEDB

\subsection{Proof of Theorems~\ref{thm:SET-rand},~\ref{thm:double-rand},~and~\ref{thm:SET-strmon}}

The proofs are almost identical to the cases without restrictions. We omit the details due to the length of supplementary material. The earliest ArXiv version of our paper contains the details.  

\subsection{Proof of Theorem~\ref{thm:SET-sens}}

We will follow a similar route as in Section~\ref{app:SET}. The
tangent space is the same as in~\eqref{eq:tangent space}. 
To avoid new symbols, we use the same notation as in Section~\ref{app:SET}.
We need the following lemma. \begin{lemma} \label{lem:omegadot}
For $\omega_{1,10}(X)$, we have 
\begin{eqnarray*}
\left.\dot{\omega}_{1,10,\theta}(X)\right\rvert _{\theta=0}=\E\left[\left\{ \frac{\omega_{1,10}(X)\psi_{S_{1}}}{p_{1}(X)}-\frac{\omega_{1,10}^{2}(X)(\psi_{S_{1}}-\psi_{S_{0}})}{p_{1}(X)}-\frac{\omega_{1,10}^{2}(X)\psi_{S_{0}}}{\epsilon_{1}(X)p_{1}(X)}\right\} \textup{s}(S\mid Z,X)\mid X\right].
\end{eqnarray*}
\end{lemma} \textit{Proof of Lemma~\ref{lem:omegadot}}. The parameter
$\omega_{1,10}(X)$ has a ratio form 
\[
\omega_{1,10}(X)=\frac{\epsilon_{1}(X)\{e_{10}(X)+e_{11}(X)\}}{\epsilon_{1}(X)e_{10}(X)+e_{11}(X)}=\frac{\epsilon_{1}(X)p_{1}(X)}{\epsilon_{1}(X)\{p_{1}(X)-p_{0}(X)\}+p_{0}(X)}.
\]
Thus, we can use Lemma \ref{lem:ratio} to calculate its path derivative
with all distributions conditional on $X$. From Lemma \ref{lem:p0dot},
the numerator satisfies 
\[
\epsilon_{1}(X)\dot{p}_{1}(X)|_{\theta=0}=\E[\epsilon_{1}(X)\{\psi_{S_{1}}-p_{1}(X)\}\textup{s}(S\mid Z,X)\mid X],
\]
and the denominator satisfies 
\begin{eqnarray*}
\epsilon_{1}(X)\{\dot{p}_{1,\theta}(X)-\dot{p}_{0,\theta}(X)\}|_{\theta=0}+\dot{p}_{0,\theta}(X)|_{\theta=0} 
=\E\left[\{\epsilon_{1}(X)(\psi_{S_{1}}-\psi_{S_{0}}-e_{10}(X))+\psi_{S_{0}}-p_{0}(X)\}\textup{s}(S\mid Z,X)\mid X\right].
\end{eqnarray*}
Using Lemma \ref{lem:ratio}, we can obtain $\dot{\omega}_{1,10,\theta}(X)|_{\theta=0}=\E\{\phi_{\omega,1,10}(V)\textup{s}(S\mid Z,X)\mid X\},$
where 
\begin{eqnarray*}
\phi_{\omega,1,10}(V) & = & \frac{\epsilon_{1}(X)\{\psi_{S_{1}}-p_{1}(X)\}-\omega_{1,10}(X)\{\epsilon_{1}(X)(\psi_{S_{1}}-\psi_{S_{0}}-e_{10}(X))+\psi_{S_{0}}-p_{0}(X)\}}{\epsilon_{1}(X)e_{10}(X)+e_{11}(X)}.
\end{eqnarray*}
The result then follows from simple algebra.

\noindent \QEDB \medskip{}

We will derive the EIF for $\tau_{10}=\mu_{1,10}-\mu_{0,10}$ and
omit the similar proofs of the EIFs for $\tau_{11}$ and $\tau_{00}$. We can
write $\mu_{1,10}=N/D$, where 
\begin{eqnarray*}
N=\E\left[\{p_{1}(X)-p_{0}(X)\}\omega_{1,10}(X)\mu_{11}(X)\right],\quad D=p_{1}-p_{0}.
\end{eqnarray*}
Lemma \ref{lem:marginalp0} implies that $\varphi_{D}(V)=(\psi_{S_{1}}-\psi_{S_{0}})-D$
is the EIF for $D$, so based on Lemma \ref{lem:ratio}, the key is
to derive the EIF for $N$. From the chain rule, we have 
\begin{eqnarray}
\left.\dot{N}_{\theta}\right\rvert _{\theta=0} & = & \left.\frac{\partial}{\partial\theta}\E_{\theta}\left[\{p_{1,\theta}(X)-p_{0,\theta}(X)\}\omega_{1,10}(X)\mu_{11,\theta}(X)\right]\right\rvert _{\theta=0}\nonumber \\
 & = & \E[\{p_{1}(X)-p_{0}(X)\}\omega_{1,10}(X)\mu_{11}(X)\textup{s}(X)] \quad (\text{Lemma}~\ref{lemma:score of E}) \nonumber \\
 & & +\left.\E_{\theta}\left[\{\dot{p}_{1,\theta}(X)-\dot{p}_{0,\theta}(X)\}\omega_{1,10,\theta}(X)\mu_{11,\theta}(X)\right]\right\rvert _{\theta=0}\nonumber \\
 & & +\left.\E_{\theta}\left[\{p_{1,\theta}(X)-p_{0,\theta}(X)\}\dot{\omega}_{1,10,\theta}(X)\mu_{11,\theta}(X)\right]\right\rvert _{\theta=0}\nonumber \\
 & & +\left.\E_{\theta}\left[\{p_{1,\theta}(X)-p_{0,\theta}(X)\}\omega_{1,10,\theta}(X)\dot{\mu}_{11,\theta}(X)\right]\right\rvert _{\theta=0}. \label{eqn:proofthm9-10-phi1}
\end{eqnarray}
Because $\E\{N\textup{s}(X)\}=0$, the first term in \eqref{eqn:proofthm9-10-phi1}
equals 
\begin{eqnarray*}
 & & \E\{\{p_{1}(X)-p_{0}(X)\}\omega_{1,10}(X)\mu_{11}(X)\textup{s}(X)\}\nonumber \\
 & = & \E\left(\left[\{p_{1}(X)-p_{0}(X)\}\omega_{1,10}(X)\mu_{11}(X)-N\right]\textup{s}(X)\}\right).\label{eqn:proofthm9-10-phi2}
\end{eqnarray*}
From Lemma~\ref{lem:p0dot}, the second term in \eqref{eqn:proofthm9-10-phi1}
reduces to 
\begin{eqnarray*}
 & & \left.\E_{\theta}\left[\{\dot{p}_{1,\theta}(X)-\dot{p}_{0,\theta}(X)\}\omega_{1,10,\theta}(X)\mu_{11,\theta}(X)\right]\right\rvert _{\theta=0}\nonumber \\
 & = & \E\left(\E\left[\{\psi_{S_{1}}-\psi_{S_{0}}-p_{1}(X)+p_{0}(X)\}\textup{s}(S\mid Z,X)\omega_{1,10}(X)\mu_{11}(X)\mid X\right]\right)\nonumber \\
 & = & \E\left[\{\psi_{S_{1}}-\psi_{S_{0}}-p_{1}(X)+p_{0}(X)\}\omega_{1,10}(X)\mu_{11}(X)\textup{s}(S\mid Z,X)\right].\label{eqn:proofthm9-10-phi3}
\end{eqnarray*}
From Lemma~\ref{lem:omegadot}, the third term in \eqref{eqn:proofthm9-10-phi1}
reduces to 
\begin{eqnarray*}
 & & \left.\E_{\theta}\left[\{p_{1,\theta}(X)-p_{0,\theta}(X)\}\dot{\omega}_{1,10,\theta}(X)\mu_{11,\theta}(X)\right]\right\rvert _{\theta=0}\nonumber \\
 & = & \E\left[\{p_{1}(X)-p_{0}(X)\}\mu_{11}(X)\right.\nonumber \\
 & & \left.\qquad\cdot\left\{ \frac{\omega_{1,10}(X)\psi_{S_{1}}}{p_{1}(X)}-\frac{\omega_{1,10}^{2}(X)(\psi_{S_{1}}-\psi_{S_{0}})}{p_{1}(X)}-\frac{\omega_{1,10}^{2}(X)\psi_{S_{0}}}{\epsilon_{1}(X)p_{1}(X)}\right\} \textup{s}(S\mid Z,X)\right]\label{eqn:proofthm9-10-phi4}
\end{eqnarray*}
From Lemma~\ref{lem:mudot}, the fourth term in \eqref{eqn:proofthm9-10-phi1}
reduces to 
\begin{eqnarray*}
 & & \left.\E_{\theta}\left[\{p_{1,\theta}(X)-p_{0,\theta}(X)\}\omega_{1,10,\theta}(X)\dot{\mu}_{11,\theta}(X)\right]\right\rvert _{\theta=0}\nonumber \\
 & = & \E\left[\{p_{1}(X)-p_{0}(X)\}\omega_{1,10}(X)\frac{\psi_{Y_{1}S_{1}}-\mu_{11}(X)\psi_{S_{1}}}{p_{1}(X)}\textup{s}(Y\mid Z,S,X)\right].\label{eqn:proofthm9-10-phi5}
\end{eqnarray*}
Plugging the above four formulas 
into~\eqref{eqn:proofthm9-10-phi1} yields 
\begin{eqnarray}
\left.\dot{N}_{\theta}\right\rvert _{\theta=0} & = & \E\left(\left[\{p_{1}(X)-p_{0}(X)\}\omega_{1,10}(X)\mu_{11}(X)-N\right]\textup{s}(X)\}\right)\nonumber \\
 & & +\E\left[\{\psi_{S_{1}}-\psi_{S_{0}}-p_{1}(X)+p_{0}(X)\}\omega_{1,10}(X)\mu_{11}(X)\textup{s}(S\mid Z,X)\right]\nonumber \\
 & & +\E\left[\{p_{1}(X)-p_{0}(X)\}\mu_{11}(X)\right.\nonumber \\
 & & \left.\quad\cdot\left\{ \frac{\omega_{1,10}(X)\psi_{S_{1}}}{p_{1}(X)}-\frac{\omega_{1,10}^{2}(X)(\psi_{S_{1}}-\psi_{S_{0}})}{p_{1}(X)}-\frac{\omega_{1,10}^{2}(X)\psi_{S_{0}}}{\epsilon_{1}(X)p_{1}(X)}\right\} \textup{s}(S\mid Z,X)\right]\nonumber \\
 & & +\E\left[\{p_{1}(X)-p_{0}(X)\}\omega_{1,10}(X)\frac{\psi_{Y_{1}S_{1}}-\mu_{11}(X)\psi_{S_{1}}}{p_{1}(X)}\textup{s}(Y\mid Z,S,X)\right].\label{eqn:proofthm9-10-phi}
\end{eqnarray}
We can verify that 
\begin{eqnarray*}
\{p_{1}(X)-p_{0}(X)\}\omega_{1,10}(X)\mu_{11}(X)-N & \in & H_{1},\\
\{\psi_{S_{1}}-\psi_{S_{0}}-p_{1}(X)+p_{0}(X)\}\omega_{1,10}(X)\mu_{11}(X) & \in & H_{3},\\
\{p_{1}(X)-p_{0}(X)\}\mu_{11}(X) \left\{\frac{\omega_{1,10}(X)\psi_{S_{1}}}{p_{1}(X)}-\frac{\omega_{1,10}^{2}(X)(\psi_{S_{1}}-\psi_{S_{0}})}{p_{1}(X)}-\frac{\omega_{1,10}^{2}(X)\psi_{S_{0}}}{\epsilon_{1}(X)p_{1}(X)}\right\} & \in & H_{3},\\
\{p_{1}(X)-p_{0}(X)\}\omega_{1,10}(X)\frac{\psi_{Y_{1}S_{1}}-\mu_{11}(X)\psi_{S_{1}}}{p_{1}(X)} & \in & H_{4}.
\end{eqnarray*}
Therefore, we can replace all the score functions in~\eqref{eqn:proofthm9-10-phi}
with $\textup{s}(V)$. As a result, the EIF for $N$ is 
\begin{eqnarray*}
\varphi_{N}(V) & = & \{p_{1}(X)-p_{0}(X)\}\omega_{1,10}(X)\mu_{11}(X)-N+\{\psi_{S_{1}}-\psi_{S_{0}}-p_{1}(X)+p_{0}(X)\}\omega_{1,10}(X)\mu_{11}(X)\\
 & & +\{p_{1}(X)-p_{0}(X)\}\mu_{11}(X)\left\{ \frac{\omega_{1,10}(X)\psi_{S_{1}}}{p_{1}(X)}-\frac{\omega_{1,10}^{2}(X)(\psi_{S_{1}}-\psi_{S_{0}})}{p_{1}(X)}-\frac{\omega_{1,10}^{2}(X)\psi_{S_{0}}}{\epsilon_{1}(X)p_{1}(X)}\right\} \\
 & & +\{p_{1}(X)-p_{0}(X)\}\omega_{1,10}(X)\frac{\psi_{Y_{1}S_{1}}-\mu_{11}(X)\psi_{S_{1}}}{p_{1}(X)}\\
 & = & \frac{e_{10}(S)\omega_{1,10}(X)}{p_{1}(X)}\psi_{Y_{1}S_{1}}-N+\{\psi_{S_{1}}-\psi_{S_{0}}\}\omega_{1,10}(X)\mu_{11}(X)\\
 & & -e_{10}(X)\mu_{11}(X)\left\{ \frac{\omega_{1,10}^{2}(X)(\psi_{S_{1}}-\psi_{S_{0}})}{p_{1}(X)}+\frac{\omega_{1,10}^{2}(X)\psi_{S_{0}}}{\epsilon_{1}(X)p_{1}(X)}\right\} \\
 & = & \frac{e_{10}(S)\omega_{1,10}(X)}{p_{1}(X)}\psi_{Y_{1}S_{1}}-N+\psi_{S_{1}}\mu_{11}(X)\left\{ \omega_{1,10}(X)-\frac{e_{10}(X)\omega_{1,10}(X)}{p_{1}(X)}\right\} \\
 & & -\psi_{S_{0}}\mu_{11}(X)\left\{ \omega_{1,10}(X)+\frac{e_{10}(X)\omega_{1,10}^{2}(X)}{\epsilon_{1}(X)p_{1}(X)}-\frac{e_{10}(X)\omega_{1,10}^{2}(X)}{p_{1}(X)}\right\} \\
 & = & \frac{e_{10}(S)\omega_{1,10}(X)}{p_{1}(X)}\psi_{Y_{1}S_{1}}-N-\frac{\omega_{1,10}^{2}(X)\mu_{11}(X)}{\epsilon_{1}(X)}\left\{ \psi_{S_{0}}-\frac{p_{0}(X)}{p_{1}(X)}\psi_{S_{1}}\right\} ,
\end{eqnarray*}
which equals $\phi'_{1,10}-N$ (see Theorem~\ref{thm:SET-sens} for
the expression of $\phi'_{1,10}$). From Lemma \ref{lem:ratio}, the
EIF for $\mu_{1,10}$ is 
\begin{eqnarray*}
\varphi_{1,10}(V)=\frac{\phi'_{1,10}}{p_{1}-p_{0}}-\frac{\mu_{1,10}(\psi_{S_{1}}-\psi_{S_{0}})}{p_{1}-p_{0}}.
\end{eqnarray*}
Similarly, we can obtain the EIF for $\mu_{0,10}$, 
\begin{eqnarray*}
\varphi_{0,10}(V)=\frac{\phi'_{0,10}}{p_{1}-p_{0}}-\frac{\mu_{0,10}(\psi_{S_{1}}-\psi_{S_{0}})}{p_{1}-p_{0}}.
\end{eqnarray*}
Therefore, the EIF for $\tau_{10}$ is 
\begin{eqnarray*}
\frac{(\phi'_{1,10}-\phi'_{0,10})-\tau_{10}(\psi_{S_{1}}-\psi_{S_{0}})}{p_{1}-p_{0}}.
\end{eqnarray*}
\QEDB

Solving $\E(\phi'_u) = 0$ for $\tau_u$ yields the following identification formulas for sensitivity analysis:
$$
\tau_{10} = \frac{\E( \phi'_{1,10}-\phi'_{0,10} ) }{ \E(\psi_{S_{1}}-\psi_{S_{0}}) },\quad 
\tau_{00} = \frac{ \E(\phi'_{1,00}-\phi'_{0,00}) }{ \E(\psi_{1-S_{1}}) }, \quad 
\tau_{11} = \frac{ \E (\phi'_{1,11}-\phi'_{0,11}) }{ \E(\psi_{S_{0}}) }. 
$$
Similar to the main text, we can construct the estimators for $\tau_u$ without principal ignorability.

\subsection{Proof of Theorem~\ref{thm:SET-sens-mon}}
\label{app:SET-sens-mon}
We will derive the EIF for $\tau_{10}$= $\mu_{1,10} - \mu_{0,10}$ and omit the similar proofs of the EIFs for $\tau_{00}$, $\tau_{11}$, and $\tau_{01}$. We can write $\mu_{1,10} = N/D$, where
\begin{eqnarray*}
N = \E\left\{\frac{p_1(X)-p_0(X)}{1-\xi(X)}\mu_{11}(X) \right\}, \quad D = \E\left\{\frac{p_1(X)-p_0(X)}{1-\xi(X)}\right\}.
\end{eqnarray*}
Lemma~\ref{lem:marginalp0} implies that
\begin{eqnarray*}
\phi_D(V) = \frac{ \psi_{S_1} - \psi_{S_0}}{1-\xi(X)}-D
\end{eqnarray*}
is the EIF for $D$, so based on Lemma~\ref{lem:ratio}, the key is to derive the EIF for $N$.
From the chain rule, we have 
\begin{eqnarray}
\left.\dot{N}_{\theta}\right\rvert _{\theta=0} & = & \left.\frac{\partial}{\partial\theta}\E_{\theta}\left\{\frac{p_{1,\theta}(X)-p_{0,\theta}(X)}{1-\xi(X)}\mu_{11,\theta}(X)\right\}\right\rvert _{\theta=0}\nonumber \\
 & = & \E\left\{ \frac{p_{1}(X)-p_{0}(X)}{1-\xi(X)}\mu_{11}(X)\textup{s}(X)\right\} \quad (\text{Lemma}~\ref{lemma:score of E}) \nonumber \\
 && +\left.\E_{\theta}\left\{\frac{\dot{p}_{1,\theta}(X)-\dot{p}_{0,\theta}(X)}{1-\xi(X)}\mu_{11,\theta}(X)\right\}\right\rvert _{\theta=0}\nonumber \\
 & & +\left.\E_{\theta}\left\{\frac{p_{1,\theta}(X)-p_{0,\theta}(X)}{1-\xi(X)}\dot{\mu}_{11,\theta}(X)\right\}\right\rvert _{\theta=0}. \label{eqn:proofthm-sens-mon-10-phi1}
\end{eqnarray}
Because $\E\{N\textup{s}(X)\}=0$, the first term in \eqref{eqn:proofthm-sens-mon-10-phi1}
equals 
\begin{eqnarray*}
\E\left\{ \frac{p_{1}(X)-p_{0}(X)}{1-\xi(X)}\mu_{11}(X)\textup{s}(X)\right\}\ =\ \E\left[\left\{\frac{p_{1}(X)-p_{0}(X)}{1-\xi(X)}\mu_{11}(X)-N\right\}\textup{s}(X)\right].\label{eqn:proofthm-sens-mon-10-phi2}
\end{eqnarray*}
From Lemma~\ref{lem:p0dot}, the second term in \eqref{eqn:proofthm-sens-mon-10-phi1}
reduces to 
\begin{eqnarray*}
 & & \left.\E_{\theta}\left\{\frac{\dot{p}_{1,\theta}(X)-\dot{p}_{0,\theta}(X)}{1-\xi(X)}\mu_{11,\theta}(X)\right\}\right\rvert _{\theta=0}\nonumber \\
 & = & \E\left[\E\left\{\frac{\psi_{S_{1}}-\psi_{S_{0}}-p_{1}(X)+p_{0}(X)}{1-\xi(X)}\textup{s}(S\mid Z,X)\mu_{11}(X)\mid X\right\}\right]\nonumber \\
 & = & \E\left\{\frac{\psi_{S_{1}}-\psi_{S_{0}}-p_{1}(X)+p_{0}(X)}{1-\xi(X)}\mu_{11}(X)\textup{s}(S\mid Z,X)\right\}.\label{eqn:proofthm-sens-mon-10-phi3}
\end{eqnarray*}
From Lemma~\ref{lem:mudot}, the third term in \eqref{eqn:proofthm-sens-mon-10-phi1}
reduces to 
\begin{eqnarray*}
 & & \left.\E_{\theta}\left\{\frac{p_{1,\theta}(X)-p_{0,\theta}(X)}{1-\xi(X)}\dot{\mu}_{11,\theta}(X)\right\}\right\rvert _{\theta=0}\nonumber \\
 & = & \E\left\{\frac{p_{1}(X)-p_{0}(X)}{1-\xi(X)}\frac{\psi_{Y_{1}S_{1}}-\mu_{11}(X)\psi_{S_{1}}}{p_{1}(X)}\textup{s}(Y\mid Z,S,X)\right\}.\label{eqn:proofthm-sens-mon-10-phi4}
\end{eqnarray*}
Plugging the above three formulas
into~\eqref{eqn:proofthm-sens-mon-10-phi1} gives 
\begin{eqnarray}
\left.\dot{N}_{\theta}\right\rvert _{\theta=0} & = & \E\left[\left\{\frac{p_{1}(X)-p_{0}(X)}{1-\xi(X)}\mu_{11}(X)-N\right\}\textup{s}(X)\right]\nonumber \\
 & & +\E\left\{\frac{\psi_{S_{1}}-\psi_{S_{0}}-p_{1}(X)+p_{0}(X)}{1-\xi(X)}\mu_{11}(X)\textup{s}(S\mid Z,X)\right\}\nonumber \\
 & & +\E\left\{\frac{p_{1}(X)-p_{0}(X)}{1-\xi(X)}\frac{\psi_{Y_{1}S_{1}}-\mu_{11}(X)\psi_{S_{1}}}{p_{1}(X)}\textup{s}(Y\mid Z,S,X)\right\}.\label{eqn:proofthm-sens-mon-10-phi}
\end{eqnarray}
We can verify that 
\begin{eqnarray*}
\left\{\frac{p_{1}(X)-p_{0}(X)}{1-\xi(X)}\mu_{11}(X)-N\right\}& \in & H_{1},\label{eqn:SET-part1}\\
\frac{\psi_{S_{1}}-\psi_{S_{0}}-p_{1}(X)+p_{0}(X)}{1-\xi(X)}\mu_{11}(X) & \in & H_{3},\label{eqn:SET-part2}\\
\frac{p_{1}(X)-p_{0}(X)}{1-\xi(X)}\frac{\psi_{Y_{1}S_{1}}-\mu_{11}(X)\psi_{S_{1}}}{p_{1}(X)} & \in & H_{4}.\label{eqn:SET-part3}
\end{eqnarray*}
Because $H_{1}$, $H_{2}$, $H_{3}$, and $H_{4}$ are orthogonal to
each other, we can write~\eqref{eqn:proofthm-sens-mon-10-phi} as 
\begin{eqnarray*}
\left.\dot{N}_{\theta}\right\rvert _{\theta=0} & = & \E\left[\left\{\frac{p_{1}(X)-p_{0}(X)}{1-\xi(X)}\mu_{11}(X)-N\right\}\textup{s}(V)\right]\nonumber \\
 & & +\E\left\{\frac{\psi_{S_{1}}-\psi_{S_{0}}-p_{1}(X)+p_{0}(X)}{1-\xi(X)}\mu_{11}(X)\textup{s}(V)\right\}\nonumber \\
 & & +\E\left\{\frac{p_{1}(X)-p_{0}(X)}{1-\xi(X)}\frac{\psi_{Y_{1}S_{1}}-\mu_{11}(X)\psi_{S_{1}}}{p_{1}(X)}\textup{s}(V)\right\}.
\end{eqnarray*}
As a result, we obtain the EIF for $N$: 
\begin{eqnarray*}
\varphi_{N}(V) & = &\frac{p_{1}(X)-p_{0}(X)}{1-\xi(X)}\mu_{11}(X)-N+\frac{\psi_{S_{1}}-\psi_{S_{0}}-p_{1}(X)+p_{0}(X)}{1-\xi(X)}\mu_{11}(X)\\
 & & +\frac{p_{1}(X)-p_{0}(X)}{1-\xi(X)}\frac{\psi_{Y_{1}S_{1}}-\mu_{11}(X)\psi_{S_{1}}}{p_{1}(X)} \\
 & = & \frac{\psi_{S_{1}}-\psi_{S_{0}}}{1-\xi(X)}\mu_{11}(X)-N+\left\{ 1-\frac{p_{0}(X)}{p_{1}(X)}\right\} \frac{ \psi_{Y_{1}S_{1}}-\mu_{11}(X)\psi_{S_{1}}}{1-\xi(X)} \\
 & = & \frac{p_{1}(X)-p_0(X)}{\{1-\xi(X)\}p_{1}(X)}\psi_{Y_{1}S_{1}}-N-\frac{\mu_{11}(X)}{1-\xi(X)}\left\{ \psi_{S_{0}}-\frac{p_{0}(X)}{p_{1}(X)}\psi_{S_{1}}\right\} ,
\end{eqnarray*}
which equals $\phi^\ast_{1,10}-N$ (see Theorem~\ref{thm:SET-sens-mon} for the
expression of $\phi^\ast_{1,10}$). From Lemma \ref{lem:ratio}, the EIF
for $\mu_{1,10}$ is 
\begin{eqnarray*}
\varphi_{1,10}(V)=\frac{1}{e_{\xi,10}} \left\{\phi^\ast_{1,10} -\frac{\mu_{1,10}(\psi_{S_{1}}-\psi_{S_{0}})}{1-\xi(X)}\right\}.
\end{eqnarray*}
Similarly, the EIF for $\mu_{0,10}$ is 
\begin{eqnarray*}
\varphi_{0,10}(V)=\frac{1}{e_{\xi,10}} \left\{\phi^\ast_{0,10} -\frac{\mu_{0,10}(\psi_{S_{1}}-\psi_{S_{0}})}{1-\xi(X)}\right\} .
\end{eqnarray*}
Therefore, the EIF for $\tau_{10}$ is 
\begin{eqnarray*}
\frac{1}{e_{\xi,10} } \left\{ (\phi^\ast_{1,10}-\phi^\ast_{0,10}) - \frac{\tau_{10}(\psi_{S_{1}}-\psi_{S_{0}})}{1-\xi(X)} \right\}.
\end{eqnarray*}
\QEDB

\section{Proofs of the multiple robustness and local efficiency}
\label{app:triple}

In this section, we prove Theorems \ref{thm:triple},~\ref{thm:triple-1},~\ref{thm:sens-dr},~and~\ref{thm:sens-mon-tr}.

\subsection{Proof of Theorem \ref{thm:triple}}

We prove the triple robustness and semiparametric efficiency
for $\widehat{\tau}_{10}$. The proofs for $\widehat{\tau}_{00}$
and $\widehat{\tau}_{11}$ are similar and hence omitted.

\noindent {\bf Proof of the triple robustness.} 
As discussed in the main text, the proof of the triple robustness reduces to the calculation of the probability limit of $\mathbb{P}_n(\widehat{\phi}_{1,10})-\E\{Y_{1}\bm{1}(U=10)\}$.
From the equivalence relationship
in~\eqref{eqn:phi110}, $\mathbb{P}_n(\widehat{\phi}_{1,10})$ can be rewritten as 
\begin{eqnarray}
\mathbb{P}_{n}\left[e_{10}(X;\widehat{\gamma})\frac{S}{p_{1}(X;\widehat{\gamma})}\frac{Z}{\pi(X;\widehat{\alpha})}\{Y-\mu_{11}(X;\widehat{\beta})\}\right]+\mathbb{P}_n \left\{\widehat{\psi}_{\mu_{11}(X;\widehat{\beta})S_{1}}\right\}-\mathbb{P}_n \left\{\widehat{\psi}_{\mu_{11}(X;\widehat{\beta})S_{0}}\right\}.\label{eqn:effscore-1-10-est-1-app}
\end{eqnarray}
The first term in~\eqref{eqn:effscore-1-10-est-1-app} is consistent for 
\begin{eqnarray*}
 & & \E\left[e_{10}(X;\gamma^{*})\frac{S}{p_{1}(X;\gamma^{*})}\frac{Z}{\pi(X;\alpha^{*})}\{Y-\mu_{11}(X; \beta^{*})\}\right]\\
 & = & \E\left(\E\left[e_{10}(X;\gamma^{*})\frac{\pr(Z=1,S=1\mid X)}{p_{1}(X;\gamma^{*})\pi(X;\alpha^{*})}\{Y-\mu_{11}(X; \beta^{*})\}\mid Z=1,S=1,X\right]\right)\quad(\LOTE)\\
 & = & \E\left[\frac{ \{p_{1}(X;\gamma^{*})-p_{0}(X;\gamma^{*})\}p_1(X)\pi(X)}{p_{1}(X;\gamma^{*})\pi(X;\alpha^{*})}\{\mu_{11}(X)-\mu_{11}(X; \beta^{*})\} \right];
\end{eqnarray*}
the second term in~\eqref{eqn:effscore-1-10-est-1-app} is consistent for 
\begin{eqnarray*}
&& \E \left[ \frac{ \{ \mu_{11}(X;\beta^{*})S_1-\mu_{11}(X;\beta^{*})p_1(X;\gamma^{*})\}\bm{1}(Z=1) }{\pi(X;\alpha^{*})} +\mu_{11}(X;\beta^{*})p_1(X;\gamma^{*})\right]\\
&=& \E \left[ \frac{ \{ \mu_{11}(X;\beta^{*})p_1(X)-\mu_{11}(X;\beta^{*})p_1(X;\gamma^{*})\}\pi(X) }{\pi(X;\alpha^{*})} +\mu_{11}(X;\beta^{*})p_1(X;\gamma^{*})\right];\quad(\LOTE)
\end{eqnarray*}
similarly, the third term in~\eqref{eqn:effscore-1-10-est-1-app} is consistent for 
\begin{eqnarray*}
&& \E \left[ \frac{ \{ \mu_{11}(X;\beta^{*})S_0-\mu_{11}(X;\beta^{*})p_0(X;\gamma^{*})\}\bm{1}(Z=0) }{1-\pi(X;\alpha^{*})} +\mu_{11}(X;\beta^{*})p_0(X;\gamma^{*})\right]\\
&=& \E \left[ \frac{ \{ \mu_{11}(X;\beta^{*})p_0(X)-\mu_{11}(X;\beta^{*})p_0(X;\gamma^{*})\}\{1-\pi(X)\} }{1-\pi(X;\alpha^{*})} +\mu_{11}(X;\beta^{*})p_0(X;\gamma^{*})\right].\quad(\LOTE)
\end{eqnarray*}

For the ease of disposition, we suppress the dependence of the functions on $X$ and write $(\pi(X),\pi(X;\alpha^*))$ as $(\pi,\pi^*)$, $(p_z(X), p_z(X;\gamma^*))$ as $(p_z,p_z^*)$, and 
$(\mu_{11}(X),\mu_{11}(X;\beta^*))$ as $(\mu_{11},\mu_{11}^*)$. Therefore, $\mathbb{P}_n(\widehat{\phi}_{1,10})-\E\{Y_{1}\bm{1}(U=10)\}$ is consistent for
\begin{eqnarray}
\nonumber&&\E\left\{\frac{\pi(p_1^*-p_0^*)p_1(\mu_{11}-\mu_{11}^*)}{\pi^* p_1^*}\right\}+\E\left\{\frac{(\mu_{11}^*p_1 -\mu_{11}^*p_1^*)\pi }{\pi^*}+\mu_{11}^*p_1^*\right\}\\
\nonumber&&-\E\left\{\frac{(\mu_{11}^*p_0 -\mu_{11}^*p_0^*)(1-\pi) }{1-\pi^*}+\mu_{11}^*p_0^*\right\} - \mu_{11}(p_1-p_0)\\
\nonumber&=& \E\left\{\frac{\pi p_1(\mu_{11}-\mu_{11}^*)}{\pi^*} +\frac{(\mu_{11}^*p_1 -\mu_{11}^*p_1^*)\pi }{\pi^*}+\mu_{11}^*p_1^*-\mu_{11}p_1 \right\}\\
\label{eqn:tr-bias1} &&- \E\left\{ \frac{\pi p_0^*p_1(\mu_{11}-\mu_{11}^*)}{\pi^* p_1^*} +\frac{(\mu_{11}^*p_0 -\mu_{11}^*p_0^*)(1-\pi) }{1-\pi^*}+\mu_{11}^*p_0^*-\mu_{11}p_0 \right\}.
\end{eqnarray}
We simplify the terms in the two expectations of~\eqref{eqn:tr-bias1} separately. The first term reduces to
\begin{eqnarray*}
\frac{\pi p_1(\mu_{11}-\mu_{11}^*)}{\pi^*} +\frac{(\mu_{11}^*p_1 -\mu_{11}^*p_1^*)\pi }{\pi^*}+\mu_{11}^*p_1^*-\mu_{11}p_1
&=&\frac{\pi p_1\mu_{11}}{\pi^*} -\frac{ \mu_{11}^*p_1^*\pi }{\pi^*}+\mu_{11}^*p_1^*-\mu_{11}p_1\\
\label{eqn:tr-bias2}&=& \frac{(\mu_{11}p_1 - \mu_{11}^*p_1^*)(\pi - \pi^*)}{\pi^*}.
\end{eqnarray*}
The second term reduces to 
\begin{eqnarray*}
\nonumber&& \frac{\pi p_0^*p_1(\mu_{11}-\mu_{11}^*)}{\pi^* p_1^*} +\frac{(\mu_{11}^*p_0 -\mu_{11}^*p_0^*)(1-\pi) }{1-\pi^*}+\mu_{11}^*p_0^*-\mu_{11}p_0 \\
\nonumber&=& \frac{(1-\pi^*) \pi p_0^*p_1(\mu_{11}-\mu_{11}^*)+(1-\pi)\pi^* p_1^*(\mu_{11}^*p_0 -\mu_{11}^*p_0^*) +\pi^* p_1^*(1-\pi^*) (\mu_{11}^*p_0^*-\mu_{11}p_0) }{\pi^* p_1^*(1-\pi^*) }\\
\nonumber&=& \frac{\pi p_0^*p_1(\mu_{11}-\mu_{11}^*)+\pi^* p_1^*(\mu_{11}^*p_0 -\mu_{11}^*p_0^*)+\pi^* p_1^* (\mu_{11}^*p_0^*-\mu_{11}p_0) }{\pi^* p_1^*(1-\pi^*) }\\
\nonumber&&- \frac{\pi p_0^*p_1(\mu_{11}-\mu_{11}^*)+\pi p_1^*(\mu_{11}^*p_0 -\mu_{11}^*p_0^*)+\pi^{*} p_1^* (\mu_{11}^*p_0^*-\mu_{11}p_0) }{ p_1^*(1-\pi^*) }\\
\nonumber&=& \frac{\pi p_0^*p_1(\mu_{11}-\mu_{11}^*)+\pi^* p_1^*(\mu_{11}^*p_0 -\mu_{11}p_0)}{\pi^* p_1^*(1-\pi^*) }\\
\nonumber&&- \frac{ (\pi p_0^*p_1-\pi^* p_0p_1^*) \mu_{11} +(\pi p_0 p_1^*-\pi p_0^* p_1^*-\pi p_0^* p_1+\pi^* p_0^* p_1^*) \mu_{11}^* }{ p_1^*(1-\pi^*) }\\
\nonumber&=& \frac{(\pi p_0^*p_1-\pi^* p_0p_1^*)(\mu_{11}-\mu_{11}^*)}{\pi^* p_1^*(1-\pi^*) } -\frac{ (\pi p_0^*p_1-\pi^* p_0p_1^*) \mu_{11}-(\pi p_0^*p_1-\pi^* p_0p_1^*) \mu_{11}^* }{ p_1^*(1-\pi^*) }\\
\nonumber&&-\frac{ (\pi p_0 p_1^*-\pi p_0^* p_1^*-\pi^* p_0 p_1^*+\pi^* p_0^* p_1^*) \mu_{11}^* }{ p_1^*(1-\pi^*) }\\
\label{eqn:tr-bias3}&=&\frac{(\pi p_0^*p_1-\pi^* p_0p_1^*)(\mu_{11}-\mu_{11}^*)}{\pi^* p_1^* } - \frac{(\pi-\pi^*)(p_0-p_0^*)\mu_{11}^*}{1-\pi^*}.
\end{eqnarray*}
Combining them yields the bias formulas in the main text. 

\noindent \textbf{Proof of the semiparametric efficiency.} Under $\Mompsta$,
we show that the influence function of $\widehat{\tau}_{10}$ is the
same as the EIF in Theorem \ref{thm:SET} and therefore $\widehat{\tau}_{10}$
achieves the local efficiency. Let $\theta$ denote the parameter
vector containing $\alpha,\beta$, and $\gamma$. Let $\theta^{*}$
be the probability limit of $\widehat{\theta}$. We first establish
that for a ratio estimator $\text{\ensuremath{\mathbb{P}_{n}N(V;\widehat{\theta})}}/\mathbb{P}_{n}D(V;\widehat{\theta})$,
by a Taylor expansion, 
\begin{equation}
\frac{\mathbb{P}_{n}N(V;\widehat{\theta})}{\mathbb{P}_{n}D(V;\widehat{\theta})}\ =\ \frac{\mathbb{P}_{n}N(V;\widehat{\theta})}{\pr\{D(V;\theta^{*})\}}-\frac{\pr\{N(V;\theta^{*})\}}{\text{[}\pr\{D(V;\theta^{*})\}]^{2}}\left[\mathbb{P}_{n}D(V;\widehat{\theta})-\pr\{D(V;\theta^{*})\}\right]+o_{\pr}(n^{-1/2}).\label{eq:tylor-ratio-1}
\end{equation}
For $\widehat{\tau}_{10}=\mathbb{P}_{n}(\widehat{\phi}_{1,10}-\widehat{\phi}_{0,10})/\mathbb{P}_{n}(\widehat{\psi}_{S_{1}}-\widehat{\psi}_{S_{0}})$,
we have $N(V;\theta)=\phi_{1,10}(\theta)-\phi_{0,10}(\theta)$ and
$D(V;\theta)=\psi_{S_{1}}(\theta)-\psi_{S_{0}}(\theta)$, where 
\begin{eqnarray*}
	\phi_{1,10}(\theta) & = & \frac{e_{10}(X;\gamma)}{p_{1}(X;\gamma)}\frac{Z}{\pi(X;\alpha)}\{Y-\mu_{11}(X;\beta)\}S+\psi_{\mu_{11}(X;\beta)S_{1}}-\psi_{\mu_{11}(X;\beta)S_{0}},\\
	\phi_{0,10}(\theta) & = & \frac{e_{10}(X;\gamma)}{1-p_{0}(X;\gamma)}\frac{1-Z}{1-\pi(X;\alpha)}\{Y-\mu_{00}(X;\beta)\}(1-S)+\psi_{\mu_{00}(X;\beta)(1-S_{0})}-\psi_{\mu_{00}(X;\beta)(1-S_{1})},\\
	\psi_{S_{z}}(\theta) & = & \frac{\{S-p_{z}(X;\gamma)\}\bm{1}(Z=z)}{\pr(Z=z\mid X;\alpha)}+p_{z}(X;\gamma)\ (z=0,1).
\end{eqnarray*}
Under $\Mompsta$, $\phi_{1,10}(\theta^{*})=\phi_{1,10}(V)$,
$\phi_{0,10}(\theta^{*})=\phi_{0,10}(V)$, $\psi_{S_{z}}(\theta^{*})=\psi_{S_{z}}$
$(z=0,1)$, $\pr\{N(V;\theta^{*})\}=\tau_{10}(p_{1}-p_{0}),$ and
$\pr\{D(V;\theta^{*})\}=p_{1}-p_{0}$ . By Taylor expansions, we have
\begin{eqnarray}
\mathbb{P}_{n}N(V;\widehat{\theta}) & = & \mathbb{P}_{n}N(V;\theta^{*})+\pr\left\{ \dot{N}(V;\theta^{*})\right\} (\widehat{\theta}-\theta)+o_{\pr}(n^{-1/2}),\label{eq:pnN}\\
\mathbb{P}_{n}D(V;\widehat{\theta}) & = & \mathbb{P}_{n}D(V;\theta^{*})+\pr\left\{ \dot{D}(V;\theta^{*})\right\} (\widehat{\theta}-\theta)+o_{\pr}(n^{-1/2}).\label{eq:pnD}
\end{eqnarray}
By some algebra, we can verify that $\pr\left\{ \dot{N}(V;\theta^{*})\right\} =\pr\left\{ \dot{D}(V;\theta^{*})\right\} =0$.
Note that (\ref{eq:pnN}) and (\ref{eq:pnD}) hold under the sufficient
conditions that (i) the propensity score and its estimator are bounded
away from zero and one and (ii) the principal scores $p_{1}(X;\gamma^{*})$
and $1-p_{0}(X;\gamma^{*})$ and their estimators are bounded away
from zero almost surely. When condition (i) is violated, $\mathbb{P}_{n}N(V;\widehat{\theta})$
and $\mathbb{P}_{n}D(V;\widehat{\theta})$ can be unbounded. When
condition (ii) is violated, $\pr\left\{ \dot{N}(V;\theta^{*})\right\} $
and $\pr\left\{ \dot{D}(V;\theta^{*})\right\} $ may be undefined. 
Combining \eqref{eq:tylor-ratio-1}, (\ref{eq:pnN}), and
(\ref{eq:pnD}), we obtain 
\begin{eqnarray*}
	\hat{\tau}_{10}-\tau_{10} & = & \frac{\mathbb{P}_{n}N(V;\widehat{\theta})}{\mathbb{P}_{n}D(V;\widehat{\theta})}-\tau_{10}\\
	& = & \frac{\mathbb{P}_{n}N(V;\widehat{\theta})}{\pr\{D(V;\theta^{*})\}}-\frac{\pr\{N(V;\theta^{*})\}}{\text{[}\pr\{D(V;\theta^{*})\}]^{2}}\left[\mathbb{P}_{n}D(V;\widehat{\theta})-\pr\{D(V;\theta^{*})\}\right]-\tau_{10}+o_{\pr}(n^{-1/2})\\
	& = & \frac{\mathbb{P}_{n}N(V;\theta^{*})}{\pr\{D(V;\theta^{*})\}}-\frac{\pr\{N(V;\theta^{*})\}}{\text{[}\pr\{D(V;\theta^{*})\}]^{2}}\left[\mathbb{P}_{n}D(V;\theta^{*})-\pr\{D(V;\theta^{*})\}\right]-\tau_{10}+o_{\pr}(n^{-1/2})\\
	& = & \frac{\mathbb{P}_{n}N(V;\theta^{*})}{\pr\{D(V;\theta^{*})\}}-\frac{\pr\{N(V;\theta^{*})\}}{\text{[}\pr\{D(V;\theta^{*})\}]^{2}}\left[\mathbb{P}_{n}D(V;\theta^{*})-\pr\{D(V;\theta^{*})\}\right]-\tau_{10}+o_{\pr}(n^{-1/2})\\
	& = & \frac{\mathbb{P}_{n}N(V;\theta^{*})}{\pr\{D(V;\theta^{*})\}}-\tau_{10}\frac{\mathbb{P}_{n}D(V;\theta^{*})}{\pr\{D(V;\theta^{*})\}}+o_{\pr}(n^{-1/2})\\
	& = & \frac{\mathbb{P}_{n}(\phi_{1,10}-\phi_{0,10})}{p_{1}-p_{0}}-\tau_{10}\frac{\mathbb{P}_{n}(\psi_{S_{1}}-\psi_{S_{0}})}{p_{1}-p_{0}}+o_{\pr}(n^{-1/2})\\
	& = & \mathbb{P}_{n}\frac{\phi_{1,10}-\phi_{0,10}-\tau_{10}(\psi_{S_{1}}-\psi_{S_{0}})}{p_{1}-p_{0}}+o_{\pr}(n^{-1/2})\\
	& = & \text{\ensuremath{\mathbb{P}_{n}}}\phi_{10}+o_{\pr}(n^{-1/2}).
\end{eqnarray*}
This completes the proof. \QEDB

\subsection{Proof of Theorem \ref{thm:triple-1} }

We show that $\widehat{\tau}_{10}$ using $\{\widehat{\pi}(x),\widehat{p}_{z}(x),\widehat{\mu}_{zs}(x)\}$
satisfying the regularity conditions (a)--(d) in Theorem \ref{thm:triple-1}
is asymptotically normal and has the influence function $\phi_{10}$
as in Theorem \ref{thm:SET}, therefore achieving the semiparametric
efficiency. The proofs for $\widehat{\tau}_{00}$ and $\widehat{\tau}_{11}$
are similar and hence omitted. Let $\theta$ denote the nuisance functions
$\{\pi(x),p_{z}(x),\mu_{zs}(x)\}$, abbreviated to $(\pi,p_{z},\mu_{zs})$
for simplicity. Let $\theta^{*}$ be the probability limit of $\widehat{\theta}$. 

For $\widehat{\tau}_{10}=\mathbb{P}_{n}(\widehat{\phi}_{1,10}-\widehat{\phi}_{0,10})/\mathbb{P}_{n}(\widehat{\psi}_{S_{1}}-\widehat{\psi}_{S_{0}})$,
it is a ratio estimator $\text{\ensuremath{\mathbb{P}_{n}N(V;\widehat{\theta})}}/\mathbb{P}_{n}D(V;\widehat{\theta})$
where $N(V;\theta)=\phi_{1,10}(\theta)-\phi_{0,10}(\theta)$ and $D(V;\theta)=\psi_{S_{1}}(\theta)-\psi_{S_{0}}(\theta)$ with
\begin{eqnarray*}
	\phi_{1,10}(\theta) & = & \frac{e_{10}(X)}{p_{1}(X)}\frac{Z}{\pi(X)}\{Y-\mu_{11}(X)\}S+\psi_{\mu_{11}(X)S_{1}}-\psi_{\mu_{11}(X)S_{0}},\\
	\phi_{0,10}(\theta) & = & \frac{e_{10}(X)}{1-p_{0}(X)}\frac{1-Z}{1-\pi(X)}\{Y-\mu_{00}(X)\}(1-S)+\psi_{\mu_{00}(X)(1-S_{0})}-\psi_{\mu_{00}(X)(1-S_{1})},\\
	\psi_{S_{z}}(\theta) & = & \frac{\{S-p_{z}(X)\}\bm{1}(Z=z)}{\pr(Z=z\mid X)}+p_{z}(X). 
\end{eqnarray*}
We continue with the Taylor expansion of a ratio estimator as in (\ref{eq:tylor-ratio-1}).
Condition (a) implies that $\theta^{*}=\{\pi(x),p_{z}(x),\mu_{zs}(x)\}$
and thus $\phi_{1,10}(\theta^{*})=\phi_{1,10}(V)$, $\phi_{0,10}(\theta^{*})=\phi_{0,10}(V)$,
and $\psi_{S_{z}}(\theta^{*})=\psi_{S_{z}}$ $(z=0,1)$. By the empirical
process theory, we have
\begin{eqnarray}
\mathbb{P}_{n}N(V;\widehat{\theta})-\pr N(V;\theta^{*}) & = & (\mathbb{P}_{n}-\pr)N(V;\widehat{\theta})+\pr\{N(V;\widehat{\theta})-N(V;\theta^{*})\}\nonumber \\
& = & (\mathbb{P}_{n}-\pr)N(V;\theta^{*})+\pr\{N(V;\widehat{\theta})-N(V;\theta^{*})\}+o_{\pr}(n^{-1/2}),\label{eq:N-expansion}
\end{eqnarray}
where the second equality follows by Condition (b). It remains to
analyze the second term $\pr\{N(V;\widehat{\theta})-N(V;\theta^{*})\}.$
Following the derivation of the bias formula, we have 
\begin{eqnarray*}
	|\pr\{N(V;\widehat{\theta})-N(V;\theta^{*})\}| & = & \left\vert \pr\left[\frac{\{\mu_{11}(X)p_{1}(X)-\widehat{\mu}_{11}(X)\widehat{p}_{1}(X)\}\{\pi(X)-\widehat{\pi}(X)\}}{\widehat{\pi}(X)}\right]\right.\\
	& & +\pr\left[\frac{\{\pi(X)\widehat{p}_{0}(X)p_{1}(X)-\widehat{\pi}(X)p_{0}(X)\widehat{p}_{1}(X)\}\{\mu_{11}(X)-\widehat{\mu}_{11}(X)\}}{\widehat{\pi}(X)\widehat{p}_{1}(X)}\right]\\
	& & \left.+\pr\left[\frac{\{\pi(X)-\widehat{\pi}(X)\}\{p_{0}(X)-\widehat{p}_{0}(X)\}\widehat{\mu}_{11}(X)}{1-\widehat{\pi}(X)}\right]\right\vert \\
	& \leq & \left\vert \pr\left[\frac{\{\mu_{11}(X)-\widehat{\mu}_{11}(X)\}p_{1}(X)\{\pi(X)-\widehat{\pi}(X)\}}{\widehat{\pi}(X)}\right]\right\vert \\
	& & +\left\vert \pr\left[\frac{\{p_{1}(X)-\widehat{p}_{1}(X)\}\widehat{\mu}_{11}(X)\{\pi(X)-\widehat{\pi}(X)\}}{\widehat{\pi}(X)}\right]\right\vert \\
	& & +\left\vert \pr\left[\frac{\{\widehat{p}_{0}(X)-p_{0}(X)\}\pi(X)p_{1}(X)\{\mu_{11}(X)-\widehat{\mu}_{11}(X)\}}{\widehat{\pi}(X)\widehat{p}_{1}(X)}\right]\right\vert \\
	& & +\left\vert \pr\left[\frac{\{\pi(X)-\widehat{\pi}(X)\}p_{1}(X)p_{0}(X)\{\mu_{11}(X)-\widehat{\mu}_{11}(X)\}}{\widehat{\pi}(X)\widehat{p}_{1}(X)}\right]\right\vert \\
	& & +\left\vert \pr\left[\frac{\{p_{1}(X)-\widehat{p}_{1}(X)\}\widehat{\pi}(X)p_{0}(X)\{\mu_{11}(X)-\widehat{\mu}_{11}(X)\}}{\widehat{\pi}(X)\widehat{p}_{1}(X)}\right]\right\vert \\
	& & +\left\vert \pr\left[\frac{\{\pi(X)-\widehat{\pi}(X)\}\{p_{0}(X)-\widehat{p}_{0}(X)\}\widehat{\mu}_{11}(X)}{1-\widehat{\pi}(X)}\right]\right\vert .
\end{eqnarray*}
By the Cauchy--Schwarz inequality and Conditions (a), (c) and (d),
it follows that for some constant $C$, we have 
\begin{eqnarray*}
	|\pr\{N(V;\widehat{\theta})-N(V;\theta^{*})\}| & \leq & C\times||\mu_{11}(X)-\widehat{\mu}_{11}(X)||_{2}\times\{||\pi(X)-\widehat{\pi}(X)||_{2}\\
	& & +||p_{1}(X)-\widehat{p}_{1}(X)||_{2}+||p_{0}(X)-\widehat{p}_{0}(X)||_{2}\}\\
	& & +C\times||\pi(X)-\widehat{\pi}(X)||_{2}\times\left\{ ||p_{1}(X)-\widehat{p}_{1}(X)||_{2}+||p_{0}(X)-\widehat{p}_{0}(X)||_{2}\right\} \\
	& = & o_{\pr}(n^{-1/2}).
\end{eqnarray*}
Continuing with (\ref{eq:N-expansion}), we have $\mathbb{P}_{n}N(V;\widehat{\theta})-\pr N(V;\theta^{*})=(\mathbb{P}_{n}-\pr)N(V;\theta^{*})+o_{\pr}(n^{-1/2}).$
Similarly, under Conditions (a)--(d), $\mathbb{P}_{n}D(V;\widehat{\theta})-\pr D(V;\theta^{*})=(\mathbb{P}_{n}-\pr)D(V;\theta^{*})+o_{\pr}(n^{-1/2}).$
Plugging these into~\eqref{eq:tylor-ratio-1}, we obtain $\hat{\tau}_{10}-\tau_{10}=\text{\ensuremath{\mathbb{P}_{n}}}\phi_{10}+o_{\pr}(n^{-1/2})$. This completes the proof. \QEDB

\subsection{Proof of Theorem \ref{thm:sens-dr}}

\label{app:sens-double} 
We prove the double robustness
for $\widehat{\tau}'_{10}$. The proofs for $\widehat{\tau}'_{00}$
and $\widehat{\tau}'_{11}$ are similar and hence omitted. 

We write $\widehat{\tau}'_{10}= \mathbb{P}_n(\widehat{\phi}'_{1,10}-\widehat{\phi}'_{0,10})/\mathbb{P}_n(\widehat{\psi}_{S_{1}}-\widehat{\psi}_{S_{0}})$.
The denominator is consistent for $\E(S_{1}-S_{0})=\pr(U=10)$ under
$\Mps$, which is a superset of $\Mpsta\cup\Momps$.
For the numerator, we will show that under $\Mps$, $\mathbb{P}_n(\widehat{\phi}_{1,10})-\E\{Y_{1}\bm{1}(U=10)\}$ has the probability limit 
\begin{eqnarray}
\label{eqn:sens-dr}
 \E \left[ \frac{ \omega_{1,10}(X)e_{10}(X)\{ \mu_{11}(X)- \mu_{11}(X;\beta^*)\}\{\pi(X)-\pi(X;\alpha^*)}{\pi(X;\alpha^*)}\right],
\end{eqnarray}
which equals $0$ if $\Mta \cup \Mom$ further holds. As a result, $\mathbb{P}_n(\widehat{\phi}'_{1,10})$ is consistent
for $\E\{Y_{1}\bm{1}(U=10)\}$ under $\Mpsta \cup\Momps$.
Similarly, we can show $\mathbb{P}_n(\widehat{\phi}'_{0,10})$ is consistent for
$\E\{Y_{0}\bm{1}(U=10)\}$ under $\Mpsta\cup\Momps$, which
leads to the double robustness of $\widehat{\tau}'_{10}$. Below we prove~\eqref{eqn:sens-dr}.

We can write $\mathbb{P}_n(\widehat{\phi}'_{1,10})$ as
\begin{eqnarray}
\label{eqn:sens-dr1}
\mathbb{P}_n\left\{\frac{\omega_{1,10}(X;\widehat{\gamma})e_{10}(X;\widehat{\gamma})}{p_{1}(X;\widehat{\gamma})}\widehat{\psi}_{Y_{1}S_{1}}\right\}-\mathbb{P}_n\left[\frac{\omega^2_{1,10}(X;\widehat{\gamma})\mu_{11}(X;\widehat{\beta})}{\epsilon_1(X)}\left\{ \widehat{\psi}_{S_{0}}-\frac{p_{0}(X;\widehat{\gamma})}{p_{1}(X;\widehat{\gamma})}\widehat{\psi}_{S_{1}}\right\}\right].
\end{eqnarray}
The first term of~\eqref{eqn:sens-dr1} is consistent for 
\begin{eqnarray*}
\nonumber &&\E \left( \frac{\omega_{1,10}(X)e_{10}(X)}{p_{1}(X)} \left[ \frac{\{ \mu_{11}(X)p_1(X)- \mu_{11}(X;\beta^*)p_1(X)\}\pi(X)}{\pi(X;\alpha^*)}+ \mu_{11}(X;\beta^*)p_1(X) \right]\right)\\
\label{eqn:sens-dr2}&=& \E \left( \omega_{1,10}(X)e_{10}(X) \left[ \frac{\{ \mu_{11}(X)- \mu_{11}(X;\beta^*)\}\pi(X)}{\pi(X;\alpha^*)}+ \mu_{11}(X;\beta^*)\right]\right).
\end{eqnarray*}
The second term of~\eqref{eqn:sens-dr1} is consistent for 
\begin{eqnarray*}
\nonumber&&\E\left(\frac{\omega^2_{1,10}(X)\mu_{11}(X;\beta^*)}{\epsilon_1(X)}\left[ \frac{\{S_0-p_0(X)\}\bm{1}(Z=0)}{1-\pi(X;\alpha^*)}+p_0(X) \right]\right)\\
\nonumber&&+-\E\left(\frac{\omega^2_{1,10}(X)\mu_{11}(X;\beta^*)p_{0}(X)}{\epsilon_1(X)p_{1}(X)} \left[ \frac{\{S_1-p_1(X)\}\bm{1}(Z=1)}{\pi(X;\alpha^*)}+p_1(X) \right] \right)\\
\nonumber&=&\E\left\{\frac{\omega^2_{1,10}(X)\mu_{11}(X;\beta^*) p_0(X)}{\epsilon_1(X)}\right\}-\E\left\{\frac{\omega^2_{1,10}(X)\mu_{11}(X;\beta^*) p_{0}(X)p_1(X)}{\epsilon_1(X)p_{1}(X)} \right\} \quad(\LOTE)\\
\label{eqn:sens-dr3} &=& 0 .
\end{eqnarray*}
These two terms, coupled with the fact that $\E\{Y_{1} \bm{1}(U=10)\} = \E\{\omega_{1,10}(X)\mu_{11}(X)e_{10}(X) \}$ from \eqref{eq:y1-10}, imply \eqref{eqn:sens-dr}. 

The proof of the semiparametric efficiency is similar to that of Theorem~\ref{thm:triple} and hence omitted. \QEDB

\subsection{Proof of Theorem \ref{thm:sens-mon-tr}}

\label{app:sens-mon-triple} 
We prove the double robustness
for $\widehat{\tau}^\ast_{10}$. The proofs for the other three estimators are similar and hence omitted. 

We write 
$$\widehat{\tau}^\ast_{10}= \mathbb{P}_n(\widehat{\phi}^\ast_{1,10}-\widehat{\phi}^\ast_{0,10})/\mathbb{P}_n\left( \frac{\widehat{\psi}_{S_{1}}-\widehat{\psi}_{S_{0}}}{1-\xi(X)} \right).$$
We can show that the denominator has the probability limit 
\begin{eqnarray*}
\E \left( \frac{  [\{p_{1}(X;\gamma^*)-p_1(X)\} - \{p_{0}(X;\gamma^*)-p_0(X)\}] \{\pi(X;\alpha^*)- \pi(X)\}  }{\{1-\xi(X)\}\pi(X;\alpha^*)}\right) + \E\left\{ \frac{p_1(X)-p_0(X)}{1-\xi(X)}\right\}.
\end{eqnarray*}
The first term is equal to zero under $\Mta \cup \Mps$ and the second term is equal to $\mathbb{P}(U=10)$ under~\eqref{eq:sensitivity-mon}. 
Therefore, the denominator is consistent for $\mathbb{P}(U=10)$ under $\Mta \cup \Mps$, which is a superset of $\Mpsta\cup\Momta\cup\Momps$.

From the derivation of the probability limit of the denominator, we see that the only difference between the derivation here and that in Theorem~\ref{thm:triple} is the additional term $1-\xi(X)$ in the expectation. Because $\xi(X)$ is known, this does not change the formulas in the proof of Theorem~\ref{thm:triple} other than adding $1-\xi(X)$ in the denominators. Therefore, we can show that 
$\mathbb{P}_n(\widehat{\phi}^\ast_{1,10})-\E\{Y_{1}\bm{1}(U=10)\}$ is consistent for
\begin{eqnarray*}
 \E\left\{\frac{(\mu_{11}p_1 - \mu_{11}^*p_1^*)(\pi - \pi^*)}{(1-\xi)\pi^*} -\frac{(\pi p_0^*p_1-\pi^* p_0p_1^*)(\mu_{11}-\mu_{11}^*)}{(1-\xi)\pi^* p_1^* } + \frac{(\pi-\pi^*)(p_0-p_0^*)\mu_{11}^*}{(1-\xi)(1-\pi^*)} \right\},
\end{eqnarray*}
where we suppress the dependence of the functions on $X$ for the ease of disposition. As a result, $\mathbb{P}_n(\widehat{\phi}^\ast_{1,10})$ is consistent
for $\E\{Y_{1}\bm{1}(U=10)\}$ under $\Mpsta\cup\Momta\cup\Momps$.
Similarly, we can show $\mathbb{P}_n(\widehat{\phi}^\ast_{0,10})$ is consistent for
$\E\{Y_{0}\bm{1}(U=10)\}$ under $\Mpsta\cup\Momta\cup\Momps$, which
leads to the triple robustness of $\widehat{\tau}^\ast_{10}$.


The proof of the semiparametric efficiency is similar to that of Theorem~\ref{thm:triple} and hence omitted. \QEDB

\end{document}